\documentclass[onecolumn, sort&compress, numbers]{els-mrw} 

\usepackage{amsmath,amssymb,amsfonts,amsthm,makeidx,graphicx}
\usepackage{txfonts}
\usepackage{helvet}

\usepackage{slashed} 
\usepackage{hyperref}

\usepackage{xcolor}

\begin{document}


\chapter{Flavour Physics Beyond the Standard Model}\label{chap1}

\author[1]{Wolfgang Altmannshofer}%
\author[2]{Peter Stangl}%

\address[1]{\orgname{University of California Santa Cruz}, \orgdiv{Department of Physics and Santa Cruz Institute for Particle Physics}, \orgaddress{Santa Cruz, CA 95064, United States}}
\address[2]{\orgname{CERN}, \orgdiv{Theoretical Physics Department}, \orgaddress{CH-1211 Geneva 23, Switzerland}}

\articletag{CERN-TH-2025-156}

\maketitle

\begin{abstract}[Abstract]
Flavour physics plays a central role in the search for physics beyond the Standard Model, posing fundamental questions whose answers may point to new physics scales far above the electroweak scale. The flavour structure of the Standard Model is strikingly hierarchical, hinting at underlying organizing principles yet to be uncovered. At the same time, flavour-changing processes rank among the most sensitive probes of new physics. In this chapter, we review how precision measurements of flavour-violating observables, in particular meson-antimeson mixing and rare decays, provide powerful indirect tests of new physics, constraining scenarios from heavy mediators to light, weakly coupled particles. We also survey theoretical frameworks proposed to explain the origin of the Standard Model flavour hierarchies. Together, these complementary aspects demonstrate how flavour physics continues to shape both experimental strategies and theoretical developments in the quest to understand the fundamental structure of particle interactions.
\end{abstract}

\begin{keywords}
Flavour Physics \sep Flavour Violation \sep Flavour-Changing Neutral Currents \sep Meson Mixing \sep Rare Decays \sep Flavour Puzzle \sep Flavour Symmetries \sep Beyond the Standard Model  \sep New Physics \sep Froggatt-Nielsen \sep Clockwork \sep Supersymmetry \sep Composite Higgs \sep Warped Extra-Dimensions
\end{keywords}


\begin{glossary}[Nomenclature]
	\begin{tabular}{@{}lp{34pc}@{}}
        2HDM & Two Higgs Doublet Model \\
        3HDM & Three Higgs Doublet Model \\
        4D & Four dimensional \\
        5D & Five dimensional\\
        ALP & Axion-Like Particle \\
        BSM & Beyond the Standard Model of particle physics \\
        CEPC & Circular Electron Positron Collider \\
        CKM & Cabibbo-Kobayashi-Maskawa \\
        CP & Charge conjugation and Parity \\
        EFT & Effective Field Theory \\
        FCC & Future Circular Collider \\
        FCNC & Flavour-Changing Neutral Current \\
        FN & Froggatt-Nielsen \\
        GIM & Glashow-Iliopoulos-Maiani \\
        IR & Infrared \\
        KK & Kaluza-Klein  \\
        LEFT & Low-Energy Effective Field Theory \\
		LHC & Large Hadron Collider \\
		MFV & Minimal Flavour Violation \\
        MSSM & Minimal Supersymmetric Standard Model \\
        NP & New Physics \\
        PDG & Particle Data Group \\
        PMNS & Pontecorvo-Maki-Nakagawa-Sakata \\
        RS & Randall-Sundrum \\
        SM & Standard Model of particle physics \\
        SMEFT & Standard Model Effective Field Theory\\
        SUSY & Supersymmetry \\
        SVD & Singular Value Decomposition \\
        UV & Ultraviolet \\
        VL & Vector-like \\
        WET & Weak Effective Theory
	\end{tabular}
\end{glossary}

\section*{Objectives}
In this chapter, the reader will:
\begin{itemize}
\item Understand the non-trivial flavour structure of the Standard Model, including the pronounced hierarchies in fermion masses and Cabibbo-Kobayashi-Maskawa mixing angles. While technically natural within the Standard Model, these patterns remain unexplained and suggest the presence of a deeper organizing principle.
\item Learn how flavour-changing processes, in particular rare decays and neutral meson mixing, serve as exceptionally sensitive probes of new physics. Owing to their strong suppression in the Standard Model, these processes allow indirect tests of new heavy particles at energy scales well beyond the reach of direct searches, providing powerful constraints on the flavour structure of possible Standard Model extensions and guiding new physics model building efforts.
\item Gain an overview of proposed solutions to the Standard Model flavour puzzle. The chapter will review theoretical approaches that aim to explain the observed fermion mass and mixing hierarchies, including frameworks based on flavour symmetries as well as dynamical mechanisms.
\end{itemize}



\section{Introduction}\label{intro}


Flavour is a peculiar feature of the Standard Model (SM) of particle physics that has its origin in the fact that the matter fields -- the SM fermions -- can be grouped into sets of similar particles that share many of their properties.
The two main sets are the quarks -- the fermions charged under the strong interaction -- and the leptons -- the fermions neutral to the strong interaction.
Each of these two sets comprises six types of particles, known as the \emph{flavours} of quarks and leptons.
The six quark flavours -- up, down, strange, charm, bottom, and top -- can be further divided into two sets of particles with the same electric charge: the three up-type quarks with electric charge $+2/3$: up, charm, and top, and the three down-type quarks with electric charge $-1/3$: down, strange, and bottom.
Similarly, the six leptons are divided into the three charged leptons with electric charge $-1$: electron, muon, and tau, and the three electrically neutral neutrinos: electron neutrino, muon neutrino, and tau neutrino.
Consequently, the SM contains three copies of fermions with identical charges: the three so-called \emph{generations} of up-type quarks, down-type quarks, charged leptons, and neutrinos.
While the charges of the particles are repeated in each generation, their masses differ substantially.
In fact, the masses of quarks and charged leptons are strongly hierarchical, spanning more than five orders of magnitude.
The masses serve as an organizing principle to assign each quark and charged lepton to one of the three generations, with the lightest ones comprising the first generation and the heaviest ones the third generation.
Neutrinos are special in the sense that they are massless in the SM, while experimentally at least two of their masses are non-zero but extremely small.
The weak interaction relates each of them to one of the charged leptons, such that the neutrinos can be assigned to the three generations even in the massless case.

In contrast to the particles' electric charges, flavour is not a conserved quantum number. The weak interaction mediates transitions between the different flavours of up- and down-type quarks and between charged leptons and neutrinos.
Consequently, the heavy quarks and leptons decay through the weak interaction into the lighter ones, and only the up quark,%
\footnote{%
While free up quarks are stable and free down quarks decay through the weak interaction, quarks are confined into hadrons -- in particular protons and neutrons -- and protons and neutrons form atomic nuclei. The stability of the confined up- and down-quarks ultimately depends on the masses and binding energies of the hadrons and nuclei.
}
the electron, and the neutrinos%
\footnote{%
In the SM, neutrinos are massless and stable. Since neutrino masses are an experimental fact, the heavier neutrinos can in principle decay to the lighter ones. But the corresponding decay widths are proportional to the extremely small neutrino masses and additionally loop suppressed, rendering even neutrinos with non-zero masses stable on cosmological time scales.
}
are stable.
The weak interaction mediating transitions between up-type and down-type quarks has a different coupling strength for each combination of quarks.
The pattern of the resulting transition rates between the three up-type and the three down-type quarks -- the so-called quark flavour mixing -- can be parametrized by a $3\times 3$ unitary matrix: the Cabibbo-Kobayashi-Maskawa (CKM) quark flavour mixing matrix~\cite{Cabibbo:1963yz, Kobayashi:1973fv}.
This matrix exhibits a clearly hierarchical structure:
the diagonal elements, corresponding to transitions within the same generation, are close to unity; the first off-diagonal elements, which parameterize transitions between adjacent generations, are suppressed by one to two orders of magnitude; and the second off-diagonal elements, corresponding to transitions between the third and the first generation, are suppressed by nearly three orders of magnitude.
An analogous mixing matrix is used to parameterize flavour transitions in the lepton sector: the Pontecorvo-Maki-Nakagawa-Sakata (PMNS) matrix~\cite{Maki:1962mu, Pontecorvo:1967fh}.
Unlike the CKM matrix, the PMNS matrix does not exhibit a pronounced hierarchy and features large mixing angles.
Moreover, due to the smallness of neutrino masses and their neutrality under the electromagnetic and strong interactions, the phenomenology of neutrinos differs significantly from that of quarks.
In the following, we will mostly focus on flavour physics in the quark sector.

Quark flavour physics explores the rich phenomenology arising from the peculiar pattern and pronounced hierarchies in quark masses and CKM mixing.
Since quarks are confined within hadrons -- with the notable exception of the top quark, which decays before hadronizing -- quark flavour transitions are studied through the transitions of mesons and baryons, in particular their decays, as well as neutral meson-antimeson oscillations.
An extensive experimental program has been developed to specifically target these processes.
Two major experimental efforts, the LHCb experiment at CERN's Large Hadron Collider (LHC)~\cite{LHCb:2008vvz, LHCb:2018roe} and the Belle II experiment at KEK~\cite{Belle-II:2018jsg}, are at the forefront. LHCb specializes in $b$-physics, investigating for example decays of heavy $B$ mesons that offer a wide variety of flavour transitions, as well as CP violation in $B_s$ mixing, CP violation in the charm sector, and the precise determination of the CKM angle $\gamma$. Meanwhile, Belle II builds upon the legacy of Belle and BaBar~\cite{BaBar:2014omp}, aiming for nearly two orders of magnitude more data, with a focus on searching for new physics (NP) in $B$ meson and tau lepton transitions, as well as precision measurements of CKM parameters. Complementary contributions come from the general-purpose LHC experiments, ATLAS and CMS, particularly in searches for rare decays with clean experimental signatures.
Beyond these large-scale efforts, several smaller specialized experiments target specific flavour-changing processes. For example, the NA62 experiment at CERN~\cite{NA62:2024pjp} and the K0TO experiment at J-PARC~\cite{KOTO:2024zbl, KOTO:2025gvq} focus on rare kaon decays $K \to \pi \nu \bar\nu$, while the MEG II experiment at PSI~\cite{MEGII:2018kmf} and the Mu2e experiment at Fermilab~\cite{Mu2e:2014fns} investigate lepton flavour violation in the muon sector.
Looking ahead, future particle colliders will further expand the scope of flavour physics. Circular electron-positron colliders operating on the $Z$ pole, such as the proposed FCC-ee~\cite{FCC:2018evy, Bernardi:2022hny} and CEPC~\cite{CEPCStudyGroup:2018ghi, CEPCPhysicsStudyGroup:2022uwl, Ai:2024nmn}, promise a significant leap forward. The sheer volume of produced $Z$ bosons will result in an unparalleled dataset for studying bottom-flavoured and charm-flavoured hadrons as well as tau leptons, offering access to flavour-changing processes beyond the reach of LHCb and Belle II.

While flavour physics is experimentally studied at low energy scales corresponding to hadronic masses, it has a particularly strong connection to physics beyond the Standard Model (BSM) at much higher energy scales.
This connection arises for two main reasons:

\begin{itemize}

 \item The peculiar pattern of quark masses and CKM mixing appears to be anything but random, featuring pronounced hierarchies that span several orders of magnitude.
 In the Standard Model, both the fermion masses and the CKM matrix arise from the Yukawa interactions of the fermions with the Higgs field and are encoded in the Yukawa coupling matrices.
 However, this is merely a parameterization: the Standard Model provides no explanation for the structure of these matrices or the origin of the observed hierarchies.
 Moreover, the Yukawa sector contains 13 of the 18 free parameters of the Standard Model, making it the largest and least understood subset of parameters in the theory.
 The unknown origin of the hierarchical spectrum of quark and charged lepton masses, along with the hierarchical structure of the CKM matrix, is known as the \emph{Standard Model flavour puzzle}.

 It is important to emphasize that the observed hierarchies are technically natural according to 't Hooft's naturalness criterion~\cite{tHooft:1979rat}: if a Yukawa coupling is set to zero, the symmetry of the theory increases.
 As a result, the hierarchies in the SM flavour parameters are stable under radiative corrections, being only logarithmically sensitive to the theory's cutoff scale (provided that new physics does not introduce significant additional sources of flavour violation beyond those already present in the Standard Model). Nevertheless, the hierarchical pattern of fermion masses and CKM mixing is strongly suggestive of a deeper underlying structure and calls for an explanation through physics beyond the Standard Model.

 \item While flavour physics experiments are conducted at low energies, their exceptional precision and the strong suppression of certain flavour transitions in the SM allow them to probe tiny deviations in the properties of known particles, potentially induced by heavy new particles with masses far beyond the reach of direct collider searches.
 In fact, new physics with generic flavour-violating interactions is strongly constrained by precision measurements in flavour physics, with generic tree-level contributions to meson mixing excluded up to energy scales of several thousand TeV.
 This makes flavour physics one of the most powerful indirect probes of physics beyond the Standard Model.
 The fact that new physics must have a highly non-generic flavour structure to evade existing constraints, yet is also required to explain the origin of the flavour structure in the SM, is known as the \emph{New Physics flavour puzzle}.
\end{itemize}
These two aspects are closely related: addressing the SM flavour puzzle typically also requires addressing the NP flavour puzzle.
In this chapter, we provide an overview of known mechanisms and strategies that aim to resolve both puzzles, and we highlight the processes most sensitive to the presence of heavy new particles at high energy scales.

In Section~\ref{sec:SM}, we review the flavour structure of the SM, the approximate accidental symmetries implied by the observed hierarchies, and the most important flavour-changing processes.
In Section~\ref{probe}, we discuss the main classes of processes for exploring BSM physics using low-energy flavour experiments and review approaches to address the NP flavour puzzle.
In Section~\ref{sec:addressing_puzzle}, we discuss general mechanisms and specific examples proposed to explain the origin of the observed flavour structure and to address the SM flavour puzzle.

\section{Flavour in the Standard Model}\label{sec:SM}

In this section, we review how the flavour structure in the Standard Model emerges from the breaking of global flavour symmetries through the Yukawa interactions that couple the fermions to the Higgs field, and how this gives rise to the fermion masses and to flavour-changing weak interactions.
We give an overview of the phenomenological consequences of these interactions and discuss the most important classes of flavour-changing processes.
This section sets the stage for the subsequent sections on flavour beyond the Standard Model.

\subsection{Flavour symmetry breaking and the CKM matrix} \label{sec:CKM}

The Standard Model organizes quarks and leptons into five representations of the gauge group $G = \text{SU}(3)_c \times \text{SU}(2)_L \times \text{U}(1)_Y$:
\begin{equation}\label{eq:fermion_fields}
Q_j = ({\mathbf 3},{\mathbf 2})_{+\frac{1}{6}} ~,\quad U_j = ({\mathbf 3},{\mathbf 1})_{+\frac{2}{3}} ~,\quad D_j = ({\mathbf 3},{\mathbf 1})_{-\frac{1}{3}} ~,\quad L_j = ({\mathbf 1},{\mathbf 2})_{-\frac{1}{2}} ~,\quad E_j = ({\mathbf 1},{\mathbf 1})_{-1} ~.
\end{equation}
The left-handed quarks $Q_j$, as well as the right-handed up-type and down-type quarks $U_j$ and $D_j$, transform as triplets under the $\text{SU}(3)_c$ gauge group, whereas the left-handed leptons $L_j$ and right-handed charged leptons $E_j$ are $\text{SU}(3)_c$ singlets. The left-handed quarks and leptons are doublets under $\text{SU}(2)_L$, while all right-handed fields are $\text{SU}(2)_L$ singlets. The hypercharges of each field are indicated by the subscripts.
Importantly, each of the five fermion representations appears in three generations, and can be viewed as a three-component vector in its associated flavour vector-space. In Eq.~\eqref{eq:fermion_fields}, this flavour structure is made explicit by the index $j = 1,2,3$.

In the Standard Model, the fermion kinetic terms and gauge couplings are flavour universal, meaning they are identical for all three generations:
\begin{equation}
\mathcal L_\text{SM} \supset  \sum_j \bigg( \bar Q_j i \slashed{D} \, Q_j + \bar U_j i \slashed{D} \, U_j + \bar D_j i \slashed{D} \, D_j + \bar L_j i \slashed{D} \, L_j + \bar E_j i \slashed{D} \, E_j \bigg)\,.
\end{equation}
As a result, any unitary transformation rotating the fermion fields in flavour space, such as
\begin{equation}\label{eq:Q_i_rotation}
 Q_j \to \sum_i U_{ji} Q_i\,
\end{equation}
with a $3\times 3$ unitary matrix $U_{ji}$, leaves the gauge-kinetic terms invariant.
This invariance reflects a large global symmetry, consisting of a $\text{U}(3)$ factor for each fermion representation:
\begin{equation}\label{eq:U(3)^5}
G_\text{flavour} = \text{U}(3)_Q \times \text{U}(3)_U \times \text{U}(3)_D \times \text{U}(3)_L \times \text{U}(3)_E ~.
\end{equation}
Each $\text{U}(3)$ factor can be decomposed into $\text{SU}(3) \times \text{U}(1)$, where the $\text{SU}(3)$ parts correspond to non-trivial flavour rotations.
The fermion fields transform as triplets under their respective $\text{U}(3)$ flavour groups:
\begin{equation}
 Q\sim ({\mathbf 3},{\mathbf 1},{\mathbf 1},{\mathbf 1},{\mathbf 1})_\text{flavour}\,,\quad
 U\sim ({\mathbf 1},{\mathbf 3},{\mathbf 1},{\mathbf 1},{\mathbf 1})_\text{flavour}\,,\quad
 D\sim ({\mathbf 1},{\mathbf 1},{\mathbf 3},{\mathbf 1},{\mathbf 1})_\text{flavour}\,,\quad
 L\sim ({\mathbf 1},{\mathbf 1},{\mathbf 1},{\mathbf 3},{\mathbf 1})_\text{flavour}\,,\quad
 E\sim ({\mathbf 1},{\mathbf 1},{\mathbf 1},{\mathbf 1},{\mathbf 3})_\text{flavour}\,,
\end{equation}
and since the covariant derivatives are singlets under $G_\text{flavour}$, the gauge-kinetic terms are invariant under $G_\text{flavour}$ transformations.

The global $G_\text{flavour}$ symmetry of the gauge-kinetic terms is not an exact symmetry of the Standard Model.
It is explicitly broken by the Yukawa couplings of the fermions to the Higgs field:
\begin{equation} \label{eq:L_Yuk}
\mathcal L_\text{SM} \supset - \sum_{j,k} \bigg( (Y_u)_{jk} \bar Q_j U_k \tilde H + (Y_d)_{jk} \bar Q_j D_k H + (Y_e)_{jk} \bar L_j E_k H \bigg) +\text{h.c.} ~,
\end{equation}
where $H$ is the Standard Model Higgs doublet, $\tilde H = i \sigma_2 H^*$ is the conjugate Higgs field, and the Yukawa matrices $Y_u$, $Y_d$, and $Y_e$ are, in general, complex $3 \times 3$ matrices. These matrices couple different fermion generations to each other and therefore break the $G_\text{flavour}$ symmetry.
This breaking is essential for flavour physics: if the flavour symmetry were exact, there would be no transitions between the different generations of quarks or leptons. Moreover, the three generations would be completely indistinguishable, and the very concept of flavour would be ill-defined. Thus, the breaking of $G_\text{flavour}$ by the Yukawa couplings is the origin of flavour physics in the Standard Model.
The language of global flavour symmetry provides a natural way to reformulate the Standard Model and New Physics flavour puzzles:

\begin{itemize}
\item The Standard Model flavour puzzle refers to the unknown origin of the breaking of the global flavour symmetry, as well as the peculiar hierarchical structure of this breaking encoded in the Yukawa matrices.

\item The New Physics flavour puzzle arises from the tension that new physics must, on the one hand, avoid introducing significant new sources of flavour symmetry breaking in order to satisfy stringent experimental constraints, and, on the other hand, provide the dynamical origin for the flavour symmetry breaking observed in the SM Yukawa sector. This implies that any breaking mechanism that is introduced not far above the electroweak scale must be minimal, in the sense that it generates only flavour structures sufficiently aligned with the SM Yukawa couplings.
\end{itemize}
To study flavour symmetry breaking, it is convenient to promote the Yukawa matrices to \emph{spurions} -- auxiliary objects that formally transform under the symmetry like fields, in such a way that the full Lagrangian remains invariant. When these spurions are set to their background values -- i.e., the actual numerical values of the Yukawa matrices -- the symmetry is broken explicitly. The Standard Model Lagrangian becomes formally invariant under $G_\text{flavour}$ if the Yukawa matrices are treated as spurions transforming as bi-triplets under the relevant $\text{U}(3)$ flavour groups:
\begin{equation} \label{eq:spurion}
 Y_u \to Y_u^{\rm spurion} \sim (\mathbf{3},\mathbf{\bar{3}},{\mathbf 1},{\mathbf 1},{\mathbf 1})_\text{flavour}\,,\quad
 Y_d \to Y_d^{\rm spurion} \sim (\mathbf{3},{\mathbf 1},\mathbf{\bar{3}},{\mathbf 1},{\mathbf 1})_\text{flavour}\,,\quad
 Y_e \to Y_e^{\rm spurion} \sim ({\mathbf 1},{\mathbf 1},{\mathbf 1},\mathbf{3},\mathbf{\bar{3}})_\text{flavour}\,.
\end{equation}
For example, $Y_u^{\rm spurion}$ transforms as a triplet under $\text{U}(3)_Q$ and an anti-triplet under $\text{U}(3)_U$. Since $\bar Q$ transforms as an anti-triplet of $\text{U}(3)_Q$ and $U$ transforms as a triplet of $\text{U}(3)_U$, the combination $Y_u^{\rm spurion}\,\bar Q\,U$ is formally invariant under $G_\text{flavour}$. Restoring the background value, $Y_u^{\rm spurion}\to Y_u$, makes it manifest how the Yukawa matrix $Y_u$ breaks the $\text{U}(3)_Q \times \text{U}(3)_U$ symmetry down to a subgroup determined by its specific numerical entries.

Since the Yukawa matrices carry flavour indices $jk$, their explicit form depends on the choice of basis in flavour space. This choice can be thought of as a coordinate system that defines what is meant by $j=1$, $j=2$, and $j=3$, and how differences between generations are described. If the flavour symmetry were exact, the only flavour-indexed couplings would be proportional to the identity matrix, implying no distinction between generations. In this case, a unitary transformation of flavour indices -- such as in Eq.~\eqref{eq:Q_i_rotation} -- would leave the form of the Lagrangian completely unchanged.
However, in the presence of non-trivial coupling matrices such as $Y_u$, $Y_d$, and $Y_e$, which break the flavour symmetry, such unitary transformations do alter the form of the Lagrangian. While a specific choice of basis is necessary for an explicit parameterization of the couplings, it is important to emphasize that all choices of flavour basis are physically equivalent. Although the form and numerical values of the Yukawa matrices vary with the basis, these changes have no observable consequences.
While we are always free to choose -- and change -- the basis in flavour space, this freedom should not be mistaken for a symmetry such as the invariance under \( G_\text{flavour} \). Choosing a basis simply means selecting a coordinate system to describe the physical differences between flavours. In contrast, the presence of a symmetry implies that no such physical differences exist in the first place.
Nevertheless, there is a relation between the two concepts. A convenient basis is usually one in which the kinetic terms take their canonical form. For this reason, although any invertible matrix could in principle be used to change basis, we typically restrict ourselves to unitary transformations that preserve the canonical kinetic terms. Therefore, when changing between bases in flavour space, we use the same unitary transformations under which the gauge-kinetic terms of the Standard Model are invariant -- not because the Lagrangian is symmetric under these transformations, but because they preserve the canonical form of the kinetic terms.

To be explicit, we express the Yukawa matrices in a generic flavour basis and then consider two commonly used specific bases. Each Yukawa matrix can be factorized using a singular value decomposition (SVD) into one diagonal and two unitary matrices:
\begin{equation}\label{eq:Yukawa_SVD}
 Y_u = U_{u_L} \hat Y_u U_{u_R}^\dagger\,,\qquad
 Y_d = U_{d_L} \hat Y_d U_{d_R}^\dagger\,,\qquad
 Y_e = U_{e_L} \hat Y_e U_{e_R}^\dagger\,,
\end{equation}
where $\hat Y_u$, $\hat Y_d$, and $\hat Y_e$ are diagonal $3\times 3$ matrices with non-negative real entries, and $U_{u_L}$, $U_{u_R}$, $U_{d_L}$, $U_{d_R}$, $U_{e_L}$, $U_{e_R}$ are unitary $3\times 3$ matrices.
This is the most general form of any complex $3\times 3$ coupling matrix. However, not all of the parameters in these expressions are physical -- they depend on the choice of flavour basis. By transforming the fermion fields, we can move to a different basis. Starting from Eq.~\eqref{eq:Yukawa_SVD}, one convenient choice is to redefine the fields as
\begin{equation}
 Q \to U_{u_L} Q\,,\quad
 U \to U_{u_R} U\,,\quad
 D \to U_{d_R} D\,,\quad
 L \to U_{e_L} L\,,\quad
 E \to U_{e_R} E\,.
\end{equation}
This leads to a basis in which the Yukawa matrices become
\begin{equation}\label{eq:Yukawa_up_aligned}
 Y_u = \hat Y_u \,,\qquad
 Y_d = U_{u_L}^\dagger U_{d_L} \hat Y_d \,,\qquad
 Y_e = \hat Y_e \,.
\end{equation}
In this so-called \emph{up-aligned} basis, $Y_u$ and $Y_e$ are diagonal, while $Y_d$ is given by the product of the unitary matrix $U_{u_L}^\dagger U_{d_L}$ and the diagonal matrix $\hat Y_d$.
The SVD of the Yukawa matrices introduces six independent unitary matrices, but only five fermion fields are available to absorb these matrices through field redefinitions. As a result, one unitary combination remains physical in any basis: the Cabibbo-Kobayashi-Maskawa (CKM) matrix~\cite{Cabibbo:1963yz, Kobayashi:1973fv}, defined as
\begin{equation}
 V_{\rm CKM} = U_{u_L}^\dagger U_{d_L}\,.
\end{equation}
We can now perform another basis transformation, starting from Eq.~\eqref{eq:Yukawa_up_aligned} and rotating only the quark doublets:
\begin{equation}
 Q \to V_{\rm CKM} Q\,,
\end{equation}
which leads to a basis in which the Yukawa matrices take the form
\begin{equation}\label{eq:Yukawa_down_aligned}
 Y_u = V_{\rm CKM}^\dagger \hat Y_u \,,\qquad
 Y_d = \hat Y_d \,,\qquad
 Y_e = \hat Y_e \,.
\end{equation}
In this \emph{down-aligned} basis $Y_d$ is diagonal, while $Y_u$ is the product of the hermitian conjugate of the CKM matrix and the diagonal matrix $\hat Y_u$.
This example illustrates how the form of the Yukawa matrices depends on the basis choice. However, the physical parameters -- such as the singular values of the Yukawa matrices and the parameters of the CKM matrix -- remain invariant under changes of basis.
We can also observe that an $\text{SU}(2)_L$-invariant transformation of the quark doublets allows one to diagonalize either the up-type or the down-type Yukawa matrix, but not both simultaneously. This reflects a misalignment between the up-aligned flavour basis, in which $Y_u$ is diagonal, and the down-aligned flavour basis, in which $Y_d$ is diagonal. This misalignment is precisely encoded in the CKM matrix.
In the lepton sector, no physical unitary matrix remains if neutrinos are massless. However, if neutrinos have masses, one can define a unitary mixing matrix in the lepton sector analogous to the CKM matrix: the Pontecorvo-Maki-Nakagawa-Sakata (PMNS) matrix~\cite{Maki:1962mu, Pontecorvo:1967fh}.

\begin{figure}[tb]
	\centering
	\includegraphics[width=\textwidth]{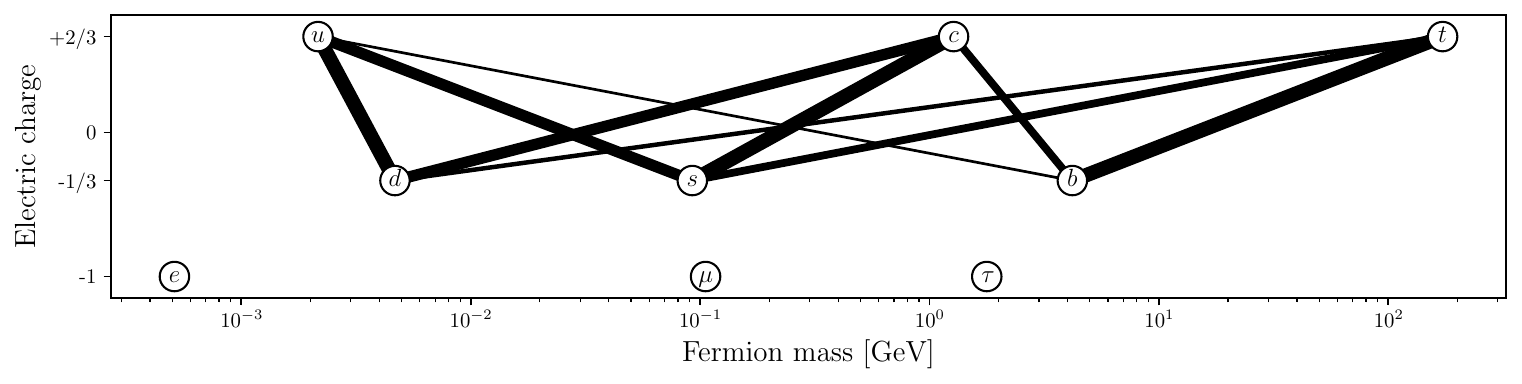}
 	\caption{Masses and electric charges of quarks and charged leptons. The lines connecting up-type quarks $i=u,c,t$ to down-type quarks $j=d,s,b$ represent the absolute values of the CKM matrix elements $|V_{ij}|$, with the thickness of each line proportional to $\log(10^3\times|V_{ij}|)$.}
 	\label{fig:fermion_masses}
 \end{figure}

During electroweak symmetry breaking, the Higgs field acquires a vacuum expectation value (vev),
\begin{equation}
 \langle H \rangle = \frac{1}{\sqrt{2}} \left(\begin{matrix} 0 \\ v \end{matrix} \right)\,,
\end{equation}
which gives rise to fermion mass matrices that are given by products of the electroweak vev $v$ and the Yukawa couplings,
\begin{equation} \label{eq:masses}
(M_u)_{jk} = \frac{v}{\sqrt{2}} (Y_u)_{jk} ~,\quad  (M_d)_{jk} = \frac{v}{\sqrt{2}} (Y_d)_{jk} ~,\quad (M_e)_{jk} = \frac{v}{\sqrt{2}} (Y_e)_{jk} ~.
\end{equation}
In the up-aligned basis, these become
\begin{equation} \label{eq:masses_up}
M_u = \text{diag}(m_u, m_c, m_t) ~,\quad  M_d = V_{\rm CKM}\,\text{diag}(m_d, m_s, m_b) ~,\quad M_e = \text{diag}(m_e, m_\mu, m_\tau) ~,
\end{equation}
while in the down-aligned basis, they take the form
\begin{equation} \label{eq:masses_down}
M_u = V_{\rm CKM}^\dagger\,\text{diag}(m_u, m_c, m_t) ~,\quad  M_d = \text{diag}(m_d, m_s, m_b) ~,\quad M_e = \text{diag}(m_e, m_\mu, m_\tau) ~,
\end{equation}
where $m_u$, $m_c$, and $m_t$ are up-, charm- and top-quark masses, $m_d$, $m_s$, and $m_b$ are the down-, strange-, and bottom-quark masses, and $m_e$, $m_\mu$, and $m_\tau$ are the masses of electron, muon, and tau.
The experimentally obtained values of the fermion masses are collected in the upper panel of Table~\ref{tab:mass} and illustrated in Fig.~\ref{fig:fermion_masses}.
They are found to be strongly hierarchical, spanning more than five orders of magnitude.
These masses are proportional to the electroweak vev, which sets their common overall scale, and to the singular values of the Yukawa matrices, which are responsible for the strongly hierarchical structure.

\renewcommand{\arraystretch}{1.5}
\begin{table}[tb]\TBL{
    \caption{Observed values of charged lepton and quark masses (upper panel) and absolute values of CKM matrix elements (lower panel).
    All values correspond to world averages taken from the PDG~\cite{ParticleDataGroup:2024cfk}. References to some of the original determinations of the masses are given in the upper panel of the table. References to determinations of the CKM matrix are given in the main text.}
    \label{tab:mass}
}{
    \begin{tabular*}{\columnwidth}{@{\extracolsep{\fill}}cccc|ccccc|cccc@{}}
        \toprule
        $m_e$ & $510.99895000(15)~\text{keV}$ & \cite{Tiesinga:2021myr} &&&
        $m_u$ & $2.16(7) ~\text{MeV}$ & \cite{Fodor:2016bgu, FermilabLattice:2018est} &&&
        $m_d$ & $4.70(7) ~\text{MeV}$ & \cite{Fodor:2016bgu, FermilabLattice:2018est} \\
        $m_{\mu}$ & $105.6583755(23)~\text{MeV}$ & \cite{Tiesinga:2021myr} &&&
        $m_c$ & $1.2730(46) ~\text{GeV}$ & \cite{Erler:2016atg, Chetyrkin:2017lif, FermilabLattice:2018est, Lytle:2018evc, Narison:2019tym, Hatton:2020qhk} &&&
        $m_s$ & $93.5(8) ~\text{MeV}$ & \cite{FermilabLattice:2018est, Lytle:2018evc} \\
        $m_\tau$ & $1.77693(9)~\text{GeV}$ & \cite{BESIII:2014srs, Belle-II:2023izd, Anashin:2023sch} &&&
        $m_t$ & $172.4(7)~\text{GeV}$ & \cite{ATLAS:2019guf, CMS:2022emx} &&&
        $m_b$ & $4.183(7) ~\text{GeV}$ & \cite{Penin:2014zaa, Beneke:2014pta, FermilabLattice:2018est, Narison:2019tym, Hatton:2020qhk} \\
        \colrule
        \multicolumn{13}{@{}c}{
            \setlength{\arraycolsep}{0.8em}
            $
            \left(\begin{array}{ccc}
                \left| V_{ud} \right| &
                \left| V_{us} \right| &
                \left| V_{ub} \right| \\
                \left| V_{cd} \right| &
                \left| V_{cs} \right| &
                \left| V_{cb} \right| \\
                \left| V_{td} \right| &
                \left| V_{ts} \right| &
                \left| V_{tb} \right|
            \end{array}\right)
            \ = \
            \left(\begin{array}{ccc}
                0.97367(31) &
                0.22431(85) &
                3.82(20) \times 10^{-3} \\
                0.221(4) &
                0.975(6) &
                4.11(12) \times 10^{-2} \\
                8.6(2) \times 10^{-3} &
                4.15(9)\times 10^{-2} &
                1.010(27)
            \end{array}\right)
            $
        } \\
        \botrule
    \end{tabular*}
}{}\end{table}
\renewcommand{\arraystretch}{1.0}

The appearance of the CKM factor in either $M_d$ or $M_u$ in Eqs.~\eqref{eq:masses_up} and~\eqref{eq:masses_down} implies that the same factor appears also inside the quark doublet when expressed in terms of the left-handed mass eigenstates $u_L$ and $d_L$:
\begin{equation}
 Q\,\Big|_{\rm up-aligned} =
 \begin{pmatrix}
 u_L \\
 V_{\rm CKM}\, d_L
 \end{pmatrix}\,,
 \qquad
 Q\,\Big|_{\rm down-aligned} =
 \begin{pmatrix}
 V_{\rm CKM}^\dagger\,u_L \\
 d_L
 \end{pmatrix}\,.
\end{equation}
The presence of the CKM matrix in one of the doublet components reflects the misalignment between the left-handed up- and down-type quark mass bases.
This misalignment has profound physical consequences that become obvious when expressing the gauge interactions in terms of the mass eigenstates. While the unitary $V_{\rm CKM}$ cancels in all neutral current interactions, which couple up-type to up-type quarks and down-type to down-type quarks, this is not the case for the charged current interactions of the $W$ boson, which couple up-type to down-type quarks.
Due to the misalignment of the up- and down-type mass eigenstates in the quark doublet, these interactions become dependent on the CKM matrix:
\begin{equation} \label{eq:cc}
\mathcal L_\text{SM} \supset \frac{g}{\sqrt{2}} \, W^+_\mu\sum_{j,k} \bigg( V_{jk}\, (\bar u_L)_j\, \gamma^\mu \, (d_L)_k\bigg) +\text{h.c.} \,,
\end{equation}
where $g$ is the $\text{SU}(2)_L$ gauge coupling, $(u_L)_j$ and $(d_L)_k$ are the left-handed up-type and down-type quark mass eigenstates, and $V_{jk}$ are the CKM matrix elements.
These charged current interactions governed by the CKM matrix are responsible for all flavour-changing phenomena in the Standard Model.

The CKM matrix is a unitary $3 \times 3$ matrix.
While a general unitary $3 \times 3$ matrix has three mixing angles and six complex phases, five of the phases of the CKM matrix can be absorbed by field redefinitions of the quark fields. Consequently, it can be expressed in terms of four physical parameters: three mixing angles $\theta_{12}$, $\theta_{23}$, $\theta_{13}$ and one CP-violating phase $\delta$:
\begin{equation}
 V_\text{CKM} = \begin{pmatrix} V_{ud} & V_{us} & V_{ub} \\ V_{cd} & V_{cs} & V_{cb} \\ V_{td} & V_{ts} & V_{tb} \end{pmatrix} = \begin{pmatrix} 1 & 0 & 0 \\ 0 & c_{23} & s_{23} \\ 0 & -s_{23} & c_{23} \end{pmatrix} \begin{pmatrix} c_{13} & 0 & s_{13} e^{-i\delta} \\ 0 & 1 & 0 \\ -s_{13} e^{i\delta} & 0 & c_{13} \end{pmatrix}  \begin{pmatrix} c_{12} & s_{12} & 0 \\ -s_{12} & c_{12} & 0 \\ 0 & 0 & 1 \end{pmatrix}~,
\end{equation}
where $c_{ij} = \cos(\theta_{ij})$ and $s_{ij} = \sin(\theta_{ij})$. In this so-called standard parameterization, the elements $V_{ud}$, $V_{us}$, $V_{cb}$, and $V_{tb}$ are real. Experimentally, the three mixing angles are found to be hierarchical:
\begin{equation}
 1 \gg \theta_{12} \gg \theta_{23} \gg \theta_{13}\,.
\end{equation}
Mixing is in general small, with that between the first and second generations being the largest, followed by that between the second and third. The strongest suppression is observed in mixing between the first and third generations.
The experimentally determined absolute values of the CKM matrix elements are collected in the lower panel of Table~\ref{tab:mass} and illustrated in Fig.~\ref{fig:fermion_masses}.
The CKM matrix elements, together with the fermion masses, fully parametrize the Yukawa matrices. They encode the breaking of the flavour symmetry and determine the phenomenology of flavour-changing processes in the Standard Model.

\subsection{Approximate flavour symmetries} \label{approx}
While the large global flavour symmetry $G_{\rm flavour}=\text{U}(3)^5$ of the SM gauge-kinetic terms is explicitly broken by the Yukawa couplings, the hierarchical structure of the Yukawa matrices implies the presence of approximate flavour symmetries that are broken only by small parameters.
In the language of spurions, a symmetry is considered to be a good approximate symmetry if the associated spurions have only small background values.
Hence, the full $G_{\rm flavour}=\text{U}(3)^5$ is clearly not a good approximate symmetry, since the spurion $Y_u^{\rm spurion}$ introduced in Eq.~\eqref{eq:spurion} has a very large background component: the top Yukawa coupling $y_t\approx 1$.

A good approximate symmetry must remain unbroken even in the presence of the top Yukawa coupling, such that the large $y_t$ does not appear in any spurion background value.
The gauge-kinetic terms, together with the top Yukawa interaction, exhibit a reduced global flavour symmetry~\cite{Feldmann:2008ja, Kagan:2009bn}
\begin{equation}
 G_\text{flavour}^{(y_t)} = \text{U}(2)_Q \times \text{U}(2)_U \times \text{U}(3)_D \times \text{U}(3)_L \times \text{U}(3)_E \times \text{U}(1)_{Q_3+U_3}\,.
\end{equation}
A minimal set of spurions transforming under $G_\text{flavour}^{(y_t)}$ and reproducing all fermion masses and CKM elements when set to their background values is given by (see e.g.~\cite{Faroughy:2020ina})
\begin{equation}\label{eq:spurions_yt}
 V_q \sim (\mathbf{2},\mathbf{1},{\mathbf 1},{\mathbf 1},{\mathbf 1})_\text{flavour}^{(y_t)}\,,\quad
 \Delta_u \sim (\mathbf{2},\mathbf{\bar{2}},{\mathbf 1},{\mathbf 1},{\mathbf 1})_\text{flavour}^{(y_t)}\,,\quad
 V_R^u \sim (\mathbf{1},\mathbf{\bar{2}},{\mathbf 1},{\mathbf 1},{\mathbf 1})_\text{flavour}^{(y_t)}\,,\quad
 \Sigma_d \sim (\mathbf{2},{\mathbf 1},\mathbf{\bar{3}},{\mathbf 1},{\mathbf 1})_\text{flavour}^{(y_t)}\,,\quad
 \Lambda_d  \sim (\mathbf{1},{\mathbf 1},\mathbf{\bar{3}},{\mathbf 1},{\mathbf 1})_\text{flavour}^{(y_t)}\,,
\end{equation}
together with $Y_e^{\rm spurion}$ defined in Eq.~\eqref{eq:spurion}.
The largest background components are those of $V_q$, $\Lambda_d$, and $Y_e^{\rm spurion}$, which are approximately $|V_{cb}|\approx 0.04$, $y_b\approx 0.02$, and $y_\tau\approx 0.01$, respectively, and therefore sufficiently small for $G_\text{flavour}^{(y_t)}$ to qualify as a good approximate symmetry.
However, since $y_b$ and $y_\tau$ appear as background components, the three largest spurions are of comparable size, so there is no clear hierarchy of symmetry breaking.
Moreover, reproducing all fermion masses and CKM elements requires alignment between the two different $\text{U}(3)_D$ anti-triplets $\Lambda_d$ and $\Sigma_d$.

These issues can be avoided by considering a symmetry that remains unbroken even in the presence of the bottom and $\tau$ Yukawa couplings, $y_b$ and $y_\tau$.
The gauge-kinetic terms, together with all third-generation Yukawa interactions, exhibit a further reduced global flavour symmetry
\begin{equation} \label{eq:U2_to_the_fifth}
 G_\text{flavour}^{(y_t,y_b,y_\tau)} = \text{U}(2)_Q \times \text{U}(2)_U \times \text{U}(2)_D \times \text{U}(2)_L \times \text{U}(2)_E \times \text{U}(1)_{Q_3+U_3+D_3} \times \text{U}(1)_{L_3+E_3}\,.
\end{equation}
A minimal set of spurions transforming under this symmetry is
\begin{equation}\label{eq:spurions_ytybytau}
 V_q \sim (\mathbf{2},\mathbf{1},{\mathbf 1},{\mathbf 1},{\mathbf 1})_\text{flavour}^{(y_t,y_b,y_\tau)}\,,\quad
 \Delta_u \sim (\mathbf{2},\mathbf{\bar{2}},{\mathbf 1},{\mathbf 1},{\mathbf 1})_\text{flavour}^{(y_t,y_b,y_\tau)}\,,\quad
 V_R^u \sim (\mathbf{1},\mathbf{\bar{2}},{\mathbf 1},{\mathbf 1},{\mathbf 1})_\text{flavour}^{(y_t,y_b,y_\tau)}\,,\quad
 \Delta_d \sim (\mathbf{2},{\mathbf 1},\mathbf{\bar{2}},{\mathbf 1},{\mathbf 1})_\text{flavour}^{(y_t,y_b,y_\tau)}\,,\quad
 V_R^d  \sim (\mathbf{1},{\mathbf 1},\mathbf{\bar{2}},{\mathbf 1},{\mathbf 1})_\text{flavour}^{(y_t,y_b,y_\tau)}\,,
\end{equation}
with analogous spurions in the lepton sector.
This symmetry treats all fermion representations uniformly and avoids the need for alignment between spurion background values.
Moreover, it leads to a clear hierarchy of symmetry-breaking background values: $|V_q| \gg |\Delta_{u,d}| \gg |V_R^{u,d}|$.
In this sense, $G_\text{flavour}^{(y_t,y_b,y_\tau)}$ can be regarded as an even better approximate symmetry than $G_\text{flavour}^{(y_t)}$.

\subsection{Flavour-changing charged currents and the determination of CKM elements}
An immediate consequence of the hierarchical structure of fermion masses and the flavour-changing charged-current $W$ interactions in Eq.~\eqref{eq:cc} is that up- and down-type quarks can decay into each other via tree-level processes mediated by the $W$ boson.

A special role is played by the top quark --  the heaviest particle in the Standard Model. Due to its large mass and the near-unity CKM element $|V_{tb}|\approx 1$, the top quark decays almost exclusively into an on-shell $W$ boson and a bottom quark. This leads to a very short lifetime of about $0.5\times 10^{-24}\,{\rm s}$~\cite{ParticleDataGroup:2024cfk}, which is too short to allow hadronization. Decays into a $W$ boson and a strange or down quark are suppressed by the small CKM elements $|V_{ts}|\approx4\times 10^{-2}$ and $|V_{td}|\approx 9\times 10^{-3}$, respectively (see Table~\ref{tab:mass}).

 \begin{figure}[b]
	\centering
	\includegraphics[width=0.44\textwidth]{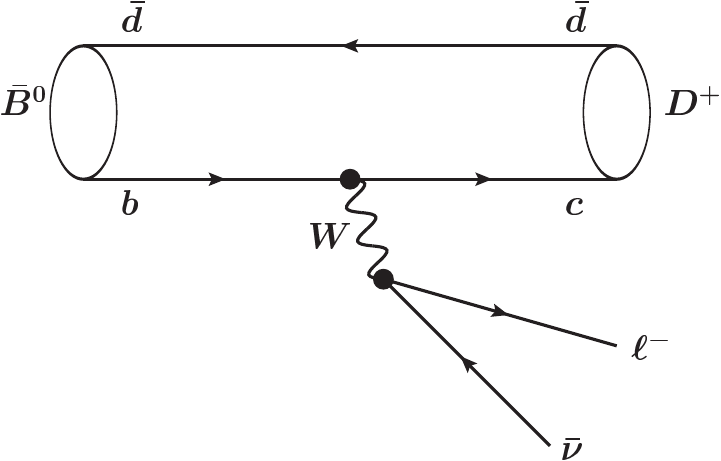} \qquad\qquad
    \raisebox{36pt}{\includegraphics[width=0.36\textwidth]{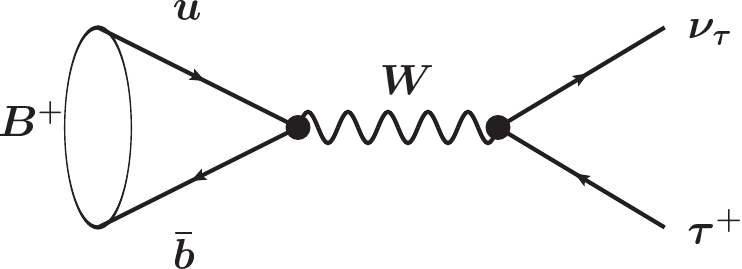}}
 	\caption{Example Feynman diagrams of flavour-changing tree-level decays of mesons. Left: the semileptonic decay $\bar B^0 \to D^+ \ell^- \bar\nu_\ell$. Right: the leptonic decay $B^+ \to \tau^+ \nu_\tau$.}
 	\label{fig:tree}
 \end{figure}

All other quarks are confined within hadrons, and their decays are observed in the decays of mesons and baryons.
Focusing on mesons -- bound states of a quark and an antiquark -- one distinguishes three types of tree-level decays:
\begin{itemize}
 \item \emph{Hadronic decays}, in which the intermediate $W$ boson produces a quark-antiquark pair, leading to two (or more) mesons in the final state.
 \item \emph{Semileptonic decays}, where one constituent quark of the meson decays into a lighter quark that forms a new meson with the second constituent quark -- the so-called spectator quark -- while the $W$ boson produces a charged lepton and a neutrino. The final state consists of a meson and a lepton pair. The example $\bar B^0 \to D^+ \ell^- \bar \nu_\ell$ is shown in the left diagram of Fig.~\ref{fig:tree}.
 \item \emph{Leptonic decays} of charged mesons, in which the quark and antiquark annihilate into a $W$ boson, which produces a charged lepton and a neutrino, resulting in a purely leptonic final state. The example $B^+ \to \tau^+ \nu_\tau$ is shown in the right diagram of Fig.~\ref{fig:tree}.
\end{itemize}
The tree-level meson decays with leptons in the final state are particularly useful for determining the magnitudes of CKM matrix elements, as their decay widths are proportional to $|V_{jk}|^2$.
Leptonic decays are theoretically the cleanest, since the only hadronic input parameter is the meson decay constant, which can be computed with high precision using lattice QCD.
Semileptonic decays, while also proportional to squared CKM elements, require knowledge of momentum-dependent hadronic form factors. These are more challenging to compute than decay constants but can still be obtained with non-perturbative methods such as lattice QCD.
Hadronic decay rates, on the other hand, depend on products of two CKM elements and involve more complicated hadronic matrix elements, making them considerably more difficult to predict reliably. As a result, they are less suitable for precision determinations of CKM parameters.

The most important tree-level processes that are used to extract the CKM matrix elements $|V_{jk}|$ are leptonic and semi-leptonic decays involving quark transitions $q_j\to q_k\, \ell\, \nu$, where $\ell = e, \mu$:
\begin{itemize}

 \item $b \to c \ell \nu$ and $b \to u \ell \nu$ transitions in semileptonic $B$ meson decays provide the most precise direct determinations of the CKM elements $|V_{cb}|$ and $|V_{ub}|$ (see~\cite{Bordone:2021oof, Bernlochner:2022ucr, Harrison:2023dzh, Aoki:2023qpa, Belle:2023bwv, Belle-II:2023okj, Leljak:2021vte, Colquhoun:2022atw, Belle:2023asa, Bolognani:2023mcf} for a selection of recent results and the PDG review~\cite{ParticleDataGroup:2024cfk} for more details).
 Their extraction follows two approaches: ``Inclusive'' decays, such as $B \to X_c \ell \nu$ and $B \to X_u \ell \nu$, where $X_c$ and $X_u$ denote sums over all hadronic final states containing a charm or up quark.
 ``Exclusive'' decays, which correspond to specific hadronic final states like $B \to D \ell \nu$ or $B \to D^* \ell \nu$ for $|V_{cb}|$ and $B \to \pi \ell \nu$ for $|V_{ub}|$. The values extracted from inclusive and exclusive decays have shown a long-standing tension at the $2\sigma - 3\sigma$ level. The PDG world averages account for this by inflating the uncertainties. Some authors prefer the inclusive determination of $|V_{cb}|$ (since exclusive decays depend on hadronic form factors that are difficult to compute precisely), while the exclusive determination of $|V_{ub}|$ is often favoured (as the inclusive decay suffers from large backgrounds due to $b\to c\ell \nu$ transitions). As of 2025, the tension remains, and one should keep in mind that it may affect theoretical predictions based on specific values of $|V_{cb}|$ and $|V_{ub}|$.

 \item $c\to s \ell \nu$ and $c\to d \ell\nu$ transitions in leptonic and semileptonic $D$ meson decays provide the best direct determinations of $|V_{cs}|$ and $|V_{cd}|$. However, these are currently less precise than the values inferred indirectly from CKM unitarity using the well-measured elements $|V_{ud}|$ and $|V_{us}|$.

 \item $d\to u \ell \nu$ and $s \to u \ell \nu$ transitions are used to determine $|V_{ud}|$ and $|V_{us}|$ with high precision.
 The most precise determination of $|V_{ud}|$ comes from superallowed nuclear beta decays~\cite{Hardy:2020qwl}, followed by measurements of the neutron lifetime~\cite{UCNt:2021pcg} and pion beta decay $\pi^+ \to \pi^0 e^+ \nu$~\cite{Pocanic:2003pf}. The PIONEER experiment~\cite{PIONEER:2022yag} aims to measure this latter decay with improved precision, potentially rivalling the current determination from nuclear beta decay.
 $|V_{us}|$ is extracted from semi-leptonic kaon decays, such as $K \to \pi \ell \nu$. In addition, ratios of leptonic kaon and pion decays, particularly $K \to \mu\nu$ and $\pi \to \mu \nu$ are used to determine the ratio $|V_{us}/V_{ud}|$ with comparable precision~\cite{Seng:2021nar, Seng:2022wcw}.
 Interestingly, the extracted values of $|V_{ud}|$, $|V_{us}|$, and $|V_{us}/V_{ud}|$ show a tension of about $3\sigma$ with CKM unitarity, which implies ${|V_{ud}|^2 + |V_{us}|^2 = 1 - |V_{ub}|^2 \simeq 1}$. Since $|V_{us}|$ and $|V_{ud}|$ are governed by the Cabibbo angle $\theta_{12}$, this discrepancy is known as the ``Cabibbo angle anomaly''~\cite{Cirigliano:2022yyo}.
\end{itemize}
The remaining three CKM matrix elements $|V_{tb}|$, $|V_{ts}|$, and $|V_{td}|$ involve the top quark and thus cannot be extracted directly from tree-level meson decays. The element $|V_{tb}|$ can be measured from the top quark decay $t \to W b$, as well as from single top production. However, a more precise value is obtained indirectly using CKM unitarity.
Similarly, $|V_{ts}|$, and $|V_{td}|$ can be inferred from unitarity, but also determined with high precision from loop-induced processes that are discussed in the following section.

\subsection{Flavour-changing neutral currents} \label{sec:FCNC}

 \begin{figure}[b]
	\centering
	\includegraphics[width=0.24\textwidth]{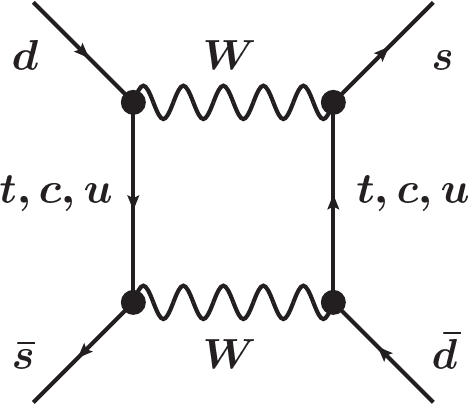} \qquad\qquad
    \includegraphics[width=0.24\textwidth]{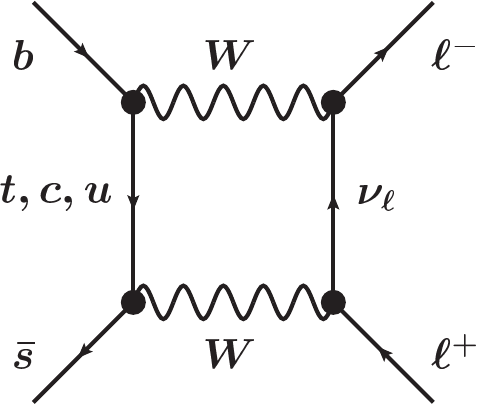}
    \qquad\qquad
    \includegraphics[width=0.28\textwidth]{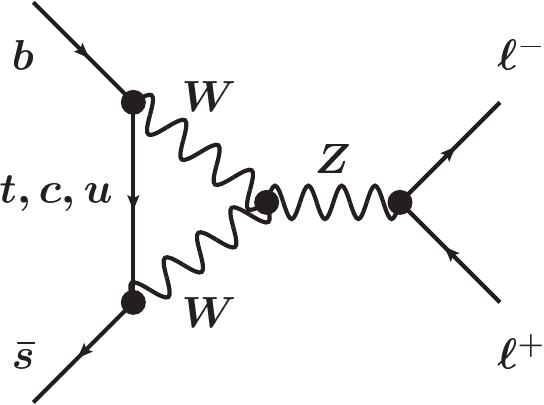}
 	\caption{Example Standard Model one-loop diagrams of flavour-changing neutral current processes. Left: a so-called box diagram contributing to kaon mixing; centre: a box diagram contributing to a rare $b \to s \ell\ell$ decay; right: a so-called penguin diagram contributing to a rare $b \to s \ell\ell$ decay.}
 	\label{fig:fcnc}
 \end{figure}

In the Standard Model, flavour-changing neutral currents (FCNCs) are absent at tree level, since the misalignment between up- and down-type quarks -- encoded in the CKM matrix -- affects only the charged-current interactions, which involve both up- and down-type quarks. In contrast, neutral current interactions that couple up-type to up-type or down-type to down-type quarks -- namely the Higgs couplings and the gauge interactions mediated by gluons, the photon, and the $Z$ boson -- remain flavour diagonal. This structure is ensured by the Glashow–Iliopoulos–Maiani (GIM) mechanism~\cite{Glashow:1970gm}, a key feature of the Standard Model.
For the neutral gauge interactions, the GIM cancellation follows from their flavour universality together with the unitarity of the CKM matrix. For the Higgs-fermion interactions, flavour diagonality results from the fact that these couplings are proportional to the fermion mass matrices and are thus, by construction, diagonal in the mass basis.

At the one-loop level, however, the situation changes. Diagrams with $W$ bosons running in the loop and involving two or more flavour-changing charged-current vertices can induce FCNCs. While the GIM mechanism ensures that contributions from different quark flavours in these loops cancel in the limit of degenerate quark masses, quark mass differences, in particular the large top quark mass, lead to incomplete cancellation and thus generate non-zero FCNCs.
Important examples include \emph{box diagrams}, such as those shown in the left and centre panels of Fig.~\ref{fig:fcnc}, and \emph{penguin diagrams}, as the one shown in the right panel of Fig.~\ref{fig:fcnc}.

These types of diagrams give rise to two classes of FCNC processes: $\Delta F = 1$ and $\Delta F = 2$ transitions, corresponding to changes in quark flavour by one and two units, respectively.
The left panel of Fig.~\ref{fig:fcnc} shows a $\Delta F = 2$ box diagram that changes an anti-strange quark into a strange quark. Since the strange flavour changes from $-1$ (anti-strange) to $+1$ (strange), this is a $\Delta S = 2$ transition.
The centre and right panels of Fig.~\ref{fig:fcnc} show $\Delta F = 1$ diagrams. In both cases, a bottom quark is converted into a strange quark. Here, the bottom flavour changes from $+1$ to $0$, and the strange flavour from $0$ to $+1$, corresponding to \( \Delta B = \Delta S = 1 \) transitions.

\subsubsection{Meson-antimeson oscillations}

The most important consequence of $\Delta F = 2$ transitions is neutral meson mixing (for a detailed introduction, see e.g.~\cite{Nierste:2009wg}).

As an example, the left panel of Fig.~\ref{fig:fcnc} shows a diagram contributing to $K^0 $-$\bar K^0$ mixing. It allows a neutral kaon, composed of an anti-strange and a down quark, to convert into its CP-conjugate state, the anti-kaon, composed of an anti-down and a strange quark,
\begin{equation}
 K^0 \sim (\bar s d)
 \quad
 \leftrightarrow
 \quad
 \bar K^0 \sim (\bar d s)\,.
\end{equation}
As a result of this mixing, the Hamiltonian describing the two-state system formed by $K^0$ and $\bar K^0$ is not diagonal. The off-diagonal terms responsible for transitions between the two states are generated by diagrams such as the one in the left panel of Fig.~\ref{fig:fcnc}. To determine the propagating mass eigenstates, one has to diagonalize the Hamiltonian.
The resulting mass eigenstates are the short-lived $K_S$ and the long-lived $K_L$, which are non-trivial linear combinations of $K^0$ and $\bar K^0$.
Importantly, mass eigenstates generally do not coincide with CP eigenstates.
In the $K^0 $-$\bar K^0$ system, the CP eigenstates $K_1$ (CP even) and $K_2$ (CP odd) are given in terms of the difference and sum of $K^0$ and $\bar K^0$,
\begin{equation}
 K_1 = \frac{1}{\sqrt{2}}(K^0 - \bar K^0)\,,
 \qquad
 K_2 = \frac{1}{\sqrt{2}}(K^0 + \bar K^0)\,.
\end{equation}
The mass eigenstates $K_S$ and $K_L$ can each be expressed in terms of one CP eigenstate plus an admixture of the other CP eigenstate proportional to a complex parameter $\bar \varepsilon$:
\begin{equation}
 K_S = \frac{1}{\sqrt{1 + |\bar \varepsilon|^2}} (K_1 + \bar \varepsilon\, \bar K_2)\,,
 \qquad
 K_L = \frac{1}{\sqrt{1 + |\bar \varepsilon|^2}} (K_2 + \bar \varepsilon\, \bar K_1)\,.
\end{equation}
In the $K^0 $-$\bar K^0$ system, this parameter is small but non-zero, $\bar \varepsilon\sim \mathcal{O}(10^{-3})$.
Since $K_S$ consists mostly of the CP-even $K_1$, it decays predominantly into the CP-even two-pion final state. In contrast, $K_L$, which consists mostly of the CP-odd $K_2$, decays primarily into the CP-odd three-pion final state. As the two-pion decay proceeds more rapidly due to the larger available phase space, this leads to a substantial difference in the lifetimes of $K_S$ and $K_L$.

There are in total four meson-antimeson systems. Three of them involve down-type quarks: the $K^0 $-$\bar K^0$, $B^0 $-$\bar B^0$, and $B_s^0 $-$\bar B_s^0$ systems. As the top quark decays before hadronizing, the $D^0 $-$\bar D^0$ system is the only one involving up-type quarks.
Since the flavour and CP eigenstates in meson-antimeson systems differ from the propagating mass eigenstates that govern time evolution, the flavour and CP states \emph{oscillate} over time.
Meson–antimeson oscillations have several important physical consequences:
\begin{itemize}
 \item \emph{CP violation}: Because the mass eigenstates differ from the CP eigenstates, meson mixing leads to CP violation. Its origin lies in the CP-violating phase $\delta$ of the CKM matrix, which enters loop diagrams such as the one shown in the left panel of Fig.~\ref{fig:fcnc}. CP-violating observables are directly sensitive to such diagrams and can be measured with high precision.

 \item \emph{Mass splitting}: Since the mass eigenstates are non-trivial linear combinations of meson and antimeson states, they differ in mass. The resulting mass difference $\Delta M$ corresponds to the oscillation frequency, which is an experimentally clean and accessible observable, also sensitive to loop-level contributions.
\end{itemize}

\subsubsection{Rare decays}

In the SM, hadron decays induced by $\Delta F=1$ FCNC transitions are strongly suppressed by both a loop factor and small CKM elements.
The relevant transitions are $b \to s$, $b \to d$, $s \to d$, and $c \to u$, which are governed by the CKM factors $|V_{tb} V_{ts}^*| \approx 4\times 10^{-2}$, $|V_{tb} V_{td}^*| \approx 9\times 10^{-3}$, $|V_{ts} V_{td}^*| \approx 4\times 10^{-4}$, and $|V_{ub} V_{cb}^*| \approx 2\times 10^{-4}$, respectively.
The CKM and loop suppression leads to branching fractions as small as $\mathcal{O}(10^{-7})$ to $\mathcal{O}(10^{-12})$, which is why these processes are referred to as \emph{rare decays}.

The phenomenologically most relevant FCNC decays are those with a single hadron or no hadron in the final state. Because the transition proceeds through a neutral current, the electric charges of the non-hadronic final states sum to zero. The most relevant non-hadronic final states are:

\begin{itemize}

\item A pair of charged leptons $\ell^+\ell^-$ with $\ell=e, \mu, \tau$.
This final state can appear in a semileptonic decay like $B \to K \mu^+\mu^-$, or in a leptonic decay like $B_s \to \mu^+\mu^-$.

The theoretical uncertainties are well under control in the leptonic decays of $B$ mesons, as the only hadronic parameters involved are the meson decay constants, which are known with good precision from lattice QCD~\cite{FlavourLatticeAveragingGroupFLAG:2024oxs}. However, leptonic decays of scalar mesons are even rarer than other rare decays due to helicity suppression: they are forbidden in the limit of massless final-state fermions and thus suppressed by the small lepton masses. The resulting tiny branching ratios pose a significant experimental challenge. By contrast, leptonic decays of $\pi$, $K$, and $D$ mesons are usually dominated by long-distance contributions, which introduce substantial theoretical uncertainties~\cite{Hoferichter:2021lct,Hoferichter:2023wiy,Burdman:2001tf}.

Semileptonic decays have considerably larger decay rates than leptonic ones, which makes them experimentally more accessible.
However, they pose additional theoretical challenges. As in semileptonic charged-current decays, the predictions rely on momentum-dependent hadronic form factors. In addition, semileptonic neutral-current decays involve non-perturbative long-distance contributions from quark-antiquark pairs, which are coupled to the $\ell^+\ell^-$ state via a virtual photon. These contributions give rise to quarkonium resonances in the dilepton spectrum.
Reasonably reliable predictions exist only away from these resonances, where the long-distance effects are estimated to be small.
Some of these theoretical issues can be avoided by considering \emph{inclusive} decays, where one sums over all hadronic final states with the same flavour quantum numbers. In particular for $B$ and $D$ mesons, the inclusive meson decay can be approximated by the partonic decay of the heavy $b$ or $c$ quark using the Heavy Quark Expansion~\cite{Chay:1990da,Bigi:1992su,Manohar:1993qn}.
While such inclusive decays are experimentally accessible at $e^+e^-$ colliders in experiments like Belle~II, studying them at hadron colliders in experiments like LHCb is very challenging.

\item A photon $\gamma$. The theoretical predictions for these \emph{radiative decays} share many features with those to $\ell^+\ell^-$. This connection arises because one contribution to the semileptonic decays stems from a virtual photon coupled to $\ell^+\ell^-$, while radiative decays involve the same photon on shell.
Consequently, exclusive radiative decays suffer from sizable hadronic uncertainties, while inclusive radiative decays are theoretically much better under control.

\item A neutrino pair $\nu \bar\nu$. While decays to neutrinos are experimentally more challenging than those involving charged leptons, their theoretical predictions are well under control. Since neutrinos cannot couple to quark-antiquark pairs via a virtual photon, these decays are free from the problematic non-perturbative long-distance contributions that affect decays to $\ell^+\ell^-$ or $\gamma$, leading to particularly clean theoretical predictions for semileptonic decays with $\nu \bar\nu$ in the final state.

\end{itemize}

\section{Probing New Physics with Flavour Violating Processes} \label{probe}

Flavour physics plays a central role in the search for physics beyond the Standard Model. The SM provides precise, testable predictions for a wide range of processes involving quark and lepton flavour transitions, many of which are strongly suppressed by approximate symmetries. This makes them exceptionally sensitive to potential contributions from new physics. In particular, FCNCs serve as powerful indirect probes of new physics.

When experimental results agree with SM expectations, they impose stringent constraints on possible new sources of flavour violation, significantly restricting the allowed parameter space for new physics models. Conversely, any observed deviation from the SM prediction would provide an indirect indication of new physics, pointing to new degrees of freedom potentially at mass scales well beyond the reach of current collider experiments.
In this way, flavour physics offers a complementary strategy to direct searches: while collider probes are limited by available collision energies, flavour observables remain sensitive to much higher mass scales, provided the new physics couples to the SM in a flavour violating way.

If the new physics lies at scales much higher than those directly probed by experiments, its effects can be systematically captured using effective field theory (EFT) techniques. The Standard Model Effective Field Theory (SMEFT)~\cite{Buchmuller:1985jz,Grzadkowski:2010es} describes possible new physics effects above the electroweak scale, encoding them in a tower of higher-dimensional operators constructed from SM fields. Below the electroweak scale, the Weak Effective Theory (WET)\footnote{%
The WET is sometimes referred to as the Low-Energy Effective Field Theory~(LEFT)~\cite{Jenkins:2017jig}, but we adopt the more traditional name, which reflects its importance as a theory of weak interactions and avoids confusion with a number of other low-energy EFTs.
}~\cite{Buchalla:1995vs,Aebischer:2017gaw,Jenkins:2017jig} takes over, after integrating out the heavy SM particles $W$, $Z$, Higgs, and top quark. This EFT framework allows for a model-independent description of heavy new physics contributions to low-energy observables, organized by the mass dimension of the effective operators and suppressed by the appropriate powers of the new physics scale.

While flavour-changing processes are highly sensitive to heavy new physics, they can also serve as valuable probes of light, feebly interacting new particles. Such particles (including axion-like particles (ALPs), dark photons,  etc.) could for example manifest in rare meson decays through signatures like missing energy events, displaced vertices, or unexpected resonances in dilepton spectra.

In the following sections, we review key classes of flavour observables that are most sensitive to new physics. Section~\ref{heavy} focuses on indirect probes of heavy new physics within the SMEFT and WET frameworks, while Section~\ref{light} turns to the search for light dark sector particles in rare flavour-violating decays. Together, these approaches illustrate the broad reach and unique discovery potential of flavour physics in the quest for physics beyond the Standard Model.
At the same time, the stringent experimental bounds from flavour physics raise the question of how new physics could evade them without being pushed to very high energy scales. In Section~\ref{sec:NP_problem} we give a brief overview of this so-called {\it New Physics Flavour Puzzle} and discuss strategies to address it.

\subsection{Probing heavy new physics with flavour} \label{heavy}

Sufficiently heavy new physics can be captured model independently using the SMEFT framework. The impact of heavy new physics is encoded in higher-dimensional operators constructed from SM fields, suppressed by inverse powers of the new physics scale $\Lambda$
\begin{equation}
\mathcal L_\text{SMEFT} = \mathcal L_\text{SM} + \frac{1}{\Lambda}\sum_i C_i^{(5)} \mathcal Q_i^{(5)} + \frac{1}{\Lambda^2}\sum_i C_i^{(6)} \mathcal Q_i^{(6)} + ~\dots ~.
\end{equation}
The $\mathcal Q_i^{(n)}$ are operators of mass dimension $n > 4$ and are invariant under the SM gauge group ${\rm SU}(3)_c \times {\rm SU}(2)_L \times {\rm U}(1)_Y$. The $C_i^{(n)}$ are the corresponding Wilson coefficients parameterizing the strength of the effective interactions.
Operators of higher dimension are increasingly suppressed by additional powers of $\Lambda$. Notably, at dimension five there exists a single operator, the so-called Weinberg operator~\cite{Weinberg:1979sa}. It violates lepton number by two units and gives neutrino masses after electroweak symmetry breaking. The leading operators that are relevant for quark and charged lepton flavour physics appear at mass dimension six.

Below the electroweak scale, after integrating out the $W$, $Z$, Higgs boson, and top quark, the appropriate EFT is the WET. In the WET, the renormalizable interactions of the light SM fields (the gluons, the photon, the leptons and the five\footnote{%
At energy scales below $m_b$, the bottom quark is integrated out, and a version of the WET with only four light quark flavours -- sometimes referred to as ``WET-4'' -- is used. Similarly, below $m_c \approx m_\tau$, the charm quark and tau lepton are integrated out, leading to ``WET-3,'' which contains three light quark flavours and two light charged lepton flavours.
} light quark flavours) remain, while the effects of heavy SM and BSM physics are captured by a set of higher-dimensional operators, analogous to those in the SMEFT but invariant under the unbroken ${\rm SU}(3)_c \times {\rm U}(1)_\text{em}$ gauge group. Among these, a subset governs flavour-changing processes at low energies.
The most important classes of observables sensitive to these higher-dimensional operators include meson-antimeson oscillations, flavour-changing neutral current decays of hadrons and leptons, and charged-current flavour-violating decays.

\subsubsection{Meson-antimeson oscillations} \label{mix}

Meson-antimeson oscillations, as introduced in Section~\ref{sec:FCNC}, are highly suppressed processes within the SM, occurring via $\Delta F = 2$ transitions. Their suppression makes them exceptionally sensitive probes of potential new physics contributions, particularly in scenarios involving new sources of flavour violation. The sensitivity of these processes can be characterized by effective interaction strengths, similar to the Fermi constant $G_F = \frac{\sqrt{2}}{8}\frac{g^2}{m_W^2}$, which characterizes the strength of effective interactions generated by tree-level $W$ boson exchange. The effective couplings can be viewed as the Wilson coefficients of dimension-six operators in the WET.

Taking into account loop suppression factors, the appropriate CKM matrix elements, and the quadratic GIM mechanism, one can estimate the SM contribution to these processes. For instance, in $B_s$, $B_d$, and neutral kaon mixing, the effective coupling strengths can be estimated as
\begin{eqnarray}
 G_{b \to s}^{\Delta F = 2} &\sim& \frac{\alpha}{4\pi} \frac{1}{s_W^2} \frac{m_t^2}{m_W^2} |V_{tb} V_{ts}^*|^2\, G_F \simeq 1.8 \times 10^{-5} \times G_F \simeq \frac{1}{(70~\text{TeV})^2} ~, \label{eq:G2_bs}  \\
 G_{b \to d}^{\Delta F = 2}  &\sim& \frac{\alpha}{4\pi} \frac{1}{s_W^2} \frac{m_t^2}{m_W^2} |V_{tb} V_{td}^*|^2\, G_F \simeq 7.1 \times 10^{-7} \times G_F \simeq \frac{1}{(350~\text{TeV})^2} ~, \label{eq:G2_bd} \\
 G_{s \to d}^{\Delta F = 2}  &\sim& \frac{\alpha}{4\pi} \frac{1}{s_W^2} \frac{m_t^2}{m_W^2} |V_{ts} V_{td}^*|^2\, G_F \simeq 1.1 \times 10^{-9} \times G_F \simeq \frac{1}{(9,000~\text{TeV})^2} ~, \label{eq:G2_sd}
\end{eqnarray}
where $\alpha=\frac{e^2}{4\pi}$ is the fine-structure constant and $s_W = \sin \theta_W$ is the sine of the weak mixing angle defined by $\cos\theta_W = \frac{m_W}{m_Z}$.
These estimates imply that NP models with tree-level, $O(1)$ flavour-changing couplings could potentially leave observable imprints even if the new physics scale lies at tens to thousands of TeV. It is important to emphasize, however, that these are generic benchmarks. The true sensitivity could be limited to lower scales if the new physics flavour-violating couplings are suppressed, or extend to even higher scales, provided that small deviations from SM predictions can be reliably disentangled through high precision measurements and high precision SM predictions.

The primary observables in meson mixing are the mass differences $\Delta M_s$, $\Delta M_d$, and $\Delta M_K$, as well as CP-violating quantities like the mixing phases in the $B_s$ and $B_d$ systems, and the parameter $\epsilon_K$ in the kaon sector. All mass differences have been measured with impressive precision~\cite{ParticleDataGroup:2024cfk, HeavyFlavorAveragingGroupHFLAV:2024ctg}, and the challenge today lies in reducing theoretical uncertainties in the SM predictions. For $B$-meson systems, the dominant uncertainties stem from lattice QCD calculations of hadronic matrix elements~\cite{FlavourLatticeAveragingGroupFLAG:2024oxs} and, to a lesser extent, CKM matrix elements, yielding an overall precision at the 10\% level. In contrast, the prediction for $\Delta M_K$ is complicated by sizable long-distance hadronic effects, which remain difficult to compute reliably~\cite{Bai:2014cva, Wang:2022lfq}.
The CP-violating observables offer complementary probes and are known experimentally with high precision~\cite{ParticleDataGroup:2024cfk, HeavyFlavorAveragingGroupHFLAV:2024ctg}. While the mixing phases in the $B$-meson systems are theoretically clean, limited predominantly by CKM uncertainties at the $O(1^\circ)$ level, the prediction for $\epsilon_K$ involves both CKM inputs and hadronic matrix elements~\cite{FlavourLatticeAveragingGroupFLAG:2024oxs}, currently leading to about 10\% precision in the SM~\cite{Buras:2024mnq}.

In addition to the neutral kaon and $B$-meson systems, neutral $D$-meson mixing provides a complementary probe of new physics in the up-quark sector~\cite{Golowich:2007ka}. Unlike $K$ and $B$ mixing, which are dominated by short-distance contributions in the SM, $D^0$–$\bar D^0$ oscillations proceed via down-type quark loops, leading to a much stronger GIM suppression. As a consequence, long-distance hadronic effects are expected to play a significant, if not dominant, role in the SM contribution, complicating precise theoretical predictions~\cite{Falk:2001hx, Lenz:2020awd, DiCarlo:2025mvt}. Experimentally, the mixing parameters $x = \Delta M_D / \Gamma_D$ and $y = \Delta \Gamma_D / (2\Gamma_D)$ have been measured with increasing precision, firmly establishing the presence of $D$-meson mixing~\cite{HeavyFlavorAveragingGroupHFLAV:2024ctg}. However, the theoretical uncertainties associated with long-distance contributions leave room for possible new physics effects at a level comparable to the SM expectation. CP violation in charm mixing offers a sensitive test of new physics. In the SM, CP-violating effects in the charm sector are predicted to be small~\cite{Kagan:2020vri}. Any significant observation of CP violation in $D^0$–$\bar D^0$ mixing would be a clear indication of physics beyond the SM. Current measurements of CP-violating observables have not yet shown significant deviations from SM expectations, but they continue to place important constraints on new physics models that contain up-sector flavour and CP violation.

Potential NP contributions to meson mixing are typically parametrized using eight independent dimension-six four-fermion operators in the WET, invariant under the unbroken ${ \rm SU}(3)_c \times {\rm U}(1)_\text{em}$ gauge symmetry. For example, in the kaon sector~\cite{Ciuchini:1997bw, Buras:2000if}
\begin{align}
\mathcal Q^{VLL} &= (\bar s \gamma_\mu P_L d)(\bar s \gamma^\mu P_L d) ~, \quad
& \mathcal Q^{VRR} &= (\bar s \gamma_\mu P_R d)(\bar s \gamma^\mu P_R d) ~, \\
\mathcal Q^{LR}_1 &= (\bar s \gamma_\mu P_L d)(\bar s \gamma^\mu P_R d) ~, \quad
&\mathcal Q^{LR}_2 &= (\bar s P_L d)(\bar s P_R d) ~, \\
\mathcal Q^{SLL}_1 &= (\bar s P_L d)(\bar s P_L d) ~, \quad
&\mathcal Q^{SRR}_1 &= (\bar s P_R d)(\bar s P_R d) ~, \\
\mathcal Q^{SLL}_2 &= (\bar s \sigma_{\mu\nu} P_L d)(\bar s \sigma^{\mu\nu} P_L d) ~, \quad
&\mathcal Q^{SRR}_2 &= (\bar s \sigma_{\mu\nu} P_R d)(\bar s \sigma^{\mu\nu} P_R d) ~.
\end{align}
Analogous operators can be defined for the other meson systems. These operators encompass all independent Lorentz and colour structures for $\Delta F = 2$ transitions.

A critical feature of meson mixing analyses is their dependence on CKM parameters. As a result, the most robust constraints on new physics are obtained from global CKM fits that include possible new physics contributions to mixing observables~\cite{UTfit:2007eik, Charles:2020dfl}. The good agreement between SM predictions and experimental measurements in these systems allows stringent limits to be placed on the corresponding Wilson coefficients. Current bounds reach scales as high as few $10^5$\,TeV from $\epsilon_K$, few $10^3$\,TeV from $B_d$ mixing, and up to $10^3$\,TeV from $B_s$ mixing, under the assumption of $O(1)$ flavour-changing couplings.

\begin{figure}[tb]
	\centering
	\includegraphics[width=0.25\textwidth]{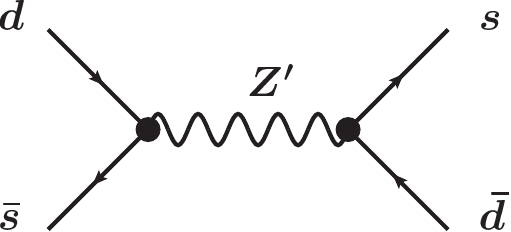} \qquad\qquad	\includegraphics[width=0.25\textwidth]{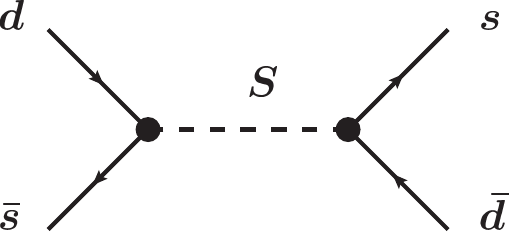} \qquad\qquad
    \includegraphics[width=0.25\textwidth]{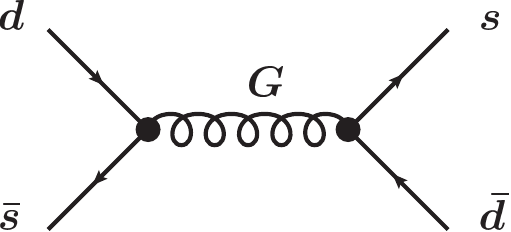} \\[16pt]
    \includegraphics[width=0.24\textwidth]{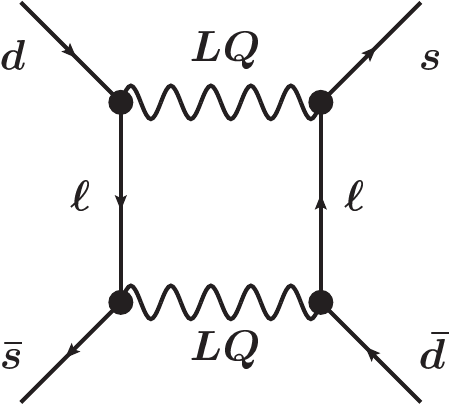} \qquad\qquad
    \includegraphics[width=0.24\textwidth]{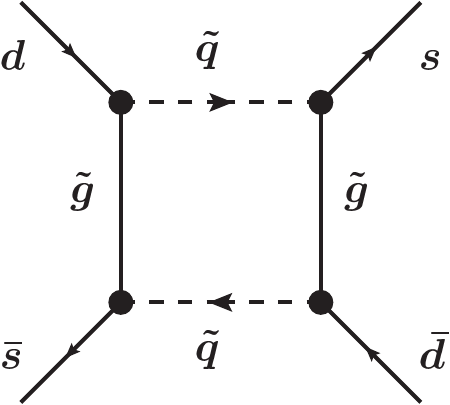}
    \qquad\qquad
    \raisebox{24pt}{\includegraphics[width=0.16\textwidth]{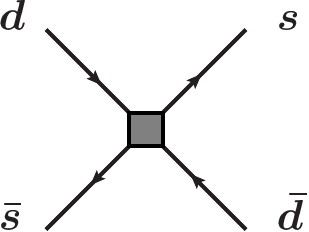}}
    \caption{Examples of new physics contributions to kaon mixing. Top row: tree-level contributions of a $Z^\prime$ gauge boson, a neutral scalar boson, and a Kaluza-Klein gluon; bottom row left and centre: 1-loop leptoquark box contributions and gluino-squark box contributions in supersymmetric models; bottom row right: an effective four-fermion operator contribution to kaon mixing.}
	\label{fig:mixing}
\end{figure}

To illustrate the connection between new physics models and the effective operator language, consider for example a $Z^\prime$ gauge boson with flavour-changing couplings to right-handed strange and down quarks (as in the diagram in the left of the top row in Fig.~\ref{fig:mixing}). Integrating out the $Z^\prime$ one finds a contribution to the Wilson coefficient of the operator $\mathcal Q^{VRR}$
\begin{equation}
g_{sd}^R Z^\prime_\mu  (\bar s \gamma^\mu P_R d) \quad \rightarrow \quad \frac{\big(g_{sd}^R\big)^2}{2 m_{Z^\prime}^2} (\bar s \gamma_\mu P_R d)(\bar s \gamma^\mu P_R d) ~ ,  \quad \Rightarrow \quad \frac{\mathcal C^{VRR}}{\Lambda^2} = \frac{\big(g_{sd}^R\big)^2}{2 m_{Z^\prime}^2} ~.
\end{equation}

Similar considerations apply to other models of heavy new physics. Many well-motivated new physics models can induce contributions to neutral meson mixing, with several more examples shown in Fig.~\ref{fig:mixing}. At tree level, these include scenarios with heavy neutral scalars, additional $Z^\prime$ gauge bosons (including vector resonances in models with partial compositeness), or Kaluza-Klein excitations of gluons in extra-dimensional frameworks. One-loop contributions can arise from models with leptoquarks or from supersymmetric extensions like the Minimal Supersymmetric Standard Model (MSSM). In the MSSM, for instance, box diagrams involving gluinos and squarks can mediate meson mixing, and in the presence of sizable flavour mixing in the squark sector, constraints on supersymmetric particle masses can reach the PeV scale~\cite{Altmannshofer:2013lfa, Isidori:2019pae}.

\subsubsection{Flavour-changing neutral current decays}

As already discussed in Section~\ref{sec:FCNC}, FCNC decays of quarks are highly suppressed in the SM, as they occur only at the loop level and are further suppressed by small CKM matrix elements. This suppression makes these processes exceptionally sensitive to virtual contributions from new heavy particles, turning them into prime indirect probes of physics beyond the SM~\cite{Blake:2016olu, Altmannshofer:2022hfs}.

A simple way to estimate the potential new physics reach of FCNC decays is to compare the effective SM couplings governing these processes to possible tree-level contributions from new physics with generic $\mathcal{O}(1)$ flavour-violating couplings. For $b \to s$, $b \to d$, and $s \to d$ transitions, one finds
\begin{eqnarray}
G_{b \to s}^{\Delta F = 1} &\sim& \frac{\alpha}{4\pi} \frac{m_t^2}{m_W^2} |V_{tb} V_{ts}^*| G_F \simeq 9.6 \times 10^{-5} \times G_F \simeq \frac{1}{(30~\text{TeV})^2} ~, \label{eq:G_bs} \\
G_{b \to d}^{\Delta F = 1} &\sim& \frac{\alpha}{4\pi} \frac{m_t^2}{m_W^2} |V_{tb} V_{td}^*| G_F \simeq 1.9 \times 10^{-5} \times G_F \simeq \frac{1}{(67~\text{TeV})^2} ~, \label{eq:G_bd} \\
G_{s \to d}^{\Delta F = 1} &\sim& \frac{\alpha}{4\pi} \frac{m_t^2}{m_W^2} |V_{ts} V_{td}^*| G_F \simeq 7.7 \times 10^{-7} \times G_F \simeq \frac{1}{(330\text{TeV})^2} ~. \label{eq:G_sd}
\end{eqnarray}
These estimates suggest that rare $B$ decays can probe new physics scales up to several tens of TeV~\cite{Altmannshofer:2017yso, DiLuzio:2017chi}, while rare kaon decays can be sensitive to mass scales well beyond 100\,TeV.

Among the most powerful FCNC probes are rare decays of kaons and $B$ mesons, where SM predictions are both precise and highly suppressed, resulting in excellent sensitivity to BSM effects. In the kaon sector, particularly clean channels include $K^+ \to \pi^+ \nu \bar \nu$ and $K_L \to \pi^0 \nu \bar \nu$~\cite{Aebischer:2022vky, DAmbrosio:2023irq}. These decays have very small theoretical uncertainties, as the relevant hadronic matrix elements can be related to well-measured $K_{\ell3}$ decays, leaving CKM parameters as the dominant source of uncertainty.

In the $B$ sector, key observables include the purely leptonic decays $B_s \to \mu^+ \mu^-$ and $B^0 \to \mu^+ \mu^-$~\cite{Altmannshofer:2017wqy, Fleischer:2017ltw}. These are theoretically exceptionally clean, with uncertainties dominated by CKM input. Owing to helicity suppression in the SM, these decays are especially sensitive to new physics that can lift this suppression, such as scalar or pseudoscalar interactions. Other important FCNC $b \to s$ decays include $B \to K^{(*)} \ell^+ \ell^-$, $B_s \to \phi \ell^+ \ell^-$, and $B \to K^{(*)} \nu \bar \nu$, where not only the branching ratios but also the rich kinematic structure, e.g. $q^2$ distributions, angular observables, and CP asymmetries, offer multiple windows for detecting BSM effects~\cite{Bobeth:2007dw, Bobeth:2008ij, Altmannshofer:2008dz, Matias:2012xw}.

A crucial distinction exists between inclusive decays, like $B \to X_s \ell^+ \ell^-$ that can be accessed in the future at Belle~II~\cite{Belle-II:2018jsg}, and where hadronic uncertainties are well under control~\cite{Lee:2006gs, Huber:2020vup, Huber:2024rbw}, and exclusive decays where high precision measurements already exist from LHCb, but where uncertainties from local and non-local form factors can be more substantial~\cite{Jager:2012uw, Lyon:2014hpa, Gubernari:2020eft, Gubernari:2022hxn, Gubernari:2023puw}. Angular analyses and the use of optimized observables allow partial mitigation of these hadronic effects, enhancing sensitivity to new physics.
The exclusive $b \to s \nu\bar\nu$ decays also play an important role in probing new physics~\cite{Altmannshofer:2009ma, Buras:2014fpa, Browder:2021hbl, Bause:2021cna, Becirevic:2023aov, Buras:2024ewl} and are a key target for Belle~II~\cite{Belle-II:2018jsg}.
Compared to decays with charged leptons in the final state, the di-neutrino modes are not affected by charm loop effects and thus theoretically cleaner.
The $b \to s \nu\bar\nu$ modes probe not only heavy new physics but also give access to light dark sectors, as decays such as $b \to s X$ (where $X$ is a neutral, invisibly decaying, or long-lived light particle) can mimic the missing-energy signature of di-neutrino processes.

Rare charm decays, such as $D^0 \to \mu^+ \mu^-$ or $D \to \pi \ell^+ \ell^-$, provide complementary probes of up-type quark FCNCs~\cite{Gisbert:2024kob}, which are even more suppressed in the SM due to an effective GIM mechanism. A key challenge is the dominance of long-distance resonance contributions, often exceeding the short-distance physics of interest by orders of magnitude. Consequently, the focus is often on so-called ``null tests'': observables that are highly suppressed in the SM by symmetries or selection rules and largely free from long-distance uncertainties. Examples include lepton universality ratios, specific angular observables, and di-neutrino modes like $D \to \pi \nu \bar \nu$~\cite{Bause:2019vpr, Bause:2020xzj}. Importantly, the new physics that could give observable effects in rare charm decays is also effectively probed by searches in di-lepton production at the LHC~\cite{Fuentes-Martin:2020lea}.

A convenient theoretical framework for analysing FCNC decays is the effective weak Hamiltonian that contains the WET operators relevant for the decays of interest. For $b \to s$ transitions, it reads
\begin{equation}
\mathcal H_\text{eff}^{\Delta F=1} = -\frac{4 G_F}{\sqrt{2}} V_{tb} V_{ts}^* \frac{\alpha}{4 \pi} \sum_i \left( \big( C_i^\text{SM} + \Delta C_i\big) \mathcal O_i + \Delta C_i^\prime \mathcal O_i^\prime \right) + \text{h.c.} ~.
\end{equation}
The most relevant dimension-six operators include:
\begin{itemize}
\item Dipole operators, dominant in radiative decays like $b \to s \gamma$, and contributing to $b \to s \ell^+ \ell^-$ transitions
\begin{equation}
\mathcal O_7 = \frac{1}{e} m_b (\bar s \sigma_{\mu\nu} P_R b) F^{\mu\nu} ~,\quad \mathcal O_7^\prime = \frac{1}{e} m_b (\bar s \sigma_{\mu\nu} P_L b) F^{\mu\nu} ~,
\end{equation}
\item Semileptonic vector and axial-vector operators, governing the $b \to s \ell^+ \ell^-$ decays
\begin{equation}
\mathcal O_9 = (\bar s \gamma_\mu P_L b)(\bar \ell \gamma^\mu \ell) ~,\quad \mathcal O_{10} = (\bar s \gamma_\mu P_L b)(\bar \ell \gamma^\mu \gamma_5 \ell) ~, \quad \mathcal O_9^\prime = (\bar s \gamma_\mu P_R b)(\bar \ell \gamma^\mu \ell) ~,\quad \mathcal O_{10}^\prime = (\bar s \gamma_\mu P_R b)(\bar \ell \gamma^\mu \gamma_5 \ell) ~,
\end{equation}
\item Scalar and pseudoscalar operators, crucial in leptonic decays such as $B_s \to \mu^+ \mu^-$
\begin{equation}
\mathcal O_S = (\bar s P_R b)(\bar \ell \ell) ~,\quad \mathcal O_P = (\bar s P_R b)(\bar \ell \gamma_5 \ell) ~, \quad \mathcal O_S^\prime = (\bar s P_L b)(\bar \ell \ell) ~,\quad \mathcal O_P^\prime = (\bar s P_L b)(\bar \ell \gamma_5 \ell) ~,
\end{equation}
\item Neutrino operators, relevant for $b \to s \nu \bar\nu$ modes
\begin{equation}
\mathcal O_\nu = (\bar s \gamma_\mu P_L b)(\bar \nu \gamma^\mu P_L \nu) ~,\quad \mathcal O_\nu^\prime = (\bar s \gamma_\mu P_R b)(\bar \nu \gamma^\mu P_L \nu) ~,
\end{equation}
\end{itemize}
The leptons in the above operators can be either electrons, muons, or taus. Lepton flavour violating interactions can also be included. The ``un-primed'' operators correspond to the operators already present in the SM, and we have expressed their Wilson coefficients as a sum of the SM contributions $C_i^\text{SM}$ and potential new physics contributions $\Delta C_i$. In contrast, the primed operators, with Wilson coefficients $\Delta C_i^\prime$, involve right-handed quark currents and represent purely new physics effects. Allowing for complex Wilson coefficients enables the study of CP-violating observables, as the imaginary parts of these coefficients can induce CP asymmetries in rare decays.

Global fits to the Wilson coefficients, using experimental data from various FCNC decay channels, offer a systematic way to search for new physics~\cite{Ciuchini:2022wbq, Greljo:2022jac, Alguero:2023jeh, Altmannshofer:2023uci, Guadagnoli:2023ddc, Hurth:2023jwr, Bordone:2024hui}.
Since 2013, persistent tensions are observed in $b \to s \mu^+ \mu^-$ observables, such as angular distributions in $B \to K^* \mu^+ \mu^-$~\cite{LHCb:2013ghj, LHCb:2020lmf, CMS:2024atz} and the branching ratios of related channels~\cite{LHCb:2014cxe, LHCb:2021zwz}. As of 2025, the origin of the tensions continues to be debated: they could for example be due to poorly understood hadronic effects, or constitute a genuine new physics effect.

\begin{figure}[tb]
	\centering
	\includegraphics[width=0.24\textwidth]{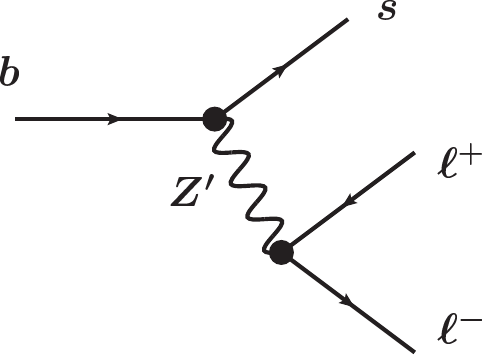} \qquad\qquad
    \raisebox{24pt}{\includegraphics[width=0.26\textwidth]{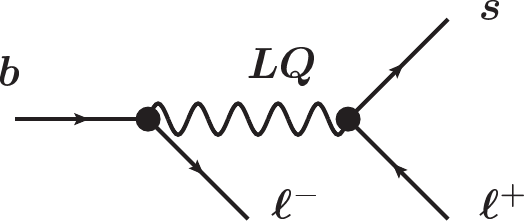}}
    \qquad\qquad
    \raisebox{24pt}{\includegraphics[width=0.16\textwidth]{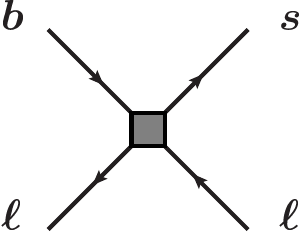}}
    \caption{Examples of new physics contributions to $b \to s \ell^+ \ell^-$ decays. Left: tree-level contribution of a $Z^\prime$ gauge boson; centre: tree-level contribution of a leptoquark; right: effective contribution of a four-fermion operator.}
	\label{fig:bsll}
\end{figure}

Several classes of BSM scenarios can lead to observable contributions in rare decays. The most prominent ones are leptoquarks and $Z^\prime$ gauge bosons that can contribute at the tree level, see Fig.~\ref{fig:bsll}. In the MSSM, one often finds sizable contributions to the decay $B_s \to \mu^+ \mu^-$ from one-loop Higgs penguin diagrams induced by flavour-violating couplings of squarks and gauginos.

\subsubsection{Flavour-changing charged current decays}

Flavour-changing charged current decays, mediated at tree level in the SM via the exchange of a $W$ boson, play a central role in the precision determination of the CKM matrix elements. Processes such as $K \to \pi \ell \nu$, $B \to D^{(*)} \ell \nu$, and $B \to \pi \ell \nu$ are used to extract $|V_{us}|$, $|V_{cb}|$, and $|V_{ub}|$, respectively. Accurate determinations of these matrix elements are vital for testing the unitarity of the CKM matrix and for providing the CKM input for SM predictions of rare decays and meson mixing observables.

While charged current decays are dominated by SM tree-level contributions, they can also serve as sensitive probes for BSM effects in observables where the SM contributions are for example helicity suppressed, or in the context of theoretically clean lepton flavour universality tests.
Notable examples include:

\begin{itemize}
\item Lepton flavour universality ratios of semileptonic decays of $B$ mesons $R_{D^{(*)}} = \text{BR}(B \to D^{(*)} \tau \nu)/\text{BR}(B \to D^{(*)} \ell \nu)$: Starting with a BaBar measurement in 2012~\cite{BaBar:2012obs}, experimental results have shown a persistent tension with SM predictions at the $\sim 3\sigma$ level, hinting at possible BSM contributions~\cite{BaBar:2013mob, Belle:2015qfa, Belle:2016dyj, Belle:2017ilt, Belle:2019rba, LHCb:2023zxo, LHCb:2023uiv, Belle-II:2024ami, Belle-II:2025yjp}. As of 2025, these tensions have not been resolved.
\item The helicity suppressed decays $B \to \tau \nu$ and $B \to \mu \nu$ which are known to be sensitive probes of charged Higgs bosons that can contribute at tree level and lift the helicity suppression.
\item Lepton flavour universality ratios in leptonic kaon and pion decays $R_{K_{\ell 2}} = \text{BR}(K \to e \nu)/\text{BR}(K \to \mu \nu)$ and $R_{\pi_{\ell 2}} = \text{BR}(\pi \to e \nu)/\text{BR}(\pi \to \mu \nu)$~\cite{Bryman:2021teu}. Existing and proposed precision measurements of these ratios~\cite{NA62:2012lny, PiENu:2015seu, PIONEER:2022yag} test the universality of lepton couplings to the $W$ boson at the per-mille level and constrain many new physics scenarios.
\end{itemize}

Although charged current decays are generally less sensitive to heavy new physics compared to loop-suppressed FCNC processes, their clean theoretical status and the availability of high-precision experimental data make them valuable components of global flavour analyses. Moreover, potential deviations from lepton universality in these decays can provide crucial complementary information to FCNC measurements in constructing a coherent picture of new physics effects in the flavour sector.

\begin{figure}[tb]
	\centering
	\includegraphics[width=0.24\textwidth]{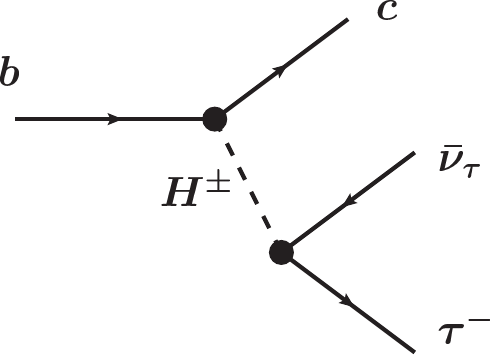} \qquad\qquad
    \raisebox{24pt}{\includegraphics[width=0.26\textwidth]{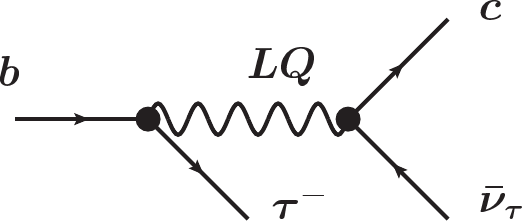}}
    \qquad\qquad
    \raisebox{24pt}{\includegraphics[width=0.16\textwidth]{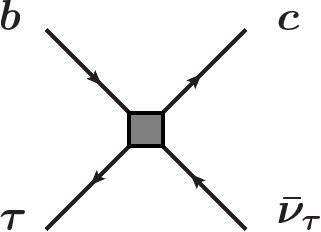}}
	\caption{Examples of new physics contributions to $b \to c \ell \nu$ decays. Left: tree-level contribution of a charged Higgs; centre: tree-level contribution of a leptoquark; right: effective contribution of a four-fermion operator.}
	\label{fig:bclnu}
\end{figure}

Several new physics models can affect charged current decays either through modifications of the $W$ boson couplings or the introduction of new charged mediators as shown in Fig.~\ref{fig:bclnu}. One class of scenarios involves charged Higgs bosons, as predicted in extended Higgs sectors of the MSSM or generic Two Higgs Doublet Models (2HDMs).
Another well-motivated class of models involves leptoquarks which can also contribute at tree level.

\subsubsection{Comparison of new physics effects in the flavour transitions}

The various classes of new physics models mentioned in the previous sections leave distinct imprints on flavour-changing processes. A qualitative comparison of their typical patterns of contributions to meson mixing, FCNC decays, and charged current decays offers valuable insight into their phenomenological profiles and the complementarity of different flavour observables. The following provides a schematic overview of some representative scenarios (see also Table~\ref{tab:NP_patterns}).

\renewcommand{\arraystretch}{1.3}
\begin{table}[b]
\TBL{
	\caption{Characteristic patterns of new physics contributions to flavour-changing processes in representative BSM scenarios}
    \label{tab:NP_patterns}
}{%
    \begin{tabular*}{\columnwidth}{@{\extracolsep\fill}lccc@{}}
        \toprule
        Model / Scenario & Meson Mixing & FCNC Decays & Charged Current Decays \\
        \colrule
        Leptoquarks & Loop  & Tree  & Tree  \\
        $Z^\prime$ bosons & Tree & Tree  & Loop  \\
        Charged Higgs bosons & Loop  & Loop  & Tree  \\
        MSSM & Loop  & Loop & Tree \\
        \botrule
	\end{tabular*}
}{}
\end{table}
\renewcommand{\arraystretch}{1.0}

\begin{itemize}
\item {\it Leptoquark Models:} Leptoquarks, which couple simultaneously to quarks and leptons, can contribute to FCNC decays and charged current decays at the tree level, leading to potentially large effects in $B$, $K$, and $D$ decays. In contrast, contributions to neutral meson mixing arise only at the loop level, typically making them a less effective probe of leptoquark effects.
\item {\it $Z^\prime$ Models}: New neutral gauge bosons, or other spin-1 resonances with flavour-violating couplings can induce tree-level contributions to both meson mixing and FCNC decays, often leading to correlated effects in processes like $B_s - \bar B_s$ mixing and $b \to s \ell^+ \ell^-$ decays. However, since $Z^\prime$ bosons do not couple to charged current transitions, they do not appreciably affect charged current decays such as $b \to c \ell \nu$ and the corresponding lepton flavour universality ratios $R_{D}$ and $R_{D^*}$.
\item {\it Charged Higgs Bosons:} Extensions of the Higgs sector, such as a 2HDM, introduce charged Higgs bosons $H^\pm$ that couple to fermions. These states primarily affect charged current decays at the tree level, in particular modifying the rates of leptonic decays like $B \to \tau \nu$ through scalar current contributions. Contributions to neutral meson mixing and FCNC decays are loop-suppressed and less relevant.
\item {\it Minimal Supersymmetric Standard Model:} In the MSSM with generic flavour-violating soft terms, sizable contributions can arise in all three categories of processes. Meson mixing can receive large loop contributions from gluino, chargino, and neutralino box diagrams. FCNC decays, $B_s \to \mu^+ \mu^-$ in particular, are affected by loop contributions from Higgs penguin diagrams. Charged current decays can also be affected through charged Higgs exchange at the tree level. The MSSM thus offers a rich and correlated pattern of effects, with specific features depending on the underlying flavour structure of the soft-breaking terms.
\item {\it Generic SMEFT/WET Analyses (see e.g.~\cite{Aebischer:2025qhh, deBlas:2025xhe}):} In a model-independent effective field theory framework, contributions to meson mixing, FCNC decays, and charged current processes are parametrized by independent higher-dimensional operators. This allows for a flexible pattern of effects, with possible correlations depending on the assumed symmetry structure (e.g., Minimal Flavour Violation, ${\rm U}(2)$ flavour symmetries, etc.). Global fits to Wilson coefficients in these frameworks can identify preferred directions in parameter space that hint at the underlying UV dynamics.
\end{itemize}
In summary, different classes of new physics scenarios exhibit characteristic hierarchies and patterns in their contributions to flavour-changing processes. These patterns enable powerful cross-checks and complementarity between various flavour observables and can help discriminate between competing models in the event of deviations from Standard Model predictions.

\subsection{Probing light new physics with flavour} \label{light}

The possibility of a dark sector composed of light, weakly coupled particles has emerged as a major focus in the search for physics beyond the Standard Model. Detecting such elusive states presents significant experimental challenges, but also offers unique opportunities. Flavour-changing processes, in particular, are powerful probes of new light degrees of freedom that interact very feebly with SM particles, and dedicated search strategies are being developed at both current and future experiments.

A generic feature of light, weakly coupled states is their long lifetimes, often leading to missing energy signatures or displaced vertices. Rare flavour-violating decays such as $K \to \pi + \text{invisible}$, $B \to K + \text{invisible}$, and $\mu \to e + \text{invisible}$ are especially sensitive to a broad range of scenarios, including light vector bosons $Z^\prime$, light scalars, light pseudoscalars like axions or axion-like particles (ALPs), and sterile neutrinos. ALPs are particularly popular candidates due to their shift symmetry, which protects their small masses~\cite{MartinCamalich:2020dfe}.
On the experimental side, NA62 and KOTO are leading searches for the charged and neutral $K \to \pi + \text{invisible}$ decays~\cite{NA62:2024pjp, KOTO:2024zbl}, while Belle~II has begun exploring $B \to K + \text{invisible}$~\cite{Belle-II:2023esi}. Similarly, lepton flavour violating processes such as $\mu \to e + \text{invisible}$ can be probed for example at Mu3e~\cite{Perrevoort:2018ttp}.

Generically speaking, rare decays into a light new physics state, $X$, such as $K \to \pi X$ or $\mu \to e X$, are exquisite probes of very feeble couplings and high new physics scales.
Compared to the SM three-body weak decays, which scale as
\begin{equation}
\Gamma(K \to \pi \ell \nu) \sim \frac{g^4}{192\pi^3} |V_{us}|^2 \frac{m_K^5}{m_W^4} ~,\quad \Gamma(B \to D \ell \nu) \sim \frac{g^4}{192\pi^3} |V_{cb}|^2 \frac{m_B^5}{m_W^4} ~,\quad \Gamma(\mu \to e \nu \bar\nu) \sim \frac{g^4}{192\pi^3} \frac{m_\mu^5}{m_W^4} ~,
\end{equation}
the corresponding exotic two-body decay rates scale as
\begin{equation}
\Gamma(K \to \pi X) \sim \frac{g_{sd}^2}{8\pi} m_K ~,\quad \Gamma(B \to K X) \sim \frac{g_{bs}^2}{8\pi} m_B~,\quad \Gamma(\mu \to e X) \sim \frac{g_{\mu e}^2}{8\pi} m_\mu ~,
\end{equation}
where we have assumed that $X$ has renormalizable flavour-changing couplings to the SM quarks and leptons: $g_{sd}$, $g_{bs}$, $g_{\mu e}$.
In such a case, the new exotic decay modes feature striking parametric enhancements of $(m_W/m_K)^4 \simeq 7\times 10^{8}$, $(m_W/m_B)^4 \simeq 5\times 10^{4}$, and $(m_W/m_\mu)^4 \simeq 3\times 10^{11}$, opening up the possibility to probe extremely feeble couplings. If the couplings of $X$ are flavour anarchic, the highest UV scale is typically probed by the lightest initial state that can decay into $X$.
In the case of ALPs that couple to the SM with shift symmetric dimension-five operators one typically has $g_{sd} \to g_{sd} m_K/f_a$, $g_{bs} \to g_{bs} m_B/f_a$, $g_{\mu e} \to g_{\mu e} m_\mu/f_a$, with the axion decay constant $f_a$. Assuming order-one flavour-violating couplings, one can thus probe extremely high UV scales: $f_a \gtrsim 10^{12}$\,GeV from kaon decays~\cite{MartinCamalich:2020dfe}, and $f_a \gtrsim 10^9$\,GeV from muon decays~\cite{Calibbi:2020jvd}.

Important complementary signatures arise if the new light state $X$ can decay visibly within the detector. In this case, rare meson or lepton decays can give rise to displaced vertices, or, if the decay is prompt, to distortions in kinematic distributions, such as anomalous peaks or shoulders in the dilepton invariant mass spectra of decays like $B \to K^{(*)} \mu^+ \mu^-$~\cite{Sala:2017ihs, Datta:2017ezo, Altmannshofer:2017bsz, Crivellin:2022obd}. Also such signatures are being actively pursued, extending sensitivity to a wide range of $X$ lifetimes and couplings.

Finally, it is interesting to note that there are well motivated flavour models that contain naturally light states with flavour-violating couplings that can be probed in the ways discussed above.
As we will discuss in Section~\ref{sec:symmetry}, one way to address the SM flavour puzzle is to use spontaneously broken ${\rm U}(1)$ flavour symmetries.
If the ${\rm U}(1)$ is a global symmetry, one expects that the spectrum of the model contains a light pseudo-Nambu-Goldstone boson, the angular mode of a so-called ``flavon'' field.
If the ${\rm U}(1)$ flavour symmetry is anomalous under QCD, the strong CP problem can also be addressed~\cite{Davidson:1981zd, Wilczek:1982rv}. In this context, the pseudo-Nambu-Goldstone boson is referred to as ``axiflavon''~\cite{Calibbi:2016hwq} or ``flaxion''~\cite{Ema:2016ops}, see also \cite{Arias-Aragon:2017eww, Bjorkeroth:2018ipq, Bonnefoy:2019lsn, DiLuzio:2023ndz}. These particles inherit flavour-violating couplings from the generation-dependent ${\rm U}(1)$ charges, making rare meson and lepton decays an ideal laboratory to search for them.

\subsection{The new physics flavour puzzle and strategies to address it} \label{sec:NP_problem}

As discussed in the previous sections, flavour-changing processes, and meson mixing in particular, offer some of the most stringent indirect probes of physics beyond the SM.
Their remarkable sensitivity to new sources of flavour and CP violation places severe constraints on the possible structure of NP interactions. In many cases the associated NP scale must be pushed to values many orders of magnitude above the electroweak scale to comply with experimental bounds.

However, there remain strong theoretical motivations for expecting NP near the TeV scale. Chief among these is the Higgs hierarchy problem, which concerns the question of why the Higgs boson mass is so much lighter than the Planck scale, despite quantum corrections that would naturally drive it to much higher values. Solutions to this problem, such as supersymmetry (SUSY) or composite Higgs models, typically involve new states that appear not far above the electroweak scale.

This tension constitutes the so-called ``new physics flavour problem’’. On the one hand, well-motivated theoretical considerations continue to suggest that new degrees of freedom should appear near the TeV scale. On the other hand, flavour physics constrains generic NP contributions to flavour-changing processes far beyond the TeV scale.
Also from a purely experimental perspective, the TeV scale is accessible to direct searches at the LHC. Sufficiently large NP couplings are required to generate detectable signatures, inevitably bringing the flavour structure of these couplings into focus. Understanding the possible patterns of flavour hierarchies in NP models is essential for a consistent and phenomenologically viable program of direct searches at colliders.

Several broad strategies exist to avoid the severe constraints from flavour-changing and CP-violating observables. These can be categorized as follows:

\begin{itemize}
\item {\it Decoupling:} The simplest approach is to assume that the new physics scale is very high, far above the electroweak scale. In this scenario, new flavour-violating effects are simply suppressed by the large NP scale. While this is phenomenologically viable, it comes at the cost of worsening the Higgs naturalness problem, and it does not reconcile the expectation of TeV-scale new physics with the absences of clear signs of this physics in flavour observables.

\item{\it Flavour universality:} A second option is to assume that the new physics couplings are flavour-universal, i.e. proportional to the identity matrix in flavour space. While this does not necessarily fully eliminate NP contributions to flavour-changing processes, it ensures that any such effects are governed by the same CKM suppression factors as those in the SM.

\item {\it Symmetry based approaches:} A third, more structured, strategy involves assuming that the new physics couplings are approximately aligned with, or derived from, the SM Yukawa couplings. Several theoretical frameworks have been developed to implement this idea, differing in the degree of alignment and the underlying symmetry assumptions. We will discuss the most prominent ones in the following.
\end{itemize}

\subsubsection{Minimal Flavour Violation}

The Minimal Flavour Violation (MFV) hypothesis postulates that the SM Yukawa couplings are the only sources of flavour and CP violation, even in the presence of new physics~\cite{Chivukula:1987py, Hall:1990ac, Buras:2000dm, DAmbrosio:2002vsn, Cirigliano:2005ck}. This is achieved by promoting the Yukawa couplings to spurion fields, as in Eq.~\eqref{eq:spurion}, such that the full SM Lagrangian becomes formally invariant under the global flavour symmetry group $G_\text{flavour}=U(3)^5$.
New physics interactions are then required to be also formally invariant under this symmetry and to depend on no spurions other than those defined in Eq.~\eqref{eq:spurion}.

In MFV scenarios, all quark flavour transitions remain governed by the hierarchical structure of the CKM matrix, significantly suppressing NP contributions to meson mixing and rare decays. This renders TeV-scale NP compatible with current flavour constraints, while predicting specific correlations among flavour observables. For instance, NP effects in $b \to s$, $b \to d$, and $s \to d$ transitions are expected to exhibit relative magnitudes dictated by the corresponding CKM factors. It is important to note that MFV does not address the origin of the SM flavour hierarchies but rather assumes their presence to constrain NP effects.

\subsubsection{{\boldmath ${\rm U}(2)^5$} flavour symmetries}

A less restrictive alternative to MFV is provided by approximate ${\rm U}(2)^5$ flavour symmetries as in Eq.~\eqref{eq:U2_to_the_fifth}. These act on the first two generations of quarks and leptons, providing protection against the most tightly constrained flavour-changing processes, while permitting greater freedom in the third-generation sector. Models based on ${\rm U}(2)^5$ were initially developed in the context of SUSY~\cite{Pomarol:1995xc, Barbieri:1995uv, Barbieri:1997tu, Barbieri:2011ci} but have since been applied to a broader range of new physics scenarios as well, see e.g.~\cite{Barbieri:2012uh, Fuentes-Martin:2019mun}.

From a phenomenological standpoint, ${\rm U}(2)^5$ models typically predict strong correlations between $b \to s$ and $b \to d$ transitions while allowing for more flexibility than MFV. The details of these correlations depend on the spurion fields that break the ${\rm U}(2)^5$ symmetry and the manner in which they are embedded within the NP sector.
${\rm U}(2)^5$ models share many similarities with models based on ``General Minimal Flavour Violation''~\cite{Feldmann:2008ja, Kagan:2009bn}, and are also an excellent starting point for flavour model building to address the SM flavour puzzle, see Section~\ref{sec:non-abelian} below.

\subsubsection{Flavour alignment}

Another viable option involves the approximate alignment of NP couplings with the SM Yukawa matrices in the fermion mass eigenstate basis. This suppresses off-diagonal flavour transitions while permitting $O(1)$ non-universal diagonal couplings. This idea has been particularly well studied in supersymmetric models, where aligning the quark Yukawa couplings and soft SUSY-breaking squark masses can naturally suppress dangerous FCNCs~\cite{Nir:1993mx}. Alignment constructions can be realized, for example, in the context of horizontal abelian flavour symmetries, see Section~\ref{sec:FN}.

However, a fundamental limitation arises for the left-handed quark doublets. In the SM itself, the CKM matrix parametrizes a misalignment between the left-handed up and down quarks. As a result, it is not possible to simultaneously align NP couplings with both up-type and down-type quark mass eigenstates unless these couplings are entirely flavour-universal. An irreducible level of flavour mixing, typically of order the Cabibbo angle, is therefore expected in either the down-type or up-type FCNC sector. Experimental constraints from neutral kaon mixing and $D^0 - \bar D^0$ mixing typically provide the strongest bounds on such models~\cite{Blum:2009sk}.

\section{Addressing the Standard Model Flavour Puzzle} \label{sec:addressing_puzzle}

As discussed in the previous sections, the hierarchical flavour structure of the SM renders flavour-changing processes highly sensitive to new physics effects. At the same time, the flavour hierarchies also call for a new physics explanation. In this section we review the most popular ways to address the Standard Model flavour puzzle, i.e. the origin of the hierarchical quark and lepton masses and the CKM mixing angles. For other recent reviews on the topic, see~\cite{Altmannshofer:2022aml, Altmannshofer:2024jyv}.

In the Standard Model, the masses of the quarks and charged leptons are determined by the product of Yukawa couplings and the Higgs vacuum expectation value, see Eq.~\eqref{eq:masses}. Therefore, there are two qualitatively different approaches to obtain a hierarchical fermion spectrum: \\

\begin{itemize}
\item[] 1.) One introduces additional subleading sources of electroweak symmetry breaking that are responsible for the light fermion masses. Although we know from measurements of the Higgs couplings to the $W$ and $Z$ bosons that the dominant source of electroweak symmetry breaking is the vacuum expectation value of the Higgs, $v \simeq v_\text{SM} \simeq 246$~GeV, the existence of additional subleading sources $v_\text{sub} \ll v$ cannot be excluded. This could, for example, be a ${\rm SU}(2)_L$ breaking condensate in a strongly interacting sector or the vacuum expectation value of another Higgs doublet. If light fermions do not couple to the Higgs but only to the subleading symmetry breaking source, their small masses can be explained
\begin{equation}
 \frac{m_\text{light}}{m_\text{heavy}} = \frac{v_\text{sub}}{v} \ll 1 ~.
\end{equation}
Such setups will be reviewed in Section~\ref{sec:subleading}. \\
\item[] 2.) If electroweak symmetry breaking proceeds as in the Standard Model, the hierarchies in the fermion spectrum need to be explained entirely through hierarchies in the Yukawa couplings
\begin{equation}
 \frac{m_\text{light}}{m_\text{heavy}} = \frac{Y_\text{light}}{Y_\text{heavy}} \ll 1 ~.
\end{equation}
There are many different mechanisms that can generate hierarchical Yukawa couplings. Most can be understood as light fermions not coupling directly to the Higgs boson but only indirectly by small mixing with other fermions. The most prominent setups are based either on flavour symmetries (see Section~\ref{sec:symmetry}) or extra dimensions (see Section~\ref{sec:geometry}). Alternatively, light fermions might not have tree-level Yukawa couplings to the Higgs at all, but they are instead induced at the loop level, thus explaining why they are small (see Section~\ref{sec:loops}).
\end{itemize}

\bigskip\noindent
In concrete models, it is possible that both approaches are at play at the same time and that the hierarchical spectrum of SM fermions is due to a combination of various mechanisms.

An interesting difference between the two approaches is the plausible range for the scale of new physics. Yukawa couplings are marginal interactions, and therefore the physics determining their structure could in principle be arbitrarily heavy. As a result, models that aim to explain hierarchical Yukawa couplings may not be directly testable, unless they involve new physics that is sufficiently light to yield experimentally observable signatures. New sources of electroweak symmetry breaking are, on the other hand, expected to involve physics not far above the electroweak scale itself.

\subsection{Subleading sources of electroweak symmetry breaking} \label{sec:subleading}

The SM flavour puzzle can be partially addressed by introducing additional sources of electroweak symmetry breaking that are providing the masses for the light flavours of fermions.
Arguably, the simplest realization of such an idea is a Two Higgs Doublet Model (2HDM) setup: one Higgs doublet has a vacuum expectation value close to the SM value $v \simeq 246$\,GeV and couples mainly to the heavy flavours, while a second Higgs doublet has a much smaller vev and couples mainly to the light generations. Variations of such a 2HDM framework have been explored, for example, in~\cite{Das:1995df, Blechman:2010cs, Altmannshofer:2015esa, Ghosh:2015gpa, Botella:2016krk, Maayergi:2025ybi}. Such models can give interesting non-standard collider phenomenology of the additional Higgs bosons~\cite{Altmannshofer:2016zrn, Altmannshofer:2018bch, Altmannshofer:2019ogm}, as the additional Higgs bosons might dominantly couple to lighter generations (in contrast to the standard assumption that Higgs bosons couple most strongly to the heaviest fermions).

Going one step further, one can consider also Three Higgs Doublet Models (3HDMs) with one Higgs doublet for each SM generation. As discussed in~\cite{Altmannshofer:2025pjj, Das:2025mqs}, such models feature an irreducible amount of flavour-violating couplings of the neutral Higgs bosons, and one finds extremely strong constraints from kaon mixing and $D^0$ meson mixing, in particular for a hierarchical set of Higgs vevs.

Pushing this idea to the extreme, one can imagine introducing a separate Higgs for each flavour of SM quarks and leptons~\cite{Porto:2007ed, Hill:2019ldq, Baek:2023cfy}.
Conceptually, such setups replace the hierarchies in the Yukawa couplings of the SM entirely with hierarchies in the mass parameters of the multi-Higgs potentials.
In contrast to hierarchies in Yukawa couplings, hierarchies in Higgs mass parameters are not necessarily radiatively stable, which poses a significant theoretical challenge.

\subsection{Hierarchical Yukawa couplings from loop effects} \label{sec:loops}

In radiative flavour models, the masses of the light SM fermions are absent at tree level and instead arise from loop corrections~\cite{Weinberg:1972ws}. The hierarchy of fermion masses is determined by loop suppression factors of order $1/16\pi^2 \sim 10^{-2}$, which approximately match the observed mass ratios of adjacent generations. This suggests a natural scenario in which the third generation fermions acquire their masses at tree level, the second generation masses arise at the one-loop level, and the first generation masses appear at the two-loop level, leading to a successful reproduction of the SM fermion spectrum
\begin{eqnarray}
\frac{m_{f_2}}{m_{f_3}} \sim \frac{1}{16\pi^2} ~, \quad \frac{m_{f_1}}{m_{f_3}} \sim \left(\frac{1}{16\pi^2}\right)^2 ~.
\end{eqnarray}
Alternative suppression patterns are also possible. For instance, both the first and second generations may receive masses at the one-loop level, ensuring a hierarchy between the third and second generations, while additional mechanisms might be at play to make the first generation significantly lighter than the second~\cite{Altmannshofer:2014qha}.

The radiative generation of fermion masses can be realized in various theoretical frameworks. A key requirement is a mechanism that prevents the light fermions from coupling to the Higgs boson at tree level. This can be enforced through flavour symmetries, geometric separation in extra-dimensional constructions, or other ways of ``sequestering'' some of the fermions from the Higgs. Additional states with flavour-violating interactions then generate loop-level couplings between the Higgs and the light fermions. The wide range of possible quantum numbers and interactions for these new fields leads to a plethora of viable models. In supersymmetric scenarios, light fermion masses can emerge from loops involving sfermions and gauginos~\cite{Banks:1987iu, Kagan:1989fp, Arkani-Hamed:1996kxn, Borzumati:1999sp, Baumgart:2014jya, Altmannshofer:2014qha}, whereas in non-supersymmetric setups, loop contributions may involve new gauge bosons, vector-like fermions, or leptoquarks~\cite{Barr:1979xt, Balakrishna:1987qd, Balakrishna:1988ks, Dobrescu:2008sz, Graham:2009gr, Baker:2020vkh}. The flexibility in model building allows radiative flavour models to accommodate a wide range of possible UV completions while providing a natural explanation for the SM flavour structure.

\subsection{Hierarchical Yukawa couplings based on symmetries} \label{sec:symmetry}

Symmetry based models that explain hierarchies in Yukawa couplings incorporate horizontal flavour symmetries, also known as family symmetries, which act across different generations of quarks and leptons. The symmetries impose constraints on the Yukawa couplings, and can naturally give hierarchical mass and mixing patterns once the symmetries are spontaneously broken by small spurions.

A distinction can be made between abelian and non-abelian flavour symmetries. Abelian symmetries typically generate multiplicative selection rules that are, in turn, responsible for hierarchical mass and mixing structures. Non-abelian symmetries can enforce, for example, degeneracies among generations before symmetry breaking. The flavour symmetries may be either gauged or global, each option leading to distinct phenomenological implications.

\subsubsection{Abelian flavour models} \label{sec:FN}

The Froggatt-Nielsen (FN) mechanism provides a natural explanation for hierarchical Yukawa couplings by introducing an abelian horizontal ${\rm U}(1)$ symmetry~\cite{Froggatt:1978nt, Davidson:1979wr, Leurer:1992wg, Leurer:1993gy}. The central idea is that SM fermions carry generation-dependent charges under the ${\rm U}(1)$ symmetry, which forbids direct Yukawa interactions at the renormalizable level. Instead, Yukawa couplings arise through higher-dimensional operators that involve a Froggatt-Nielsen field (or flavon field) $\Phi$, which is also charged under the ${\rm U}(1)$
\begin{equation}
\mathcal L \supset - \sum_{j,k} \left( (\tilde Y_u)_{jk} \bar Q_j U_k \tilde H \left(\frac{\Phi}{\Lambda}\right)^{Q_{Q_j}-Q_{U_k}+Q_H} + (\tilde Y_d)_{jk} \bar Q_j D_k H \left(\frac{\Phi}{\Lambda}\right)^{Q_{Q_j}-Q_{D_k}-Q_H} +(\tilde Y_e)_{jk}  \bar L_j E_k H \left(\frac{\Phi}{\Lambda}\right)^{Q_{L_j}-Q_{E_k}-Q_H} \right) + \text{h.c.} ~.
\end{equation}
Here, $\tilde Y_{u,d,e}$ are ``proto'' Yukawa couplings, usually assumed to be of $\mathcal O(1)$, and $Q_{Q_j}$, $Q_{L_j}$, $Q_{U_k}$, $Q_{D_k}$, $Q_{E_k}$ represent the ${\rm U}(1)$ charges of quarks and leptons and $Q_H$ is the charge of the Higgs field. Without loss of generality, the flavon charge is set to $Q_\Phi = 1$. When $\Phi$ acquires a vacuum expectation value, the flavour symmetry is spontaneously broken. Provided that $\langle \Phi \rangle$ is smaller than the cut-off scale $\Lambda$, the resulting effective Yukawa couplings are suppressed by powers of the small parameter $\varepsilon = \langle \Phi \rangle/\Lambda \ll 1 $
\begin{equation}
 (Y_u)_{jk} = (\tilde Y_u)_{jk} \varepsilon^{Q_{Q_j} - Q_{U_k}+Q_H} ~, \quad  (Y_d)_{jk} = (\tilde Y_d)_{jk}  \varepsilon^{Q_{Q_j} - Q_{D_k}-Q_H} ~, \quad   (Y_e)_{jk} = (\tilde Y_e)_{jk}  \varepsilon^{Q_{L_j} - Q_{E_k}- Q_H} ~.
\end{equation}
Correspondingly, fermion mass ratios and the CKM matrix elements can be expressed as powers of the spurion
\begin{equation}
 \frac{m_{u_j}}{m_{u_k}} \sim \varepsilon^{Q_{Q_j} - Q_{U_j} - Q_{Q_k} + Q_{U_k}} ~, \quad \frac{m_{d_j}}{m_{d_k}} \sim \varepsilon^{Q_{Q_j} - Q_{D_j} - Q_{Q_k} + Q_{D_k}} ~, \quad \frac{m_{e_j}}{m_{e_k}} \sim \varepsilon^{Q_{L_j} - Q_{E_j} - Q_{L_k} + Q_{E_k}} ~, \quad |V_{u_j d_k}| \sim \varepsilon^{|Q_{Q_j} - Q_{Q_k}|}~.
\end{equation}
By assigning the appropriate ${\rm U}(1)$ charges, the Froggatt-Nielsen mechanism can successfully reproduce the observed mass hierarchies and CKM mixing angles using a single expansion parameter $\varepsilon$. The hierarchical structure of the CKM matrix suggests that one should identify $\varepsilon$ with the sine of the Cabibbo angle $\lambda \sim 0.2$. With this choice, the following constraints on the ${\rm U}(1)$ charges of the left-handed quarks arise
\begin{equation}
 Q_{Q_1} - Q_{Q_3} = 3~, \quad  Q_{Q_2} - Q_{Q_3} = 2~, \quad   Q_{Q_1} - Q_{Q_2} = 1~.
\end{equation}
Notably, these three conditions are self-consistent. Furthermore, the observed ratios of fermion masses, combined with the fact that the top quark mass is near the electroweak scale, suggest the charge assignments of the ``master model'' proposed in~\cite{Leurer:1993gy, Ben-Hamo:1994dha}
\begin{equation}
Q_Q = \begin{pmatrix} 3 \\ 2 \\ 0 \end{pmatrix}, \quad
Q_U = \begin{pmatrix} -3 \\ -1 \\ 0 \end{pmatrix}, \quad
Q_D = \begin{pmatrix} -3 \\ -2 \\ -2 \end{pmatrix}, \quad
Q_L = \begin{pmatrix} 3 \\ 3 \\ 3 \end{pmatrix}, \quad
Q_E = \begin{pmatrix} -5 \\ -2 \\ 0 \end{pmatrix}, \quad Q_H = 0~.
\end{equation}
While this is a well-motivated set of charges, also other viable charge assignments exist. Small variations in the charges can be compensated by $O(1)$ shifts in the proto-Yukawa couplings and slight adjustments to the expansion parameter $\varepsilon$~\cite{Fedele:2020fvh, Cornella:2023zme}.
The charge assignments of the master model lead to the following hierarchical structure of the Yukawa couplings
\begin{equation}
    Y_u \sim \begin{pmatrix}
    \lambda^6 & \lambda^4 & \lambda^3 \\
    \lambda^5 & \lambda^3 & \lambda^2 \\
    \lambda^3 & \lambda^1 & 1 \end{pmatrix} , \quad
    Y_d \sim \begin{pmatrix}
    \lambda^6 & \lambda^5 & \lambda^5 \\
    \lambda^5 & \lambda^4 & \lambda^4 \\
    \lambda^3 & \lambda^2 & \lambda^2 \end{pmatrix} , \quad
    Y_e \sim \begin{pmatrix}
    \lambda^8 & \lambda^5 & \lambda^3 \\
    \lambda^8 & \lambda^5 & \lambda^3 \\
    \lambda^8 & \lambda^5 & \lambda^3 \end{pmatrix} .
\end{equation}
Here, we have omitted the explicit dependence on the proto-Yukawa couplings.

A natural UV completion of the higher-dimensional operators in Froggatt-Nielsen models involves chains of vector-like fermions that mediate the flavour symmetry breaking to the SM fermions.
Chains of vector-like fermions provide a general mechanism for generating hierarchical Yukawa couplings, which is described in more detail in Section~\ref{sec:fermion_mixing}. In the context of Froggatt-Nielsen models, the scale $\Lambda$ corresponds to the masses of the vector-like fermions, whereas the proto-Yukawa couplings arise from their interactions. The powers $\varepsilon^{Q_f}$ can be interpreted as effective mixing angles of the SM fermions with those vector-like fermions that couple to the Higgs boson. The FN framework provides a sufficient number of $O(1)$ parameters to accurately reproduce the observed fermion masses and mixings.

If the ${\rm U}(1)$ flavour symmetry is gauged, this setup predicts a $Z^\prime$ gauge boson with a mass of the order $\langle \Phi \rangle$. Such a $Z^\prime$ generically features flavour-changing couplings, which could have observable effects in low energy flavour-changing processes, see Section~\ref{heavy}. Conversely, if the ${\rm U}(1)$ is global, one instead expects a light pseudo-Nambu-Goldstone boson, potentially leading to new phenomenology in rare decays of kaon or $B$ mesons, see Section~\ref{light}. However, experimental signatures at low energies are not guaranteed. FN models usually feature a decoupling limit. Taking both $\langle \Phi \rangle$ and $\Lambda$ to high scales while keeping $\varepsilon = \langle \Phi \rangle / \Lambda$ constant allows them to address the SM flavour puzzle without introducing excessive FCNCs at low energies.

\subsubsection{Non-abelian flavour symmetries} \label{sec:non-abelian}

Non-abelian flavour symmetries, ${\rm U}(2)$ symmetries in particular, are a popular alternative to the Froggatt-Nielsen models discussed above. They are based on the observation that ${\rm U}(2)^5$ as in Eq.~\eqref{eq:U2_to_the_fifth} is an excellent approximate flavour symmetry of the Standard Model.

In a ${\rm U}(2)$ flavour model, one distinguishes the first two generations, which transform as a doublet, from the third generation, which is a singlet under ${\rm U}(2)$.
In the symmetric limit, only the third generation is massive and the CKM matrix is trivial. On the one hand, this is a reasonable zeroth order starting point for flavour model building; on the other hand, it also suppresses FCNCs among the first two generations.

The hierarchical flavour structure of the Yukawa couplings typically arises through a sequence of symmetry breaking steps. First, the ${\rm U}(2) = {\rm SU}(2)\times {\rm U}(1)$ symmetry is broken to an intermediate ${\rm U}(1)$ symmetry and the second generation acquires a small Yukawa coupling. In a second stage, also the remaining ${\rm U}(1)$ that acts on the first generation is broken, resulting in an even smaller Yukawa coupling proportional to the product of two symmetry breaking spurions.

Models based on ${\rm U}(2)$ flavour symmetries can effectively suppress new physics effects in low energy processes that involve flavour change between the first and second generation and thus largely avoid the stringent constraints that arise, for example, from kaon mixing and $D^0$ meson mixing.

While ${\rm U}(2)$ flavour symmetries were originally introduced in the context of supersymmetric models~\cite{Pomarol:1995xc, Barbieri:1995uv, Barbieri:1997tu, Barbieri:2011ci} they have a much wider applicability~\cite{Barbieri:2015yvd, Antusch:2023shi, Greljo:2024zrj}.

\subsection{Hierarchical Yukawa couplings from fermion mixing} \label{sec:fermion_mixing}

Small Yukawa couplings can also emerge from mass mixing between a chiral fermion and heavy vector-like (VL) fermions, all transforming in the same gauge representation as a given SM fermion.
In such a setup, the mass-mixing terms lead to off-diagonal entries in the combined mass matrix of the chiral and VL fermions.
To determine the mass eigenstates, this matrix has to be diagonalized.
Since the initial Lagrangian contains a chiral fermion, the diagonalization yields one vanishing eigenvalue, corresponding to a massless state that can be identified with an SM fermion. The remaining non-zero eigenvalues correspond to heavy NP states.
Because the initial mass matrix is not diagonal, all mass eigenstates -- including the SM fermion -- are linear combinations of the initial chiral and VL fermions, with coefficients that can be strongly hierarchical and depend on ratios of masses and mixing parameters in the initial Lagrangian.
As a result, non-hierarchical Yukawa couplings of the fermions in the initial Lagrangian can give rise to hierarchical Yukawa couplings of the fermion mass eigenstates.

As an example, we consider a chiral fermion $f_L^\prime$ mixing with a VL fermion $F^\prime$,
\begin{equation}\label{eq:single_fermion_mixing}
 \mathcal{L} \supset \Delta\, \bar f_L^\prime F_R^\prime - m\, \bar F_L^\prime F_R^\prime  +\mathrm{h.c.}\,,
\end{equation}
where $\Delta$ is the strength of the mixing, $m$ is the mass of the VL fermion, and we denote fields in the initial Lagrangian (before mass diagonalization) with a prime.
Diagonalizing the mass matrix then yields the mass eigenstates
\begin{equation}\label{eq:single_fermion_mixing_mass_eigenstates}
 f_L = \cos \theta\, f_L^\prime + \sin \theta\, F_L^\prime\,,
 \qquad
 F_L = \cos \theta\, F_L^\prime - \sin \theta\, f_L^\prime\,,
 \qquad
 F_R = F_R^\prime\,,
\end{equation}
with
\begin{equation}\label{eq:fermion_mixing_defs}
 \cos \theta = \frac{m}{M}\,,
 \qquad
 \sin \theta = \frac{\Delta}{M}\,,
 \qquad
 \text{and}
 \qquad
 M = \sqrt{m^2 + \Delta^2}\,.
\end{equation}
The field $f_L$ is a massless chiral fermion, which we refer to as the \emph{zero mode}, and $F$ is a heavy VL fermion with mass $M$.
To see how a small Yukawa coupling can arise in the mass basis, consider an order-one Yukawa coupling of $F^\prime$ to another chiral fermion $f_R$ and a Higgs field $H$,
\begin{equation}
 \mathcal{L} \supset -Y^\prime\, \bar F_L^\prime\, f_R\, H\,,
\end{equation}
with $Y^\prime = \mathcal{O}(1)$.
After diagonalizing the fermion mass matrix, this leads to a Yukawa coupling between the two chiral fermions,
\begin{equation}
 \mathcal{L} \supset -Y\, \bar f_L\, f_R\, H
 \qquad
 \text{with}
 \qquad
 Y = \sin \theta\, Y^\prime\,.
\end{equation}
We see that the Yukawa coupling $Y$ can be suppressed compared to $Y^\prime$ if $\sin \theta$ is small, which is the case in particular when the mixing strength is small compared to the mass of the VL fermion, $\Delta \ll m \approx M$.

As a second example, we consider a chiral fermion $\psi_L^{\prime (0)}$ mixing with a chain of $N$ VL fermions $\psi^{\prime (j)}$,
\begin{equation}\label{eq:fermion_chain}
 \mathcal{L} \supset \sum_{j=1}^N\left( \Delta_j\, \bar \psi_L^{\prime(j-1)} \psi_R^{\prime(j)} - m_j\, \bar \psi_L^{\prime(j)} \psi_R^{\prime(j)}
 \right)  +\mathrm{h.c.}\,,
\end{equation}
where we again denote fields in the initial Lagrangian (before mass diagonalization) with a prime, and we refer to them as \emph{chain fields}.
The quantities $\Delta_j$ and $m_j$ are the mixing strengths and VL masses, respectively, which may differ for each $j$.
The mass matrix depending on the $2 N$ parameters $\Delta_j$ and $m_j$ can be diagonalized in terms of an orthonormal basis of mass eigenstates,
\begin{equation}\label{eq:chain_decomp}
\psi_{L,R}^{(k)} = \sum_{j=0}^N \, a_{L,R}^{(k),(j)}\, \psi_{L,R}^{\prime(j)}\,,
\end{equation}
which expresses the mass eigenstates $\psi_{L,R}^{(k)}$ as linear combinations of the chain fields $\psi_{L,R}^{\prime(j)}$.
The orthonormal coefficients $a_{L,R}^{(k),(j)}$ specify how much the $j$-th chain field contributes to the $k$-th mass eigenstate.
Since the Lagrangian in Eq.~\eqref{eq:fermion_chain} contains $N+1$ left-handed and $N$ right-handed chiral fermions, the spectrum of mass eigenstates consists of $N$ massive Dirac fermions along with a single left-handed massless zero mode.
The coefficients of the zero mode take a particularly simple form if the ratio of mixing and mass parameters is constant along the chain, i.e.\ $\frac{\Delta_j}{m_j} = \chi$ for all~$j$~\cite{Bai:2009ij,Burdman:2012sb}:
\begin{equation}\label{eq:zero_mode_coeff}
 a_{L}^{(0),(j)} = \frac{\chi^j}{\mathcal{N}}
 \qquad
 \text{with}
 \qquad
 \mathcal{N}  = \sqrt{\sum_{\ell=0}^N \chi^{2 \ell}} = \sqrt{\frac{1-\chi^{2(N+1)}}{1-\chi^2}}\,.
\end{equation}
For $N=1$, the normalization factor is $\mathcal{N} = \sqrt{1+\chi^2}=\frac{M}{m}$ (with $M$ defined in Eq.~\eqref{eq:fermion_mixing_defs}), and we immediately recover the results from the previous example by identifying $\psi_L^{(\prime)(0)} = f_L^{(\prime)}$ and $\psi^{(\prime)(1)} = F^{(\prime)}$.
For $N > 1$, we observe that the composition of the zero mode in terms of the chain fields becomes hierarchical if $\chi\ll 1$ or $\chi \gg 1$:
\begin{itemize}
 \item In the first case, $\chi \ll 1$, the normalization factor becomes $\mathcal{N}\approx (1-\chi^2)^{-1/2}\approx 1$. The contribution of the $j$-th chain field to the zero mode is then approximately $\chi^j$, i.e.\ the zero mode consists primarily of the $0$-th chain field, which has an $\mathcal{O}(1)$ coefficient, while the $j$-th field is suppressed by the $j$-th power of the small parameter $\chi$, with maximal suppression $\chi^N$ for the $N$-th field.
 This hierarchical structure of the coefficients can be interpreted as a \emph{localization} of the zero mode on the chain.
 For $\chi \ll 1$, the zero mode is localized towards the $j=0$ end of the chain.

 \item In the second case, $\chi \gg 1$, we find $\mathcal{N} \approx \chi^N$, and the contribution of the $j$-th chain field is approximately $\chi^{j-N}$.
 The zero mode is then localized towards the $j = N$ end of the chain and the $j=0$ field is suppressed by $1/\chi^N$.

 \item If instead $\chi \approx 1$, then all $N+1$ chain fields contribute equally to the zero mode, each with coefficient $(N+1)^{-1/2}$; in this case, the zero mode is maximally delocalized along the chain.
\end{itemize}
The strong suppression of the contributions to the zero mode from one side of the fermion chain if either $\chi\ll 1$ or $\chi\gg 1$ can generate small Yukawa couplings if the Higgs field couples to the suppressed chain field. In other words, the induced Yukawa coupling is small when an order-one Yukawa interaction is present at one end of the chain, but the zero mode is localized towards the opposite end.

The general mechanism for generating small Yukawa couplings through fermion mixing, as described above, is realized in different ways across various models. In the following, we discuss concrete examples: clockwork models, warped extra dimensions, and partial compositeness.

\subsubsection{Flavour clockwork models} \label{sec:clockwork}

Flavour clockwork models employ the clockwork mechanism~\cite{Giudice:2016yja} to generate the hierarchical structures observed in fermion masses and mixing angles.
This mechanism is based on chains of fields with nearest-neighbour interactions, which in the case of fermions is equivalent to Eq.~\eqref{eq:fermion_chain}.
For each flavour of quarks and leptons, these models introduce chains of $2N + 1$ chiral fermions, governed by the following mass Lagrangian~\cite{Giudice:2016yja}
\begin{equation}
    \mathcal L \supset - m \sum_{A=1}^{N}  \left( \overline{\psi}_L^{(A)} \psi_R^{(A)} - \chi   \overline{\psi}_L^{(A-1)} \psi_R^{(A)} \right) + \mathrm{h.c.}\,,
\end{equation}
which is equal to Eq.~\eqref{eq:fermion_chain} with $m_j=m$ and $\frac{\Delta_j}{m_j}=\chi$.
Upon diagonalizing the corresponding mass matrix, the spectrum consists of $N$ massive Dirac fermions along with a single massless chiral zero mode.
For values $\chi > 1$, the zero mode
is localized towards the $A=N$ end of the chain and
exhibits an exponentially suppressed overlap with the original $A=0$ fermion, scaling as $1/\chi^N$, where $\chi$ is the so-called clockwork gear ratio.
This suppression naturally generates exponential hierarchies in the couplings of the zero mode. Applying this mechanism to Standard Model Yukawa couplings results in
\begin{equation}
 (Y_u)_{jk} = (\tilde Y_u)_{jk} \chi^{-N_q^j -N_u^k} ~, \quad  (Y_d)_{jk} = (\tilde Y_d)_{jk} \chi^{-N_q^j -N_d^k} ~, \quad   (Y_e)_{jk} = (\tilde Y_e)_{jk} \chi^{-N_\ell^j -N_e^k} ~,
\end{equation}
where $N_f^i$ denotes the length of the clockwork chain for each SM fermion
and the $\tilde Y_f=\mathcal{O}(1)$ are Yukawa couplings between the $A=0$ chain fields.
This framework has been successfully applied to explain the hierarchical structure of quark, lepton, and neutrino masses, see e.g.~\cite{vonGersdorff:2017iym, Patel:2017pct, Ibarra:2017tju, Alonso:2018bcg, Hong:2019bki, AbreudeSouza:2019ixc, vonGersdorff:2020ods, Kang:2020cxo, Babu:2020tnf}.

Flavour clockwork models share many similarities with the Froggatt-Nielsen setups discussed in Section~\ref{sec:FN} and the extra-dimensional frameworks in Section~\ref{sec:geometry}. The flavour-dependent lengths of the clockwork chains effectively act as the ${\rm U}(1)$ charges in Froggatt-Nielsen models, while the inverse clockwork gear ratio, $1/\chi$, plays the role of the symmetry-breaking spurion $\varepsilon = \langle \phi \rangle / M \ll 1$. This analogy has inspired the development of ``inverted'' Froggatt-Nielsen models, where the expansion parameter is given instead by $M/\langle \phi \rangle \ll 1$~\cite{Alonso:2018bcg, Smolkovic:2019jow}. Notably, such constructions are anomaly-free, allowing the horizontal ${\rm U}(1)$ symmetries to be gauged. The associated $Z^\prime$ gauge bosons could, in principle, be light, making them accessible to searches in precision flavour experiments, astrophysics, and beam dump experiments.

Combining clockwork flavour models with supersymmetry also leads to interesting phenomenological consequences~\cite{Altmannshofer:2021qwx}. In the simplest scenario, where the zero modes of the clockwork chains correspond to the fermions and sfermions of the Minimal Supersymmetric Standard Model (MSSM), the clockwork mechanism predicts a characteristic flavour structure for the soft SUSY-breaking sfermion masses. This structure closely resembles that of minimal supersymmetric ${\rm U}(1)$ Froggatt-Nielsen models, and implies large flavour mixing among the first and second generation squarks, roughly of the order of the Cabibbo angle. Therefore, the stringent constraints from kaon oscillations imply that either the squarks or gluinos have to be heavier than $\mathcal O(1\,\text{PeV})$~\cite{Altmannshofer:2013lfa}.

\subsubsection{Warped extra dimensions}\label{sec:geometry}

Small Yukawa couplings can also emerge in models with extra spatial dimensions.
The effective four-dimensional~(4D) description of a field in a higher dimensional space-time contains a massless zero mode and an infinite tower of heavy resonances -- the so-called Kaluza-Klein~(KK) excitations -- whose properties depend on the localization of the field in the extra dimensions and are encoded in KK wave functions.
If Standard Model fields are localized at different positions along an extra-dimensional coordinate, the effective 4D couplings of their zero modes and KK excitations are determined by the degree of overlap between the corresponding KK wave functions. When these wave functions are peaked at separated locations, the overlap can be exponentially suppressed, leading to naturally small Yukawa couplings of the zero modes and providing a geometric explanation for the observed fermion mass hierarchy~\cite{Arkani-Hamed:1999ylh}.

Interestingly, discretizing the extra-dimensional coordinate results in a chain of 4D fields with nearest-neighbour interactions, similar to those in Eq.~\eqref{eq:fermion_chain}.
Diagonalizing the mass matrix of the chain fields yields a zero mode localized on the chain, along with a tower of heavy mass eigenstates.
The generation of small Yukawa couplings in models with extra dimensions via small wave function overlaps can thus be understood as a continuum version of fermion mixing in a chain of $N$ 4D fermions, and is described by Eq.~\eqref{eq:fermion_chain} in the $N\to \infty$ limit.
The mass and mixing parameters of the 4D chain are then determined by the geometry of the extra dimension.

Generic predictions of such scenarios are flavour-non-universal and flavour-violating couplings between SM fermions and the KK excitations of gauge fields. These couplings arise due to the non-trivial wave function profiles of the fermions and gauge bosons in the extra dimension. If the KK modes are sufficiently light, they can give measurable deviations in low energy flavour observables, providing potential signatures that can be tested in precision experiments.

The most studied example in which such a geometric mechanism explains the Yukawa hierarchies is the Randall-Sundrum~(RS) model~\cite{Randall:1999ee}, which features a warped extra dimension.
The extra dimension $y$ is compactified on an interval $0 \leq y \leq L$ and the metric is given by
\begin{equation}
 ds^2 = e^{-2 k y} ~ \eta_{\mu\nu} dx^\mu dx^\nu - dy^2~.
\end{equation}
The warp factor $e^{-2 ky}$ contains the curvature scale $k$, assumed to be of the order of the Planck scale $k \sim M_\text{pl}$.
The boundaries of the extra dimension are the infrared (IR) brane at $y = L$ and the ultraviolet (UV) brane at $y= 0$. The warp factor exponentially suppresses energy scales such that $k e^{- kL}\sim \mathcal O(1\,\text{TeV})$. If the Higgs boson is confined to the IR brane it is naturally light and the Higgs hierarchy problem is addressed.

The action of a fermion field $\Psi(x,y)$ in the five-dimensional (5D) bulk can be written as~\cite{Bai:2009ij}
\begin{equation}\label{eq:5D_fermions}
 S_5 \supset \int\,d^4 x \int_0^L\,dy\ \left\{\,
 \bar\Psi\, i\slashed\partial\, \Psi
 + e^{-k y}\, \bar\Psi \gamma_5 \left(\partial_y -\tfrac{k}{2}\right)\Psi
 - e^{-k y}\, m_\Psi\, \bar\Psi \Psi
 \,\right\}\,,
\end{equation}
where $\partial_\mu$ denotes the partial derivatives with respect to the 4D coordinates $x$, $\partial_y$ the extra-dimensional derivative, and $m_\Psi$ the fermion bulk mass.
To obtain a left-handed KK zero mode, we impose the boundary conditions $\Psi_R(x,0)=\Psi_R(x,L)=0$.

The extra dimension can be discretized on a lattice -- a procedure also known as dimensional deconstruction~\cite{Arkani-Hamed:2001kyx,Hill:2000mu} -- using the fields at $y = \epsilon \,j$ with the discrete index $j\in[0,N]$ and constant lattice spacing $\epsilon=L/N$.
The resulting 4D action is equivalent to Eq.~\eqref{eq:5D_fermions} in the $\epsilon\to0, N\to \infty$ limit (keeping $L$ fixed), and can be written as~\cite{Bai:2009ij}
\begin{equation}
 S_4 \supset  \int\,d^4 x\, \Bigg\{
\bar\psi_L^{\prime(0)}\,i \slashed\partial\,\psi_L^{\prime(0)}
 + \sum_{j=1}^N \bigg(
\bar\psi^{\prime(j)}\,i \slashed\partial\,\psi^{\prime(j)}
-
\frac{e^{-k\epsilon j}}{\epsilon}\, \bar \psi_L^{\prime(j-1)} \psi_R^{\prime(j)}
+
\frac{e^{-k\epsilon \left(j+c_\Psi-1/2\right)}}{\epsilon}\, \bar \psi_L^{\prime(j)} \psi_R^{\prime(j)}
 \bigg)
 \Bigg\}+\mathcal{O}(\epsilon)\,,
\end{equation}
where we have defined the dimensionless bulk mass parameter $c_\Psi = \frac{m_\Psi}{k}$ and the 4D fields $\psi^{\prime (j)}(x)=\sqrt{\epsilon}\, \Psi(x,\epsilon j)$.
We have used the freedom to add $\mathcal{O}(\epsilon)$ terms that vanish in the continuum limit to express the mass and mixing\footnote{%
The mixing term coupling a field at site $j-1$ to one at site $j$ arises from the extra-dimensional derivative,
\begin{equation}
 \partial_y \Psi(x,y)= \lim_{\epsilon\to 0}\frac{\Psi(x,\epsilon j)-\Psi(x,\epsilon (j-1))}{\epsilon}\,.
\end{equation}
} terms in a convenient form.
The Lagrangian is equal to that of the fermion chain in Eq.~\eqref{eq:fermion_chain}, with
\begin{equation}
 \Delta_j = -\frac{e^{-k\epsilon j}}{\epsilon}
 \qquad
 \text{and}
 \qquad
 m_j = -\frac{e^{-k\epsilon \left(j+c_\Psi-1/2\right)}}{\epsilon}\,.
\end{equation}
While both the mass and mixing parameters depend on $j$, their ratios are independent of $j$:
\begin{equation}
\frac{\Delta_j}{m_j} = e^{k\epsilon \left(c_\Psi-1/2\right)} = \chi\,,
\end{equation}
and the coefficients of the zero mode are given by Eq.~\eqref{eq:zero_mode_coeff}.
The size of $\chi$ -- and thus the localization of the zero mode -- depends on the bulk mass parameter $c_\Psi$. For $c_\Psi<1/2$, the zero mode is localized towards the UV brane at $j=0$ and exponentially suppressed on the IR brane at $j=N$.
On the other hand, in the case of $c_\Psi > 1/2$, one finds the opposite: the zero mode is localized towards the IR brane and suppressed on the UV brane.

If the Higgs is localized on the IR brane in order to address the Higgs hierarchy problem, the Yukawa interactions between the Higgs and the zero modes (which are identified with the SM fields) depend on the overlap of the zero modes with the IR brane.
This overlap is proportional to the $j=N$ coefficient of the zero mode, $a_{L,R}^{(0),(N)}$, and can be hierarchical for appropriate values of $c_\Psi$.
We denote the zero mode identified with an SM fermion~$f$ as $\psi_f^{(0)}$, and the corresponding chain coefficients as $a_f^{(0),(j)}$.
The overlap of the SM fermion~$f$ with the Higgs at the IR brane in the continuum limit is further denoted by~$\xi_f$:
\begin{equation}
 \xi_f \sim \lim_{\epsilon\to 0}\, a_f^{(0),(N)}\Big|_{N=L/\epsilon}\,.
\end{equation}
The 4D Yukawa couplings are then given by a combination of 5D ``proto'' Yukawa couplings~$\tilde Y_f$, which can all be of~$\mathcal O(1)$, and hierarchical overlap factors~$\xi_f$:
\begin{equation} \label{eq:RS_Yukawa}
 (Y_u)_{jk} = (\tilde Y_u)_{jk} \xi_{q_j} \xi_{u_k} ~, \quad  (Y_d)_{jk} = (\tilde Y_d)_{jk}  \xi_{q_j} \xi_{d_k} ~, \quad   (Y_e)_{jk} = (\tilde Y_e)_{jk}  \xi_{\ell_j} \xi_{e_k} ~.
\end{equation}
Comparing to the Froggatt-Nielsen case discussed in Section~\ref{sec:FN}, we see that the role of the ${\rm U}(1)$ flavour charges is now played by the flavour dependent fermion bulk masses.
Exponential hierarchies in the Yukawa couplings arise from order-one differences in the bulk masses~\cite{Grossman:1999ra, Gherghetta:2000qt, Huber:2000ie}.

The flavour-dependent, extra-dimensional profiles of the SM fermions lead in general to a rich flavour phenomenology~\cite{Agashe:2004cp, Blanke:2008zb}.
In particular, one finds that the KK excitation of the SM gauge bosons, the gluons in particular, have flavour violating couplings and can contribute at the tree-level to meson mixing, see Section~\ref{mix}.
In fact, in many cases, the constraints from meson mixing push the new physics scale to $\mathcal{O}(10)$\,TeV or above creating a serious little hierarchy problem.
It is possible to invoke flavour symmetries to avoid too large flavour-changing neutral currents and to keep the new physics scale around $1$\,TeV~\cite{Cacciapaglia:2007fw}. The downside of such models is that one gives up on the attractive mechanism of fermion localization to explain the SM flavour hierarchies.

\subsubsection{Partial compositeness}

The idea behind partial compositeness is that the SM fermions are mixtures of elementary states and composite states that are part of a strongly coupled sector~\cite{Kaplan:1991dc}. Schematically, the mixing occurs via operators of the form
\begin{equation}
\mathcal L \supset \frac{c}{\Lambda^{d_f - 5/2}} \bar f \mathcal O_f ~+~\text{h.c.} ~,
\end{equation}
where $f$ denotes an elementary fermion, $\mathcal O_f$ is a composite fermionic operator with the appropriate quantum numbers, and $d_f$ is its scaling dimension.
This Lagrangian resembles the one in Eq.~\ref{eq:single_fermion_mixing} and diagonalization of the fermion mass matrix leads to SM fermions corresponding to the massless state in Eq.~\ref{eq:single_fermion_mixing_mass_eigenstates}.
The elementary states do not couple directly to the Higgs; their interactions with the Higgs are instead mediated through their mixing with the composite sector. The strength of this mixing is governed by the scaling dimension of the composite fermionic operators. The resulting effective Yukawa couplings of the SM fermions are a combination of $O(1)$ couplings of the composite states to the Higgs and hierarchical mixing angles.
The structure closely mirrors the situation in warped extra dimensions, as encapsulated in Eq.~\eqref{eq:RS_Yukawa}, with the role of the fermion overlap factors replaced by the mixing angles. In fact, according to the AdS/CFT correspondence, models with warped extra dimensions and models with partial compositeness describe dual realizations of analogous physics.

Partial compositeness is most naturally realized within composite Higgs models (see e.g.\cite{Bellazzini:2014yua, Panico:2015jxa}), where the Higgs boson appears as a pseudo-Nambu–Goldstone boson of a spontaneously broken global symmetry in a strongly interacting sector not far above the electroweak scale. The phenomenology is very similar to the one of models with warped extra dimensions. Instead of KK excitations of SM gauge fields, composite vector resonances of the strongly interacting sector can mediate flavour-changing neutral currents at the tree level. Too large FCNCs can be avoided, for example, by imposing flavour symmetries~\cite{Csaki:2008zd, Redi:2011zi, Niehoff:2015iaa, Niehoff:2016zso, Glioti:2024hye, Agashe:2025tge}.

\subsection{Other recent developments} \label{sec:recent}

In recent years, several novel approaches have been explored in flavour model building. In the following subsections, we briefly review the most prominent directions.

\subsubsection{Flavour deconstruction}

Flavour deconstruction refers to the idea that a gauge group $G$ is extended to a product of three generation specific gauge groups $G_1 \times G_2 \times G_3$ in the UV.
Such a deconstruction can involve the entire SM gauge group ${\rm SU}(3) \times {\rm SU}(2) \times {\rm U}(1)$, only a subgroup, or suitable extensions.
The concept was originally proposed to accommodate lepton flavour non-universality of the weak interactions~\cite{Li:1981nk} and has gained renewed popularity in recent years, particularly in response to experimental hints for lepton flavour universality violation in $B$ decays\footnote{Some of these anomalies have since been weakened or disappeared~\cite{LHCb:2022qnv}.} (see e.g.~\cite{Bordone:2017bld, Davighi:2023iks, Davighi:2023evx, Davighi:2023xqn, Barbieri:2023qpf, Capdevila:2024gki, Fuentes-Martin:2024fpx, FernandezNavarro:2024hnv}).
A reunification of the gauge group may occur at even higher energy scales in the UV~\cite{Davighi:2022fer}.

In models featuring flavour deconstruction, fermion fields from each generation are charged under their respective gauge group factor. Typically, the Higgs boson is charged only under the third generation group, $G_3$, so that only third generation fermions acquire mass at leading order, leaving the first and second generations initially massless.
Generating masses for the lighter generations requires breaking the deconstructed gauge symmetry, usually through a two-step process. First, the gauge factors of the first two generations are broken to their diagonal subgroup via the vacuum expectation value $v_{12}$ of a scalar link field: $G_1 \times G_2 \to G_{12}$. Subsequently, the remaining symmetry is broken at a lower scale down to a universal SM-like gauge group by the vev $v_{23}$ of another scalar link field: $G_{12} \times G_3 \to G$. In this setup, the mass ratios of the fermion generations are controlled by the symmetry breaking scales and associated cutoff scales
\begin{equation}
 \frac{m_{f_2}}{m_{f_3}} \sim \frac{v_{23}}{\Lambda_{23}} ~, \quad  \frac{m_{f_1}}{m_{f_3}} \sim \frac{v_{23}}{\Lambda_{23}} \frac{v_{12}}{\Lambda_{12}} ~.
\end{equation}
A realistic mass spectrum requires $v_{23}/\Lambda_{23} \sim v_{12}/\Lambda_{12} \sim 10^{-2}$. The vev $v_{23}$ could be as low at $\sim 1$\,TeV, while $v_{12}$ is typically constrained to be at least $\sim 100$\,TeV to satisfy stringent flavour constraints. Importantly, this multiscale structure does not destabilize the electroweak scale, remaining consistent with the principle of finite naturalness~\cite{Farina:2013mla}.
Moreover, the multiscale structure largely decouples the new physics that could potentially generate dangerous FCNC transitions between the first and second generation. Indeed, for a large $v_{12}$, approximate ${\rm U}(2)$ flavour symmetries emerge, protecting the setups from flavour constraints. The TeV scale new physics of such models interacts predominantly with the third generation and might give interesting signals in $B$ decays and $\tau$ decays.

Flavour deconstruction can also be realized within broader new physics frameworks. It has been investigated, for instance, in extra-dimensional models and in setups featuring composite sectors~\cite{Fuentes-Martin:2020bnh,Fuentes-Martin:2022xnb, Covone:2024elw}, where it provides an interesting mechanism for flavour hierarchies and controlled flavour violation.

\subsubsection{Modular and eclectic flavour symmetries}

Over the past several years, modular flavour symmetries have emerged as a popular framework in flavour model building, particularly in the context of neutrino mass models~\cite{Feruglio:2017spp, Ding:2023htn}. Modular flavour symmetries generalize traditional discrete symmetries, providing greater flexibility in achieving neutrino masses and mixings that agree with experimental observations, even within minimal models. Although consistent modular transformations typically necessitate supersymmetry, recent studies have also begun exploring non-supersymmetric realizations~\cite{Ding:2020zxw, Qu:2024rns, Ding:2024inn}.

The basic idea of models with modular flavour symmetries is that Yukawa couplings are promoted to so-called modular forms, $Y(\tau)$, which depend on a complex modulus $\tau$. When the vev of $\tau$ is near a symmetry point, hierarchical structures can emerge in the Yukawa couplings~\cite{Okada:2020ukr, Feruglio:2021dte, Novichkov:2021evw, Chen:2025tby}. The hierarchies are dictated by the ``weight'' and ``level'' of the modular form, while finite modular groups -- typically isomorphic to permutation groups -- impose additional constraints on the Yukawa structure.

Compared to Froggatt-Nielsen models, modular flavour symmetries lead to more constrained patterns of Yukawa couplings, enhancing their predictive power. Another distinction is that models based on modular symmetries generally require a smaller field content. They do not rely on flavon fields for spontaneous flavour symmetry breaking and eliminate the need for an extended sector of vector-like matter, which in Froggatt-Nielsen models mediates the breaking of flavour symmetries to the SM fermions.

Achieving fully realistic fermion masses and mixings often requires additional ingredients beyond the basic features of modular flavour symmetries. One approach is to introduce a ``weighton'' field, which suppresses Yukawa couplings according to the assigned modular weights of SM fermions~\cite{King:2020qaj}. Additional flexibility in matching observed masses and mixing angles can arise from modifications to the K{\"a}hler potential~\cite{Chen:2019ewa} or by expressing Yukawa couplings as linear combinations of multiple modular forms. Example models that successfully reproduce realistic quark and lepton masses and mixing patterns can be found in~\cite{Lu:2019vgm, Okada:2019uoy, Liu:2020akv}.

A related development in this direction of flavour model building is the concept of eclectic flavour symmetries~\cite{Baur:2019kwi, Baur:2019iai, Nilles:2020nnc, Nilles:2020kgo, Baur:2022hma}. These symmetries non-trivially combine modular flavour symmetries, conventional discrete flavour symmetries, and CP-like symmetries within string-motivated top-down models. Another recent development explores links between the flavour puzzle and the strong CP problem, offering new perspectives on both phenomena using modular symmetries~\cite{Feruglio:2024ytl}.

\subsubsection{Flavour and generalized symmetries}

Generalized symmetries extend the conventional concept of symmetries~\cite{Gaiotto:2014kfa}. Higher-form global symmetries, for instance, do not act on point-like objects such as particles but on extended structures like lines, or surfaces. Meanwhile, non-invertible global symmetries differ from traditional symmetries in that their product operation does not necessarily adhere to group structure requirements -- in particular, elements may lack an inverse.
Despite these differences, higher-form and non-invertible global symmetries share many characteristics with standard global symmetries, including their ability to impose ``selection rules''. Research into the implications of such symmetries is ongoing, and they have begun to find applications in particle physics phenomenology, as reviewed in~\cite{Brennan:2023mmt}.

In the context of flavour model building, one particularly intriguing aspect is that non-invertible chiral symmetries can naturally generate exponential hierarchies through non-perturbative effects. This property presents potential applications in the construction of flavour models~\cite{Cordova:2022ieu, Cordova:2022fhg}.
Furthermore, string-inspired non-invertible symmetries offer a framework for generating various Yukawa matrix textures~\cite{Kobayashi:2024yqq, Kobayashi:2024cvp}. Notably, this includes configurations featuring texture zeros in both the quark and lepton sectors, which serve as promising starting points for further flavour model building. Finally, as shown in~\cite{Cordova:2022qtz}, the conventional SM flavour symmetries are intertwined with a one-form magnetic hypercharge symmetry. This interplay imposes constraints on possible vertical gauge unification schemes and could have relevant implications for horizontal flavour symmetries and gauge-flavour unification. Such insights may offer new directions for constructing more unified and predictive frameworks in flavour physics.

\section{Conclusions}
\label{sec:conclusions}


Flavour physics plays a central role in the landscape of particle physics: while traditional flavour probes are experimentally studied at low energies, they offer access to fundamental questions whose origins likely reside at much higher scales. In this chapter, we have explored how the peculiar structure of flavour in the Standard Model (manifested in the pronounced hierarchies of fermion masses and mixing angles) points towards physics beyond the SM and how precision measurements of flavour-changing processes provide exceptionally sensitive probes of new physics.

We have revisited how flavour arises within the Standard Model, focusing on the structure of the Yukawa couplings that break the global flavour symmetries and give rise to the quark mixing encoded in the CKM matrix. This structure leads to the suppression of flavour-changing neutral currents and governs the patterns of meson mixing and rare decays. The pronounced hierarchies in fermion masses and mixing angles represent one of the most striking and least understood features of the SM, commonly referred to as the SM flavour puzzle. While technically natural in the sense of 't Hooft's criterion, this observed pattern hints at a deeper organizing principle yet to be uncovered.

Building on this, we turned to the powerful role of flavour-violating processes as indirect probes of new physics. Meson-antimeson mixing and rare decays, owing to their suppression within the SM and the availability of increasingly precise experimental data, provide sensitive tests of physics at scales far beyond the direct reach of current colliders. We surveyed how different processes constrain various classes of new physics scenarios, including models with new heavy mediators such as $Z^\prime$ bosons and leptoquarks, as well as the growing interest in light, weakly coupled states that could reveal themselves through rare flavour processes. The absence of new physics signals in flavour observables has far-reaching implications for the expected flavour structure of any TeV-scale extensions of the SM. Either new flavour structures must be extremely suppressed, or aligned with the known pattern of SM Yukawa couplings. The frameworks of Minimal Flavour Violation, ${\rm U}(2)^5$ flavour symmetries, and flavour alignment represent the principal theoretical strategies for achieving this.

Finally, we reviewed a broad spectrum of theoretical ideas proposed to address the origin of the Yukawa couplings and their hierarchies, seeking to resolve the SM flavour puzzle. These include frameworks based on additional sources of electroweak symmetry breaking, abelian and non-abelian flavour symmetries, mechanisms involving extra spatial dimensions, radiative generation of masses, the clockwork mechanism, and more. Each scenario carries distinct phenomenological implications, particularly in terms of expected signals in flavour-violating observables, thus linking model building directly to experimental tests.

Taken together, these aspects highlight the dual role of flavour physics in the quest for physics beyond the SM. On the one hand, it offers important hints about the organizing principles of fermion masses and mixings. On the other hand, it imposes some of the most stringent constraints on the flavour structure of any proposed new physics.

Looking ahead, the interplay between theoretical developments and upcoming experimental advances -- including the high-luminosity phases of LHCb and Belle~II, as well as future $Z$-factories like FCC-ee or CEPC -- promises to further sharpen the picture. Whether through the discovery of deviations in precision measurements or the emergence of new guiding principles for Yukawa couplings, flavour physics will remain an essential tool in deciphering the fundamental structure of particle interactions and in the effort to discover what lies beyond the Standard Model.

\begin{ack}[Acknowledgments]%
 The research of WA is supported by the U.S. Department of Energy grant number DE-SC0010107.
\end{ack}

\seealso{Recent pedagogical introductions to flavour physics can be found for example in the following lecture notes: \cite{Grossman:2017thq, Zupan:2019uoi, Silvestrini:2019sey, Gori:2019ybw, Altmannshofer:2024ykf, Isidori:2025iyu}}

\bibliographystyle{Numbered-Style} 
\bibliography{reference}

\begin{thebibliography*}{100}
\providecommand{\bibtype}[1]{}
\providecommand{\url}[1]{{\tt #1}}
\providecommand{\urlprefix}{URL }
\expandafter\ifx\csname urlstyle\endcsname\relax
  \providecommand{\doi}[1]{doi:\discretionary{}{}{}#1}\else
  \providecommand{\doi}{doi:\discretionary{}{}{}\begingroup
  \urlstyle{rm}\Url}\fi
\providecommand{\bibinfo}[2]{#2}
\providecommand{\eprint}[2][]{\url{#2}}
\makeatletter\def\@biblabel#1{\bibinfo{label}{[#1]}}\makeatother

\bibtype{Article}%
\bibitem{Cabibbo:1963yz}
\bibinfo{author}{Nicola Cabibbo}, \bibinfo{title}{{Unitary Symmetry and
  Leptonic Decays}}, \bibinfo{journal}{Phys. Rev. Lett.} \bibinfo{volume}{10}
  (\bibinfo{year}{1963}) \bibinfo{pages}{531--533},
  \bibinfo{doi}{\doi{10.1103/PhysRevLett.10.531}}.

\bibtype{Article}%
\bibitem{Kobayashi:1973fv}
\bibinfo{author}{Makoto Kobayashi}, \bibinfo{author}{Toshihide Maskawa},
  \bibinfo{title}{{CP Violation in the Renormalizable Theory of Weak
  Interaction}}, \bibinfo{journal}{Prog. Theor. Phys.} \bibinfo{volume}{49}
  (\bibinfo{year}{1973}) \bibinfo{pages}{652--657},
  \bibinfo{doi}{\doi{10.1143/PTP.49.652}}.

\bibtype{Article}%
\bibitem{Maki:1962mu}
\bibinfo{author}{Ziro Maki}, \bibinfo{author}{Masami Nakagawa},
  \bibinfo{author}{Shoichi Sakata}, \bibinfo{title}{{Remarks on the unified
  model of elementary particles}}, \bibinfo{journal}{Prog. Theor. Phys.}
  \bibinfo{volume}{28} (\bibinfo{year}{1962}) \bibinfo{pages}{870--880},
  \bibinfo{doi}{\doi{10.1143/PTP.28.870}}.

\bibtype{Article}%
\bibitem{Pontecorvo:1967fh}
\bibinfo{author}{B. Pontecorvo}, \bibinfo{title}{{Neutrino Experiments and the
  Problem of Conservation of Leptonic Charge}}, \bibinfo{journal}{Zh. Eksp.
  Teor. Fiz.} \bibinfo{volume}{53} (\bibinfo{year}{1967})
  \bibinfo{pages}{1717--1725}.

\bibtype{Article}%
\bibitem{LHCb:2008vvz}
\bibinfo{author}{A.~Augusto Alves, Jr.}, et al.
  (\bibinfo{collaboration}{LHCb}), \bibinfo{title}{{The LHCb Detector at the
  LHC}}, \bibinfo{journal}{JINST} \bibinfo{volume}{3} (\bibinfo{year}{2008})
  \bibinfo{pages}{S08005}, \bibinfo{doi}{\doi{10.1088/1748-0221/3/08/S08005}}.

\bibtype{Article}%
\bibitem{LHCb:2018roe}
\bibinfo{author}{Roel Aaij}, et al. (\bibinfo{collaboration}{LHCb}),
  \bibinfo{title}{{Physics case for an LHCb Upgrade II - Opportunities in
  flavour physics, and beyond, in the HL-LHC era}}  (\bibinfo{year}{2018}).
  \href{http://arxiv.org/abs/1808.08865}{\tt arXiv:1808.08865}.

\bibtype{Article}%
\bibitem{Belle-II:2018jsg}
\bibinfo{author}{W. Altmannshofer}, et al. (\bibinfo{collaboration}{Belle-II}),
  \bibinfo{title}{{The Belle II Physics Book}}, \bibinfo{journal}{PTEP}
  \bibinfo{volume}{2019} (\bibinfo{number}{12}) (\bibinfo{year}{2019})
  \bibinfo{pages}{123C01}, \bibinfo{doi}{\doi{10.1093/ptep/ptz106}},
  \bibinfo{note}{[Erratum: PTEP 2020, 029201 (2020)]}.
  \href{http://arxiv.org/abs/1808.10567}{\tt arXiv:1808.10567}.

\bibtype{Article}%
\bibitem{BaBar:2014omp}
\bibinfo{author}{A.~J. Bevan}, et al. (\bibinfo{collaboration}{BaBar, Belle}),
  \bibinfo{title}{{The Physics of the B Factories}}, \bibinfo{journal}{Eur.
  Phys. J. C} \bibinfo{volume}{74} (\bibinfo{year}{2014})
  \bibinfo{pages}{3026}, \bibinfo{doi}{\doi{10.1140/epjc/s10052-014-3026-9}}.
  \href{http://arxiv.org/abs/1406.6311}{\tt arXiv:1406.6311}.

\bibtype{Article}%
\bibitem{NA62:2024pjp}
\bibinfo{author}{Eduardo Cortina~Gil}, et al. (\bibinfo{collaboration}{NA62}),
  \bibinfo{title}{{Observation of the $ {K}^{+}\to {\pi}^{+}\nu \overline{\nu}
  $ decay and measurement of its branching ratio}}, \bibinfo{journal}{JHEP}
  \bibinfo{volume}{02} (\bibinfo{year}{2025}) \bibinfo{pages}{191},
  \bibinfo{doi}{\doi{10.1007/JHEP02(2025)191}}.
  \href{http://arxiv.org/abs/2412.12015}{\tt arXiv:2412.12015}.

\bibtype{Article}%
\bibitem{KOTO:2024zbl}
\bibinfo{author}{J.~K. Ahn}, et al. (\bibinfo{collaboration}{KOTO}),
  \bibinfo{title}{{Search for the $K_L \to \pi^0 \nu \bar\nu$ Decay at the
  J-PARC KOTO Experiment}}, \bibinfo{journal}{Phys. Rev. Lett.}
  \bibinfo{volume}{134} (\bibinfo{number}{8}) (\bibinfo{year}{2025})
  \bibinfo{pages}{081802}, \bibinfo{doi}{\doi{10.1103/PhysRevLett.134.081802}}.
  \href{http://arxiv.org/abs/2411.11237}{\tt arXiv:2411.11237}.

\bibtype{Article}%
\bibitem{KOTO:2025gvq}
\bibinfo{author}{John Fry}, et al. (\bibinfo{collaboration}{KOTO}),
  \bibinfo{title}{{Proposal of the KOTO II experiment}}
  (\bibinfo{year}{2025}). \href{http://arxiv.org/abs/2501.14827}{\tt
  arXiv:2501.14827}.

\bibtype{Article}%
\bibitem{MEGII:2018kmf}
\bibinfo{author}{A.~M. Baldini}, et al. (\bibinfo{collaboration}{MEG II}),
  \bibinfo{title}{{The design of the MEG II experiment}},
  \bibinfo{journal}{Eur. Phys. J. C} \bibinfo{volume}{78} (\bibinfo{number}{5})
  (\bibinfo{year}{2018}) \bibinfo{pages}{380},
  \bibinfo{doi}{\doi{10.1140/epjc/s10052-018-5845-6}}.
  \href{http://arxiv.org/abs/1801.04688}{\tt arXiv:1801.04688}.

\bibtype{Article}%
\bibitem{Mu2e:2014fns}
\bibinfo{author}{L. Bartoszek}, et al. (\bibinfo{collaboration}{Mu2e}),
  \bibinfo{title}{{Mu2e Technical Design Report}}  (\bibinfo{year}{2014}),
  \bibinfo{doi}{\doi{10.2172/1172555}}.
  \href{http://arxiv.org/abs/1501.05241}{\tt arXiv:1501.05241}.

\bibtype{Article}%
\bibitem{FCC:2018evy}
\bibinfo{author}{A. Abada}, et al. (\bibinfo{collaboration}{FCC}),
  \bibinfo{title}{{FCC-ee: The Lepton Collider}: {Future Circular Collider
  Conceptual Design Report Volume 2}}, \bibinfo{journal}{Eur. Phys. J. ST}
  \bibinfo{volume}{228} (\bibinfo{number}{2}) (\bibinfo{year}{2019})
  \bibinfo{pages}{261--623}, \bibinfo{doi}{\doi{10.1140/epjst/e2019-900045-4}}.

\bibtype{Article}%
\bibitem{Bernardi:2022hny}
\bibinfo{author}{G. Bernardi}, et al., \bibinfo{title}{{The Future Circular
  Collider: a Summary for the US 2021 Snowmass Process}}
  (\bibinfo{year}{2022}). \href{http://arxiv.org/abs/2203.06520}{\tt
  arXiv:2203.06520}.

\bibtype{Article}%
\bibitem{CEPCStudyGroup:2018ghi}
\bibinfo{author}{Mingyi Dong}, et al. (\bibinfo{collaboration}{CEPC Study
  Group}), \bibinfo{title}{{CEPC Conceptual Design Report: Volume 2 - Physics
  \& Detector}}  (\bibinfo{year}{2018}).
  \href{http://arxiv.org/abs/1811.10545}{\tt arXiv:1811.10545}.

\bibtype{Inproceedings}%
\bibitem{CEPCPhysicsStudyGroup:2022uwl}
\bibinfo{author}{Huajie Cheng}, et al. (\bibinfo{collaboration}{CEPC Physics
  Study Group}), \bibinfo{title}{{The Physics potential of the CEPC. Prepared
  for the US Snowmass Community Planning Exercise (Snowmass 2021)}}, in:
  \bibinfo{booktitle}{{Snowmass 2021}} \bibinfo{year}{2022}.
  \href{http://arxiv.org/abs/2205.08553}{\tt arXiv:2205.08553}.

\bibtype{Article}%
\bibitem{Ai:2024nmn}
\bibinfo{author}{Xiaocong Ai}, et al., \bibinfo{title}{{Flavor Physics at CEPC:
  a General Perspective}}  (\bibinfo{year}{2024}).
  \href{http://arxiv.org/abs/2412.19743}{\tt arXiv:2412.19743}.

\bibtype{Article}%
\bibitem{tHooft:1979rat}
\bibinfo{author}{Gerard 't Hooft}, \bibinfo{title}{{Naturalness, chiral
  symmetry, and spontaneous chiral symmetry breaking}}, \bibinfo{journal}{NATO
  Sci. Ser. B} \bibinfo{volume}{59} (\bibinfo{year}{1980})
  \bibinfo{pages}{135--157}, \bibinfo{doi}{\doi{10.1007/978-1-4684-7571-5_9}}.

\bibtype{Article}%
\bibitem{ParticleDataGroup:2024cfk}
\bibinfo{author}{S. Navas}, et al. (\bibinfo{collaboration}{Particle Data
  Group}), \bibinfo{title}{{Review of particle physics}},
  \bibinfo{journal}{Phys. Rev. D} \bibinfo{volume}{110} (\bibinfo{number}{3})
  (\bibinfo{year}{2024}) \bibinfo{pages}{030001},
  \bibinfo{doi}{\doi{10.1103/PhysRevD.110.030001}}.

\bibtype{Article}%
\bibitem{Tiesinga:2021myr}
\bibinfo{author}{Eite Tiesinga}, \bibinfo{author}{Peter~J. Mohr},
  \bibinfo{author}{David~B. Newell}, \bibinfo{author}{Barry~N. Taylor},
  \bibinfo{title}{{CODATA recommended values of the fundamental physical
  constants: 2018*}}, \bibinfo{journal}{Rev. Mod. Phys.} \bibinfo{volume}{93}
  (\bibinfo{number}{2}) (\bibinfo{year}{2021}) \bibinfo{pages}{025010},
  \bibinfo{doi}{\doi{10.1103/RevModPhys.93.025010}}.

\bibtype{Article}%
\bibitem{Fodor:2016bgu}
\bibinfo{author}{Z. Fodor}, \bibinfo{author}{C. Hoelbling}, \bibinfo{author}{S.
  Krieg}, \bibinfo{author}{L. Lellouch}, \bibinfo{author}{Th. Lippert},
  \bibinfo{author}{A. Portelli}, \bibinfo{author}{A. Sastre},
  \bibinfo{author}{K.~K. Szabo}, \bibinfo{author}{L. Varnhorst},
  \bibinfo{title}{{Up and down quark masses and corrections to Dashen's theorem
  from lattice QCD and quenched QED}}, \bibinfo{journal}{Phys. Rev. Lett.}
  \bibinfo{volume}{117} (\bibinfo{number}{8}) (\bibinfo{year}{2016})
  \bibinfo{pages}{082001}, \bibinfo{doi}{\doi{10.1103/PhysRevLett.117.082001}}.
  \href{http://arxiv.org/abs/1604.07112}{\tt arXiv:1604.07112}.

\bibtype{Article}%
\bibitem{FermilabLattice:2018est}
\bibinfo{author}{A. Bazavov}, et al. (\bibinfo{collaboration}{Fermilab Lattice,
  MILC, TUMQCD}), \bibinfo{title}{{Up-, down-, strange-, charm-, and
  bottom-quark masses from four-flavor lattice QCD}}, \bibinfo{journal}{Phys.
  Rev. D} \bibinfo{volume}{98} (\bibinfo{number}{5}) (\bibinfo{year}{2018})
  \bibinfo{pages}{054517}, \bibinfo{doi}{\doi{10.1103/PhysRevD.98.054517}}.
  \href{http://arxiv.org/abs/1802.04248}{\tt arXiv:1802.04248}.

\bibtype{Article}%
\bibitem{Erler:2016atg}
\bibinfo{author}{Jens Erler}, \bibinfo{author}{Pere Masjuan},
  \bibinfo{author}{Hubert Spiesberger}, \bibinfo{title}{{Charm Quark Mass with
  Calibrated Uncertainty}}, \bibinfo{journal}{Eur. Phys. J. C}
  \bibinfo{volume}{77} (\bibinfo{number}{2}) (\bibinfo{year}{2017})
  \bibinfo{pages}{99}, \bibinfo{doi}{\doi{10.1140/epjc/s10052-017-4667-2}}.
  \href{http://arxiv.org/abs/1610.08531}{\tt arXiv:1610.08531}.

\bibtype{Article}%
\bibitem{Chetyrkin:2017lif}
\bibinfo{author}{Konstantin~G. Chetyrkin}, \bibinfo{author}{Johann~H. Kuhn},
  \bibinfo{author}{Andreas Maier}, \bibinfo{author}{Philipp Maierhofer},
  \bibinfo{author}{Peter Marquard}, \bibinfo{author}{Matthias Steinhauser},
  \bibinfo{author}{Christian Sturm}, \bibinfo{title}{{Addendum to ``Charm and
  bottom quark masses: An update''}}  (\bibinfo{year}{2017}),
  \bibinfo{doi}{\doi{10.1103/PhysRevD.96.116007}}, \bibinfo{note}{[Addendum:
  Phys.Rev.D 96, 116007 (2017)]}. \href{http://arxiv.org/abs/1710.04249}{\tt
  arXiv:1710.04249}.

\bibtype{Article}%
\bibitem{Lytle:2018evc}
\bibinfo{author}{A.~T. Lytle}, \bibinfo{author}{C.~T.~H. Davies},
  \bibinfo{author}{D. Hatton}, \bibinfo{author}{G.~P. Lepage},
  \bibinfo{author}{C. Sturm} (\bibinfo{collaboration}{HPQCD}),
  \bibinfo{title}{{Determination of quark masses from $\mathbf{n_f=4}$ lattice
  QCD and the RI-SMOM intermediate scheme}}, \bibinfo{journal}{Phys. Rev. D}
  \bibinfo{volume}{98} (\bibinfo{number}{1}) (\bibinfo{year}{2018})
  \bibinfo{pages}{014513}, \bibinfo{doi}{\doi{10.1103/PhysRevD.98.014513}}.
  \href{http://arxiv.org/abs/1805.06225}{\tt arXiv:1805.06225}.

\bibtype{Article}%
\bibitem{Narison:2019tym}
\bibinfo{author}{Stephan Narison}, \bibinfo{title}{{$\overline m_c$ and
  $\overline m_b$ from $M_{Bc}$ and improved estimates of $f_{Bc}$ and
  $f_{Bc(2S)}$}}, \bibinfo{journal}{Phys. Lett. B} \bibinfo{volume}{802}
  (\bibinfo{year}{2020}) \bibinfo{pages}{135221},
  \bibinfo{doi}{\doi{10.1016/j.physletb.2020.135221}}.
  \href{http://arxiv.org/abs/1906.03614}{\tt arXiv:1906.03614}.

\bibtype{Article}%
\bibitem{Hatton:2020qhk}
\bibinfo{author}{D. Hatton}, \bibinfo{author}{C.~T.~H. Davies},
  \bibinfo{author}{B. Galloway}, \bibinfo{author}{J. Koponen},
  \bibinfo{author}{G.~P. Lepage}, \bibinfo{author}{A.~T. Lytle}
  (\bibinfo{collaboration}{HPQCD}), \bibinfo{title}{{Charmonium properties from
  lattice $QCD$+QED : Hyperfine splitting, $J/\psi$ leptonic width, charm quark
  mass, and $a^c_\mu$}}, \bibinfo{journal}{Phys. Rev. D} \bibinfo{volume}{102}
  (\bibinfo{number}{5}) (\bibinfo{year}{2020}) \bibinfo{pages}{054511},
  \bibinfo{doi}{\doi{10.1103/PhysRevD.102.054511}}.
  \href{http://arxiv.org/abs/2005.01845}{\tt arXiv:2005.01845}.

\bibtype{Article}%
\bibitem{BESIII:2014srs}
\bibinfo{author}{M. Ablikim}, et al. (\bibinfo{collaboration}{BESIII}),
  \bibinfo{title}{{Precision measurement of the mass of the $\tau$ lepton}},
  \bibinfo{journal}{Phys. Rev. D} \bibinfo{volume}{90} (\bibinfo{number}{1})
  (\bibinfo{year}{2014}) \bibinfo{pages}{012001},
  \bibinfo{doi}{\doi{10.1103/PhysRevD.90.012001}}.
  \href{http://arxiv.org/abs/1405.1076}{\tt arXiv:1405.1076}.

\bibtype{Article}%
\bibitem{Belle-II:2023izd}
\bibinfo{author}{I. Adachi}, et al. (\bibinfo{collaboration}{Belle-II}),
  \bibinfo{title}{{Measurement of the $\tau$ lepton mass with the Belle II
  experiment}}, \bibinfo{journal}{Phys. Rev. D} \bibinfo{volume}{108}
  (\bibinfo{number}{3}) (\bibinfo{year}{2023}) \bibinfo{pages}{032006},
  \bibinfo{doi}{\doi{10.1103/PhysRevD.108.032006}}.
  \href{http://arxiv.org/abs/2305.19116}{\tt arXiv:2305.19116}.

\bibtype{Article}%
\bibitem{Anashin:2023sch}
\bibinfo{author}{V.~V. Anashin}, et al., \bibinfo{title}{{Experiments with the
  KEDR Detector at the $e^+ e^-$ Collider VEPP-4M in the Energy Range $\sqrt s
  = 1.84-3.88$\, GeV}}, \bibinfo{journal}{Phys. Part. Nucl.}
  \bibinfo{volume}{54} (\bibinfo{number}{1}) (\bibinfo{year}{2023})
  \bibinfo{pages}{185--226}, \bibinfo{doi}{\doi{10.1134/S1063779623010033}}.

\bibtype{Article}%
\bibitem{ATLAS:2019guf}
\bibinfo{author}{Georges Aad}, et al. (\bibinfo{collaboration}{ATLAS}),
  \bibinfo{title}{{Measurement of the top-quark mass in $t\bar{t}+1$-jet events
  collected with the ATLAS detector in $pp$ collisions at $\sqrt{s}=8$ TeV}},
  \bibinfo{journal}{JHEP} \bibinfo{volume}{11} (\bibinfo{year}{2019})
  \bibinfo{pages}{150}, \bibinfo{doi}{\doi{10.1007/JHEP11(2019)150}}.
  \href{http://arxiv.org/abs/1905.02302}{\tt arXiv:1905.02302}.

\bibtype{Article}%
\bibitem{CMS:2022emx}
\bibinfo{author}{Armen Tumasyan}, et al. (\bibinfo{collaboration}{CMS}),
  \bibinfo{title}{{Measurement of the top quark pole mass using $
  \textrm{t}\overline{\textrm{t}} $+jet events in the dilepton final state in
  proton-proton collisions at $ \sqrt{s} $ = 13 TeV}}, \bibinfo{journal}{JHEP}
  \bibinfo{volume}{07} (\bibinfo{year}{2023}) \bibinfo{pages}{077},
  \bibinfo{doi}{\doi{10.1007/JHEP07(2023)077}}.
  \href{http://arxiv.org/abs/2207.02270}{\tt arXiv:2207.02270}.

\bibtype{Article}%
\bibitem{Penin:2014zaa}
\bibinfo{author}{Alexander~A. Penin}, \bibinfo{author}{Nikolai Zerf},
  \bibinfo{title}{{Bottom Quark Mass from $\Upsilon$ Sum Rules to ${\cal
  O}(\alpha_s^3)$}}, \bibinfo{journal}{JHEP} \bibinfo{volume}{04}
  (\bibinfo{year}{2014}) \bibinfo{pages}{120},
  \bibinfo{doi}{\doi{10.1007/JHEP04(2014)120}}.
  \href{http://arxiv.org/abs/1401.7035}{\tt arXiv:1401.7035}.

\bibtype{Article}%
\bibitem{Beneke:2014pta}
\bibinfo{author}{M. Beneke}, \bibinfo{author}{A. Maier}, \bibinfo{author}{J.
  Piclum}, \bibinfo{author}{T. Rauh}, \bibinfo{title}{{The bottom-quark mass
  from non-relativistic sum rules at NNNLO}}, \bibinfo{journal}{Nucl. Phys. B}
  \bibinfo{volume}{891} (\bibinfo{year}{2015}) \bibinfo{pages}{42--72},
  \bibinfo{doi}{\doi{10.1016/j.nuclphysb.2014.12.001}}.
  \href{http://arxiv.org/abs/1411.3132}{\tt arXiv:1411.3132}.

\bibtype{Article}%
\bibitem{Feldmann:2008ja}
\bibinfo{author}{Thorsten Feldmann}, \bibinfo{author}{Thomas Mannel},
  \bibinfo{title}{{Large Top Mass and Non-Linear Representation of Flavour
  Symmetry}}, \bibinfo{journal}{Phys. Rev. Lett.} \bibinfo{volume}{100}
  (\bibinfo{year}{2008}) \bibinfo{pages}{171601},
  \bibinfo{doi}{\doi{10.1103/PhysRevLett.100.171601}}.
  \href{http://arxiv.org/abs/0801.1802}{\tt arXiv:0801.1802}.

\bibtype{Article}%
\bibitem{Kagan:2009bn}
\bibinfo{author}{Alexander~L. Kagan}, \bibinfo{author}{Gilad Perez},
  \bibinfo{author}{Tomer Volansky}, \bibinfo{author}{Jure Zupan},
  \bibinfo{title}{{General Minimal Flavor Violation}}, \bibinfo{journal}{Phys.
  Rev. D} \bibinfo{volume}{80} (\bibinfo{year}{2009}) \bibinfo{pages}{076002},
  \bibinfo{doi}{\doi{10.1103/PhysRevD.80.076002}}.
  \href{http://arxiv.org/abs/0903.1794}{\tt arXiv:0903.1794}.

\bibtype{Article}%
\bibitem{Faroughy:2020ina}
\bibinfo{author}{Darius~A. Faroughy}, \bibinfo{author}{Gino Isidori},
  \bibinfo{author}{Felix Wilsch}, \bibinfo{author}{Kei Yamamoto},
  \bibinfo{title}{{Flavour symmetries in the SMEFT}}, \bibinfo{journal}{JHEP}
  \bibinfo{volume}{08} (\bibinfo{year}{2020}) \bibinfo{pages}{166},
  \bibinfo{doi}{\doi{10.1007/JHEP08(2020)166}}.
  \href{http://arxiv.org/abs/2005.05366}{\tt arXiv:2005.05366}.

\bibtype{Article}%
\bibitem{Bordone:2021oof}
\bibinfo{author}{Marzia Bordone}, \bibinfo{author}{Bernat Capdevila},
  \bibinfo{author}{Paolo Gambino}, \bibinfo{title}{{Three loop calculations and
  inclusive Vcb}}, \bibinfo{journal}{Phys. Lett. B} \bibinfo{volume}{822}
  (\bibinfo{year}{2021}) \bibinfo{pages}{136679},
  \bibinfo{doi}{\doi{10.1016/j.physletb.2021.136679}}.
  \href{http://arxiv.org/abs/2107.00604}{\tt arXiv:2107.00604}.

\bibtype{Article}%
\bibitem{Bernlochner:2022ucr}
\bibinfo{author}{Florian Bernlochner}, \bibinfo{author}{Matteo Fael},
  \bibinfo{author}{Kevin Olschewsky}, \bibinfo{author}{Eric Persson},
  \bibinfo{author}{Raynette van Tonder}, \bibinfo{author}{K.~Keri Vos},
  \bibinfo{author}{Maximilian Welsch}, \bibinfo{title}{{First extraction of
  inclusive V$_{cb}$ from q$^{2}$ moments}}, \bibinfo{journal}{JHEP}
  \bibinfo{volume}{10} (\bibinfo{year}{2022}) \bibinfo{pages}{068},
  \bibinfo{doi}{\doi{10.1007/JHEP10(2022)068}}.
  \href{http://arxiv.org/abs/2205.10274}{\tt arXiv:2205.10274}.

\bibtype{Article}%
\bibitem{Harrison:2023dzh}
\bibinfo{author}{Judd Harrison}, \bibinfo{author}{Christine T.~H. Davies}
  (\bibinfo{collaboration}{HPQCD, (HPQCD Collaboration)}), \bibinfo{title}{{$B
  \to D^*$ and $B_s \to D^*_s$ vector, axial-vector and tensor form factors for
  the full $q^2$ range from lattice QCD}}, \bibinfo{journal}{Phys. Rev. D}
  \bibinfo{volume}{109} (\bibinfo{number}{9}) (\bibinfo{year}{2024})
  \bibinfo{pages}{094515}, \bibinfo{doi}{\doi{10.1103/PhysRevD.109.094515}}.
  \href{http://arxiv.org/abs/2304.03137}{\tt arXiv:2304.03137}.

\bibtype{Article}%
\bibitem{Aoki:2023qpa}
\bibinfo{author}{Y. Aoki}, \bibinfo{author}{B. Colquhoun}, \bibinfo{author}{H.
  Fukaya}, \bibinfo{author}{S. Hashimoto}, \bibinfo{author}{T. Kaneko},
  \bibinfo{author}{R. Kellermann}, \bibinfo{author}{J. Koponen},
  \bibinfo{author}{E. Kou} (\bibinfo{collaboration}{JLQCD}),
  \bibinfo{title}{{$B \to D^* \ell \nu_\ell$ semileptonic form factors from
  lattice QCD with M{\"o}bius domain-wall quarks}}, \bibinfo{journal}{Phys.
  Rev. D} \bibinfo{volume}{109} (\bibinfo{number}{7}) (\bibinfo{year}{2024})
  \bibinfo{pages}{074503}, \bibinfo{doi}{\doi{10.1103/PhysRevD.109.074503}}.
  \href{http://arxiv.org/abs/2306.05657}{\tt arXiv:2306.05657}.

\bibtype{Article}%
\bibitem{Belle:2023bwv}
\bibinfo{author}{M.~T. Prim}, et al. (\bibinfo{collaboration}{Belle}),
  \bibinfo{title}{{Measurement of differential distributions of $B \to D^* \ell
  \nu_\ell$ and implications on $|V_{cb}|$}}, \bibinfo{journal}{Phys. Rev. D}
  \bibinfo{volume}{108} (\bibinfo{number}{1}) (\bibinfo{year}{2023})
  \bibinfo{pages}{012002}, \bibinfo{doi}{\doi{10.1103/PhysRevD.108.012002}}.
  \href{http://arxiv.org/abs/2301.07529}{\tt arXiv:2301.07529}.

\bibtype{Article}%
\bibitem{Belle-II:2023okj}
\bibinfo{author}{I. Adachi}, et al. (\bibinfo{collaboration}{Belle-II}),
  \bibinfo{title}{{Determination of $|V_{cb}|$ using $B^0 \to D^{*+} \ell^-
  \nu_\ell$ decays with Belle II}}, \bibinfo{journal}{Phys. Rev. D}
  \bibinfo{volume}{108} (\bibinfo{number}{9}) (\bibinfo{year}{2023})
  \bibinfo{pages}{092013}, \bibinfo{doi}{\doi{10.1103/PhysRevD.108.092013}}.
  \href{http://arxiv.org/abs/2310.01170}{\tt arXiv:2310.01170}.

\bibtype{Article}%
\bibitem{Leljak:2021vte}
\bibinfo{author}{Domagoj Leljak}, \bibinfo{author}{Bla{\v{z}}enka Meli{\'c}},
  \bibinfo{author}{Danny van Dyk}, \bibinfo{title}{{The $ \overline{B} \to \pi
  $ form factors from QCD and their impact on $|V_{ub}|$}},
  \bibinfo{journal}{JHEP} \bibinfo{volume}{07} (\bibinfo{year}{2021})
  \bibinfo{pages}{036}, \bibinfo{doi}{\doi{10.1007/JHEP07(2021)036}}.
  \href{http://arxiv.org/abs/2102.07233}{\tt arXiv:2102.07233}.

\bibtype{Article}%
\bibitem{Colquhoun:2022atw}
\bibinfo{author}{Brian Colquhoun}, \bibinfo{author}{Shoji Hashimoto},
  \bibinfo{author}{Takashi Kaneko}, \bibinfo{author}{Jonna Koponen}
  (\bibinfo{collaboration}{JLQCD}), \bibinfo{title}{{Form factors of $B \to \pi
  \ell \nu$ and a determination of $|V_{ub}|$ with M{\"o}bius domain-wall
  fermions}}, \bibinfo{journal}{Phys. Rev. D} \bibinfo{volume}{106}
  (\bibinfo{number}{5}) (\bibinfo{year}{2022}) \bibinfo{pages}{054502},
  \bibinfo{doi}{\doi{10.1103/PhysRevD.106.054502}}.
  \href{http://arxiv.org/abs/2203.04938}{\tt arXiv:2203.04938}.

\bibtype{Article}%
\bibitem{Belle:2023asa}
\bibinfo{author}{M. Hohmann}, et al. (\bibinfo{collaboration}{Belle}),
  \bibinfo{title}{{Measurement of the ratio of partial branching fractions of
  inclusive $B \to X_u \ell\nu$ to $B \to X_c \ell\nu$ and the ratio of their
  spectra with hadronic tagging}}, \bibinfo{journal}{Phys. Rev. D}
  \bibinfo{volume}{111} (\bibinfo{number}{9}) (\bibinfo{year}{2025})
  \bibinfo{pages}{092016}, \bibinfo{doi}{\doi{10.1103/PhysRevD.111.092016}}.
  \href{http://arxiv.org/abs/2311.00458}{\tt arXiv:2311.00458}.

\bibtype{Article}%
\bibitem{Bolognani:2023mcf}
\bibinfo{author}{Carolina Bolognani}, \bibinfo{author}{Danny van Dyk},
  \bibinfo{author}{K.~Keri Vos}, \bibinfo{title}{{New determination of
  $|V_{ub}/V_{cb}|$ from $B_s^0\to \lbrace K^-, D_s^- \rbrace \mu^+\nu$}},
  \bibinfo{journal}{JHEP} \bibinfo{volume}{11} (\bibinfo{year}{2023})
  \bibinfo{pages}{082}, \bibinfo{doi}{\doi{10.1007/JHEP11(2023)082}}.
  \href{http://arxiv.org/abs/2308.04347}{\tt arXiv:2308.04347}.

\bibtype{Article}%
\bibitem{Hardy:2020qwl}
\bibinfo{author}{J.~C. Hardy}, \bibinfo{author}{I.~S. Towner},
  \bibinfo{title}{{Superallowed $0^+ \to 0^+$ nuclear $\beta$ decays: 2020
  critical survey, with implications for V$_{ud}$ and CKM unitarity}},
  \bibinfo{journal}{Phys. Rev. C} \bibinfo{volume}{102} (\bibinfo{number}{4})
  (\bibinfo{year}{2020}) \bibinfo{pages}{045501},
  \bibinfo{doi}{\doi{10.1103/PhysRevC.102.045501}}.

\bibtype{Article}%
\bibitem{UCNt:2021pcg}
\bibinfo{author}{F.~M. Gonzalez}, et al.
  (\bibinfo{collaboration}{UCN{\ensuremath{\tau}}}), \bibinfo{title}{{Improved
  neutron lifetime measurement with UCN$\tau$}}, \bibinfo{journal}{Phys. Rev.
  Lett.} \bibinfo{volume}{127} (\bibinfo{number}{16}) (\bibinfo{year}{2021})
  \bibinfo{pages}{162501}, \bibinfo{doi}{\doi{10.1103/PhysRevLett.127.162501}}.
  \href{http://arxiv.org/abs/2106.10375}{\tt arXiv:2106.10375}.

\bibtype{Article}%
\bibitem{Pocanic:2003pf}
\bibinfo{author}{D. Pocanic}, et al., \bibinfo{title}{{Precise measurement of
  the $\pi^+ \to \pi^0 e^+ \nu$ branching ratio}}, \bibinfo{journal}{Phys. Rev.
  Lett.} \bibinfo{volume}{93} (\bibinfo{year}{2004}) \bibinfo{pages}{181803},
  \bibinfo{doi}{\doi{10.1103/PhysRevLett.93.181803}}.
  \href{http://arxiv.org/abs/hep-ex/0312030}{\tt arXiv:hep-ex/0312030}.

\bibtype{Article}%
\bibitem{PIONEER:2022yag}
\bibinfo{author}{W. Altmannshofer}, et al. (\bibinfo{collaboration}{PIONEER}),
  \bibinfo{title}{{PIONEER: Studies of Rare Pion Decays}}
  (\bibinfo{year}{2022}). \href{http://arxiv.org/abs/2203.01981}{\tt
  arXiv:2203.01981}.

\bibtype{Article}%
\bibitem{Seng:2021nar}
\bibinfo{author}{Chien-Yeah Seng}, \bibinfo{author}{Daniel Galviz},
  \bibinfo{author}{William~J. Marciano}, \bibinfo{author}{Ulf-G. Mei{\ss}ner},
  \bibinfo{title}{{Update on $|V_{us}|$ and $|V_{us}/V_{ud}|$ from semileptonic
  kaon and pion decays}}, \bibinfo{journal}{Phys. Rev. D} \bibinfo{volume}{105}
  (\bibinfo{number}{1}) (\bibinfo{year}{2022}) \bibinfo{pages}{013005},
  \bibinfo{doi}{\doi{10.1103/PhysRevD.105.013005}}.
  \href{http://arxiv.org/abs/2107.14708}{\tt arXiv:2107.14708}.

\bibtype{Article}%
\bibitem{Seng:2022wcw}
\bibinfo{author}{Chien-Yeah Seng}, \bibinfo{author}{Daniel Galviz},
  \bibinfo{author}{Mikhail Gorchtein}, \bibinfo{author}{Ulf-G. Mei{\ss}ner},
  \bibinfo{title}{{Complete theory of radiative corrections to $K_{\ell 3}$
  decays and the $V_{us}$ update}}, \bibinfo{journal}{JHEP}
  \bibinfo{volume}{07} (\bibinfo{year}{2022}) \bibinfo{pages}{071},
  \bibinfo{doi}{\doi{10.1007/JHEP07(2022)071}}.
  \href{http://arxiv.org/abs/2203.05217}{\tt arXiv:2203.05217}.

\bibtype{Article}%
\bibitem{Cirigliano:2022yyo}
\bibinfo{author}{Vincenzo Cirigliano}, \bibinfo{author}{Andreas Crivellin},
  \bibinfo{author}{Martin Hoferichter}, \bibinfo{author}{Matthew Moulson},
  \bibinfo{title}{{Scrutinizing CKM unitarity with a new measurement of the
  $K_{\mu 3}/K_{\mu 2}$ branching fraction}}, \bibinfo{journal}{Phys. Lett. B}
  \bibinfo{volume}{838} (\bibinfo{year}{2023}) \bibinfo{pages}{137748},
  \bibinfo{doi}{\doi{10.1016/j.physletb.2023.137748}}.
  \href{http://arxiv.org/abs/2208.11707}{\tt arXiv:2208.11707}.

\bibtype{Article}%
\bibitem{Glashow:1970gm}
\bibinfo{author}{S.~L. Glashow}, \bibinfo{author}{J. Iliopoulos},
  \bibinfo{author}{L. Maiani}, \bibinfo{title}{{Weak Interactions with
  Lepton-Hadron Symmetry}}, \bibinfo{journal}{Phys. Rev. D} \bibinfo{volume}{2}
  (\bibinfo{year}{1970}) \bibinfo{pages}{1285--1292},
  \bibinfo{doi}{\doi{10.1103/PhysRevD.2.1285}}.

\bibtype{Inproceedings}%
\bibitem{Nierste:2009wg}
\bibinfo{author}{Ulrich Nierste}, \bibinfo{title}{{Three Lectures on Meson
  Mixing and CKM phenomenology}}, in: \bibinfo{booktitle}{{Helmholz
  International Summer School on Heavy Quark Physics}} \bibinfo{year}{2009},
  pp. \bibinfo{pages}{1--38}. \href{http://arxiv.org/abs/0904.1869}{\tt
  arXiv:0904.1869}.

\bibtype{Article}%
\bibitem{FlavourLatticeAveragingGroupFLAG:2024oxs}
\bibinfo{author}{Y. Aoki}, et al. (\bibinfo{collaboration}{Flavour Lattice
  Averaging Group (FLAG)}), \bibinfo{title}{{FLAG Review 2024}}
  (\bibinfo{year}{2024}). \href{http://arxiv.org/abs/2411.04268}{\tt
  arXiv:2411.04268}.

\bibtype{Article}%
\bibitem{Hoferichter:2021lct}
\bibinfo{author}{Martin Hoferichter}, \bibinfo{author}{Bai-Long Hoid},
  \bibinfo{author}{Bastian Kubis}, \bibinfo{author}{Jan L{\"u}dtke},
  \bibinfo{title}{{Improved Standard-Model prediction for $\pi^0\to e^+e^-$}},
  \bibinfo{journal}{Phys. Rev. Lett.} \bibinfo{volume}{128}
  (\bibinfo{number}{17}) (\bibinfo{year}{2022}) \bibinfo{pages}{172004},
  \bibinfo{doi}{\doi{10.1103/PhysRevLett.128.172004}}.
  \href{http://arxiv.org/abs/2105.04563}{\tt arXiv:2105.04563}.

\bibtype{Article}%
\bibitem{Hoferichter:2023wiy}
\bibinfo{author}{Martin Hoferichter}, \bibinfo{author}{Bai-Long Hoid},
  \bibinfo{author}{Jacobo~Ruiz de Elvira}, \bibinfo{title}{{Improved
  Standard-Model prediction for $K_{L} \to \ell^+\ell^-$}},
  \bibinfo{journal}{JHEP} \bibinfo{volume}{04} (\bibinfo{year}{2024})
  \bibinfo{pages}{071}, \bibinfo{doi}{\doi{10.1007/JHEP04(2024)071}}.
  \href{http://arxiv.org/abs/2310.17689}{\tt arXiv:2310.17689}.

\bibtype{Article}%
\bibitem{Burdman:2001tf}
\bibinfo{author}{Gustavo Burdman}, \bibinfo{author}{Eugene Golowich},
  \bibinfo{author}{JoAnne~L. Hewett}, \bibinfo{author}{Sandip Pakvasa},
  \bibinfo{title}{{Rare charm decays in the standard model and beyond}},
  \bibinfo{journal}{Phys. Rev. D} \bibinfo{volume}{66} (\bibinfo{year}{2002})
  \bibinfo{pages}{014009}, \bibinfo{doi}{\doi{10.1103/PhysRevD.66.014009}}.
  \href{http://arxiv.org/abs/hep-ph/0112235}{\tt arXiv:hep-ph/0112235}.

\bibtype{Article}%
\bibitem{Chay:1990da}
\bibinfo{author}{Junegone Chay}, \bibinfo{author}{Howard Georgi},
  \bibinfo{author}{Benjamin Grinstein}, \bibinfo{title}{{Lepton energy
  distributions in heavy meson decays from QCD}}, \bibinfo{journal}{Phys. Lett.
  B} \bibinfo{volume}{247} (\bibinfo{year}{1990}) \bibinfo{pages}{399--405},
  \bibinfo{doi}{\doi{10.1016/0370-2693(90)90916-T}}.

\bibtype{Article}%
\bibitem{Bigi:1992su}
\bibinfo{author}{Ikaros I.~Y. Bigi}, \bibinfo{author}{N.~G. Uraltsev},
  \bibinfo{author}{A.~I. Vainshtein}, \bibinfo{title}{{Nonperturbative
  corrections to inclusive beauty and charm decays: QCD versus phenomenological
  models}}, \bibinfo{journal}{Phys. Lett. B} \bibinfo{volume}{293}
  (\bibinfo{year}{1992}) \bibinfo{pages}{430--436},
  \bibinfo{doi}{\doi{10.1016/0370-2693(92)90908-M}}, \bibinfo{note}{[Erratum:
  Phys.Lett.B 297, 477--477 (1992)]}.
  \href{http://arxiv.org/abs/hep-ph/9207214}{\tt arXiv:hep-ph/9207214}.

\bibtype{Article}%
\bibitem{Manohar:1993qn}
\bibinfo{author}{Aneesh~V. Manohar}, \bibinfo{author}{Mark~B. Wise},
  \bibinfo{title}{{Inclusive semileptonic B and polarized Lambda(b) decays from
  QCD}}, \bibinfo{journal}{Phys. Rev. D} \bibinfo{volume}{49}
  (\bibinfo{year}{1994}) \bibinfo{pages}{1310--1329},
  \bibinfo{doi}{\doi{10.1103/PhysRevD.49.1310}}.
  \href{http://arxiv.org/abs/hep-ph/9308246}{\tt arXiv:hep-ph/9308246}.

\bibtype{Article}%
\bibitem{Buchmuller:1985jz}
\bibinfo{author}{W. Buchmuller}, \bibinfo{author}{D. Wyler},
  \bibinfo{title}{{Effective Lagrangian Analysis of New Interactions and Flavor
  Conservation}}, \bibinfo{journal}{Nucl. Phys. B} \bibinfo{volume}{268}
  (\bibinfo{year}{1986}) \bibinfo{pages}{621--653},
  \bibinfo{doi}{\doi{10.1016/0550-3213(86)90262-2}}.

\bibtype{Article}%
\bibitem{Grzadkowski:2010es}
\bibinfo{author}{B. Grzadkowski}, \bibinfo{author}{M. Iskrzynski},
  \bibinfo{author}{M. Misiak}, \bibinfo{author}{J. Rosiek},
  \bibinfo{title}{{Dimension-Six Terms in the Standard Model Lagrangian}},
  \bibinfo{journal}{JHEP} \bibinfo{volume}{10} (\bibinfo{year}{2010})
  \bibinfo{pages}{085}, \bibinfo{doi}{\doi{10.1007/JHEP10(2010)085}}.
  \href{http://arxiv.org/abs/1008.4884}{\tt arXiv:1008.4884}.

\bibtype{Article}%
\bibitem{Jenkins:2017jig}
\bibinfo{author}{Elizabeth~E. Jenkins}, \bibinfo{author}{Aneesh~V. Manohar},
  \bibinfo{author}{Peter Stoffer}, \bibinfo{title}{{Low-Energy Effective Field
  Theory below the Electroweak Scale: Operators and Matching}},
  \bibinfo{journal}{JHEP} \bibinfo{volume}{03} (\bibinfo{year}{2018})
  \bibinfo{pages}{016}, \bibinfo{doi}{\doi{10.1007/JHEP03(2018)016}},
  \bibinfo{note}{[Erratum: JHEP 12, 043 (2023)]}.
  \href{http://arxiv.org/abs/1709.04486}{\tt arXiv:1709.04486}.

\bibtype{Article}%
\bibitem{Buchalla:1995vs}
\bibinfo{author}{Gerhard Buchalla}, \bibinfo{author}{Andrzej~J. Buras},
  \bibinfo{author}{Markus~E. Lautenbacher}, \bibinfo{title}{{Weak decays beyond
  leading logarithms}}, \bibinfo{journal}{Rev. Mod. Phys.} \bibinfo{volume}{68}
  (\bibinfo{year}{1996}) \bibinfo{pages}{1125--1144},
  \bibinfo{doi}{\doi{10.1103/RevModPhys.68.1125}}.
  \href{http://arxiv.org/abs/hep-ph/9512380}{\tt arXiv:hep-ph/9512380}.

\bibtype{Article}%
\bibitem{Aebischer:2017gaw}
\bibinfo{author}{Jason Aebischer}, \bibinfo{author}{Matteo Fael},
  \bibinfo{author}{Christoph Greub}, \bibinfo{author}{Javier Virto},
  \bibinfo{title}{{B physics Beyond the Standard Model at One Loop: Complete
  Renormalization Group Evolution below the Electroweak Scale}},
  \bibinfo{journal}{JHEP} \bibinfo{volume}{09} (\bibinfo{year}{2017})
  \bibinfo{pages}{158}, \bibinfo{doi}{\doi{10.1007/JHEP09(2017)158}}.
  \href{http://arxiv.org/abs/1704.06639}{\tt arXiv:1704.06639}.

\bibtype{Article}%
\bibitem{Weinberg:1979sa}
\bibinfo{author}{Steven Weinberg}, \bibinfo{title}{{Baryon and Lepton
  Nonconserving Processes}}, \bibinfo{journal}{Phys. Rev. Lett.}
  \bibinfo{volume}{43} (\bibinfo{year}{1979}) \bibinfo{pages}{1566--1570},
  \bibinfo{doi}{\doi{10.1103/PhysRevLett.43.1566}}.

\bibtype{Article}%
\bibitem{HeavyFlavorAveragingGroupHFLAV:2024ctg}
\bibinfo{author}{Swagato Banerjee}, et al. (\bibinfo{collaboration}{Heavy
  Flavor Averaging Group (HFLAV)}), \bibinfo{title}{{Averages of $b$-hadron,
  $c$-hadron, and $\tau$-lepton properties as of 2023}}
  (\bibinfo{year}{2024}). \href{http://arxiv.org/abs/2411.18639}{\tt
  arXiv:2411.18639}.

\bibtype{Article}%
\bibitem{Bai:2014cva}
\bibinfo{author}{Z. Bai}, \bibinfo{author}{N.~H. Christ}, \bibinfo{author}{T.
  Izubuchi}, \bibinfo{author}{C.~T. Sachrajda}, \bibinfo{author}{A. Soni},
  \bibinfo{author}{J. Yu}, \bibinfo{title}{{$K_L-K_S$ Mass Difference from
  Lattice QCD}}, \bibinfo{journal}{Phys. Rev. Lett.} \bibinfo{volume}{113}
  (\bibinfo{year}{2014}) \bibinfo{pages}{112003},
  \bibinfo{doi}{\doi{10.1103/PhysRevLett.113.112003}}.
  \href{http://arxiv.org/abs/1406.0916}{\tt arXiv:1406.0916}.

\bibtype{Article}%
\bibitem{Wang:2022lfq}
\bibinfo{author}{Bigeng Wang}, \bibinfo{title}{{Calculating $\Delta m_K$ with
  lattice QCD}}, \bibinfo{journal}{PoS} \bibinfo{volume}{LATTICE2021}
  (\bibinfo{year}{2022}) \bibinfo{pages}{141},
  \bibinfo{doi}{\doi{10.22323/1.396.0141}}.
  \href{http://arxiv.org/abs/2301.01387}{\tt arXiv:2301.01387}.

\bibtype{Article}%
\bibitem{Buras:2024mnq}
\bibinfo{author}{Andrzej~J. Buras}, \bibinfo{author}{Peter Stangl},
  \bibinfo{title}{{On the interplay of constraints from $B_{s},$D,~ and K meson
  mixing in $Z^\prime $ models with implications for $b\!\rightarrow \! s \nu
  {\bar{\nu }}$ transitions}}, \bibinfo{journal}{Eur. Phys. J. C}
  \bibinfo{volume}{85} (\bibinfo{number}{5}) (\bibinfo{year}{2025})
  \bibinfo{pages}{519}, \bibinfo{doi}{\doi{10.1140/epjc/s10052-025-14168-z}}.
  \href{http://arxiv.org/abs/2412.14254}{\tt arXiv:2412.14254}.

\bibtype{Article}%
\bibitem{Golowich:2007ka}
\bibinfo{author}{Eugene Golowich}, \bibinfo{author}{JoAnne Hewett},
  \bibinfo{author}{Sandip Pakvasa}, \bibinfo{author}{Alexey~A. Petrov},
  \bibinfo{title}{{Implications of $D^0$ - $\bar{D}^0$ Mixing for New
  Physics}}, \bibinfo{journal}{Phys. Rev. D} \bibinfo{volume}{76}
  (\bibinfo{year}{2007}) \bibinfo{pages}{095009},
  \bibinfo{doi}{\doi{10.1103/PhysRevD.76.095009}}.
  \href{http://arxiv.org/abs/0705.3650}{\tt arXiv:0705.3650}.

\bibtype{Article}%
\bibitem{Falk:2001hx}
\bibinfo{author}{Adam~F. Falk}, \bibinfo{author}{Yuval Grossman},
  \bibinfo{author}{Zoltan Ligeti}, \bibinfo{author}{Alexey~A. Petrov},
  \bibinfo{title}{{SU(3) breaking and D0 - anti-D0 mixing}},
  \bibinfo{journal}{Phys. Rev. D} \bibinfo{volume}{65} (\bibinfo{year}{2002})
  \bibinfo{pages}{054034}, \bibinfo{doi}{\doi{10.1103/PhysRevD.65.054034}}.
  \href{http://arxiv.org/abs/hep-ph/0110317}{\tt arXiv:hep-ph/0110317}.

\bibtype{Article}%
\bibitem{Lenz:2020awd}
\bibinfo{author}{Alexander Lenz}, \bibinfo{author}{Guy Wilkinson},
  \bibinfo{title}{{Mixing and CP Violation in the Charm System}},
  \bibinfo{journal}{Ann. Rev. Nucl. Part. Sci.} \bibinfo{volume}{71}
  (\bibinfo{year}{2021}) \bibinfo{pages}{59--85},
  \bibinfo{doi}{\doi{10.1146/annurev-nucl-102419-124613}}.
  \href{http://arxiv.org/abs/2011.04443}{\tt arXiv:2011.04443}.

\bibtype{Article}%
\bibitem{DiCarlo:2025mvt}
\bibinfo{author}{Matteo Di~Carlo}, \bibinfo{author}{Felix Erben},
  \bibinfo{author}{Maxwell~T. Hansen}, \bibinfo{title}{{Long distance
  contributions to neutral $D$-meson mixing from lattice QCD}}
  (\bibinfo{year}{2025}). \href{http://arxiv.org/abs/2504.16189}{\tt
  arXiv:2504.16189}.

\bibtype{Article}%
\bibitem{Kagan:2020vri}
\bibinfo{author}{Alexander~L. Kagan}, \bibinfo{author}{Luca Silvestrini},
  \bibinfo{title}{{Dispersive and absorptive $CP$ violation in $D^0-
  \overline{D^0}$ mixing}}, \bibinfo{journal}{Phys. Rev. D}
  \bibinfo{volume}{103} (\bibinfo{number}{5}) (\bibinfo{year}{2021})
  \bibinfo{pages}{053008}, \bibinfo{doi}{\doi{10.1103/PhysRevD.103.053008}}.
  \href{http://arxiv.org/abs/2001.07207}{\tt arXiv:2001.07207}.

\bibtype{Article}%
\bibitem{Ciuchini:1997bw}
\bibinfo{author}{Marco Ciuchini}, \bibinfo{author}{E. Franco},
  \bibinfo{author}{V. Lubicz}, \bibinfo{author}{G. Martinelli},
  \bibinfo{author}{I. Scimemi}, \bibinfo{author}{L. Silvestrini},
  \bibinfo{title}{{Next-to-leading order QCD corrections to Delta F = 2
  effective Hamiltonians}}, \bibinfo{journal}{Nucl. Phys. B}
  \bibinfo{volume}{523} (\bibinfo{year}{1998}) \bibinfo{pages}{501--525},
  \bibinfo{doi}{\doi{10.1016/S0550-3213(98)00161-8}}.
  \href{http://arxiv.org/abs/hep-ph/9711402}{\tt arXiv:hep-ph/9711402}.

\bibtype{Article}%
\bibitem{Buras:2000if}
\bibinfo{author}{Andrzej~J. Buras}, \bibinfo{author}{Mikolaj Misiak},
  \bibinfo{author}{Joerg Urban}, \bibinfo{title}{{Two loop QCD anomalous
  dimensions of flavor changing four quark operators within and beyond the
  standard model}}, \bibinfo{journal}{Nucl. Phys. B} \bibinfo{volume}{586}
  (\bibinfo{year}{2000}) \bibinfo{pages}{397--426},
  \bibinfo{doi}{\doi{10.1016/S0550-3213(00)00437-5}}.
  \href{http://arxiv.org/abs/hep-ph/0005183}{\tt arXiv:hep-ph/0005183}.

\bibtype{Article}%
\bibitem{UTfit:2007eik}
\bibinfo{author}{M. Bona}, et al. (\bibinfo{collaboration}{UTfit}),
  \bibinfo{title}{{Model-independent constraints on $\Delta F=2$ operators and
  the scale of new physics}}, \bibinfo{journal}{JHEP} \bibinfo{volume}{03}
  (\bibinfo{year}{2008}) \bibinfo{pages}{049},
  \bibinfo{doi}{\doi{10.1088/1126-6708/2008/03/049}}.
  \href{http://arxiv.org/abs/0707.0636}{\tt arXiv:0707.0636}.

\bibtype{Article}%
\bibitem{Charles:2020dfl}
\bibinfo{author}{J\'er\^ome Charles}, \bibinfo{author}{S\'ebastien
  Descotes-Genon}, \bibinfo{author}{Zoltan Ligeti}, \bibinfo{author}{St\'ephane
  Monteil}, \bibinfo{author}{Michele Papucci}, \bibinfo{author}{Karim
  Trabelsi}, \bibinfo{author}{Luiz Vale~Silva}, \bibinfo{title}{{New physics in
  $B$ meson mixing: future sensitivity and limitations}},
  \bibinfo{journal}{Phys. Rev. D} \bibinfo{volume}{102} (\bibinfo{number}{5})
  (\bibinfo{year}{2020}) \bibinfo{pages}{056023},
  \bibinfo{doi}{\doi{10.1103/PhysRevD.102.056023}}.
  \href{http://arxiv.org/abs/2006.04824}{\tt arXiv:2006.04824}.

\bibtype{Article}%
\bibitem{Altmannshofer:2013lfa}
\bibinfo{author}{Wolfgang Altmannshofer}, \bibinfo{author}{Roni Harnik},
  \bibinfo{author}{Jure Zupan}, \bibinfo{title}{{Low Energy Probes of PeV Scale
  Sfermions}}, \bibinfo{journal}{JHEP} \bibinfo{volume}{11}
  (\bibinfo{year}{2013}) \bibinfo{pages}{202},
  \bibinfo{doi}{\doi{10.1007/JHEP11(2013)202}}.
  \href{http://arxiv.org/abs/1308.3653}{\tt arXiv:1308.3653}.

\bibtype{Article}%
\bibitem{Isidori:2019pae}
\bibinfo{author}{Gino Isidori}, \bibinfo{author}{Sokratis Trifinopoulos},
  \bibinfo{title}{{Exploring the flavour structure of the high-scale MSSM}},
  \bibinfo{journal}{Eur. Phys. J. C} \bibinfo{volume}{80} (\bibinfo{number}{3})
  (\bibinfo{year}{2020}) \bibinfo{pages}{291},
  \bibinfo{doi}{\doi{10.1140/epjc/s10052-020-7821-1}}.
  \href{http://arxiv.org/abs/1912.09940}{\tt arXiv:1912.09940}.

\bibtype{Article}%
\bibitem{Blake:2016olu}
\bibinfo{author}{Thomas Blake}, \bibinfo{author}{Gaia Lanfranchi},
  \bibinfo{author}{David~M. Straub}, \bibinfo{title}{{Rare $B$ Decays as Tests
  of the Standard Model}}, \bibinfo{journal}{Prog. Part. Nucl. Phys.}
  \bibinfo{volume}{92} (\bibinfo{year}{2017}) \bibinfo{pages}{50--91},
  \bibinfo{doi}{\doi{10.1016/j.ppnp.2016.10.001}}.
  \href{http://arxiv.org/abs/1606.00916}{\tt arXiv:1606.00916}.

\bibtype{Inproceedings}%
\bibitem{Altmannshofer:2022hfs}
\bibinfo{author}{Wolfgang Altmannshofer}, \bibinfo{author}{Flavio Archilli},
  \bibinfo{title}{{Rare decays of b and c hadrons}}, in:
  \bibinfo{booktitle}{{Snowmass 2021}} \bibinfo{year}{2022}.
  \href{http://arxiv.org/abs/2206.11331}{\tt arXiv:2206.11331}.

\bibtype{Article}%
\bibitem{Altmannshofer:2017yso}
\bibinfo{author}{Wolfgang Altmannshofer}, \bibinfo{author}{Peter Stangl},
  \bibinfo{author}{David~M. Straub}, \bibinfo{title}{{Interpreting Hints for
  Lepton Flavor Universality Violation}}, \bibinfo{journal}{Phys. Rev. D}
  \bibinfo{volume}{96} (\bibinfo{number}{5}) (\bibinfo{year}{2017})
  \bibinfo{pages}{055008}, \bibinfo{doi}{\doi{10.1103/PhysRevD.96.055008}}.
  \href{http://arxiv.org/abs/1704.05435}{\tt arXiv:1704.05435}.

\bibtype{Article}%
\bibitem{DiLuzio:2017chi}
\bibinfo{author}{Luca Di~Luzio}, \bibinfo{author}{Marco Nardecchia},
  \bibinfo{title}{{What is the scale of new physics behind the $B$-flavour
  anomalies?}}, \bibinfo{journal}{Eur. Phys. J. C} \bibinfo{volume}{77}
  (\bibinfo{number}{8}) (\bibinfo{year}{2017}) \bibinfo{pages}{536},
  \bibinfo{doi}{\doi{10.1140/epjc/s10052-017-5118-9}}.
  \href{http://arxiv.org/abs/1706.01868}{\tt arXiv:1706.01868}.

\bibtype{Inproceedings}%
\bibitem{Aebischer:2022vky}
\bibinfo{author}{Jason Aebischer}, \bibinfo{author}{Andrzej~J. Buras},
  \bibinfo{author}{Jacky Kumar}, \bibinfo{title}{{On the Importance of Rare
  Kaon Decays: A Snowmass 2021 White Paper}}, in: \bibinfo{booktitle}{{Snowmass
  2021}} \bibinfo{year}{2022}. \href{http://arxiv.org/abs/2203.09524}{\tt
  arXiv:2203.09524}.

\bibtype{Article}%
\bibitem{DAmbrosio:2023irq}
\bibinfo{author}{G. D'Ambrosio}, \bibinfo{author}{F. Mahmoudi},
  \bibinfo{author}{S. Neshatpour}, \bibinfo{title}{{Beyond the Standard Model
  prospects for kaon physics at future experiments}}, \bibinfo{journal}{JHEP}
  \bibinfo{volume}{02} (\bibinfo{year}{2024}) \bibinfo{pages}{166},
  \bibinfo{doi}{\doi{10.1007/JHEP02(2024)166}}.
  \href{http://arxiv.org/abs/2311.04878}{\tt arXiv:2311.04878}.

\bibtype{Article}%
\bibitem{Altmannshofer:2017wqy}
\bibinfo{author}{Wolfgang Altmannshofer}, \bibinfo{author}{Christoph Niehoff},
  \bibinfo{author}{David~M. Straub}, \bibinfo{title}{{$B_s\to\mu^+\mu^-$ as
  current and future probe of new physics}}, \bibinfo{journal}{JHEP}
  \bibinfo{volume}{05} (\bibinfo{year}{2017}) \bibinfo{pages}{076},
  \bibinfo{doi}{\doi{10.1007/JHEP05(2017)076}}.
  \href{http://arxiv.org/abs/1702.05498}{\tt arXiv:1702.05498}.

\bibtype{Article}%
\bibitem{Fleischer:2017ltw}
\bibinfo{author}{Robert Fleischer}, \bibinfo{author}{Ruben Jaarsma},
  \bibinfo{author}{Gilberto Tetlalmatzi-Xolocotzi}, \bibinfo{title}{{In Pursuit
  of New Physics with $B^0_{s,d}\to\ell^+\ell^-$}}, \bibinfo{journal}{JHEP}
  \bibinfo{volume}{05} (\bibinfo{year}{2017}) \bibinfo{pages}{156},
  \bibinfo{doi}{\doi{10.1007/JHEP05(2017)156}}.
  \href{http://arxiv.org/abs/1703.10160}{\tt arXiv:1703.10160}.

\bibtype{Article}%
\bibitem{Bobeth:2007dw}
\bibinfo{author}{Christoph Bobeth}, \bibinfo{author}{Gudrun Hiller},
  \bibinfo{author}{Giorgi Piranishvili}, \bibinfo{title}{{Angular distributions
  of $\bar{B} \to \bar{K} \ell^+\ell^-$ decays}}, \bibinfo{journal}{JHEP}
  \bibinfo{volume}{12} (\bibinfo{year}{2007}) \bibinfo{pages}{040},
  \bibinfo{doi}{\doi{10.1088/1126-6708/2007/12/040}}.
  \href{http://arxiv.org/abs/0709.4174}{\tt arXiv:0709.4174}.

\bibtype{Article}%
\bibitem{Bobeth:2008ij}
\bibinfo{author}{Christoph Bobeth}, \bibinfo{author}{Gudrun Hiller},
  \bibinfo{author}{Giorgi Piranishvili}, \bibinfo{title}{{CP Asymmetries in bar
  $B \to \bar{K}^* (\to \bar{K} \pi) \bar{\ell} \ell$ and Untagged $\bar{B}_s$,
  $B_s \to \phi (\to K^{+} K^-) \bar{\ell} \ell$ Decays at NLO}},
  \bibinfo{journal}{JHEP} \bibinfo{volume}{07} (\bibinfo{year}{2008})
  \bibinfo{pages}{106}, \bibinfo{doi}{\doi{10.1088/1126-6708/2008/07/106}}.
  \href{http://arxiv.org/abs/0805.2525}{\tt arXiv:0805.2525}.

\bibtype{Article}%
\bibitem{Altmannshofer:2008dz}
\bibinfo{author}{Wolfgang Altmannshofer}, \bibinfo{author}{Patricia Ball},
  \bibinfo{author}{Aoife Bharucha}, \bibinfo{author}{Andrzej~J. Buras},
  \bibinfo{author}{David~M. Straub}, \bibinfo{author}{Michael Wick},
  \bibinfo{title}{{Symmetries and Asymmetries of $B \to K^{*} \mu^{+} \mu^{-}$
  Decays in the Standard Model and Beyond}}, \bibinfo{journal}{JHEP}
  \bibinfo{volume}{01} (\bibinfo{year}{2009}) \bibinfo{pages}{019},
  \bibinfo{doi}{\doi{10.1088/1126-6708/2009/01/019}}.
  \href{http://arxiv.org/abs/0811.1214}{\tt arXiv:0811.1214}.

\bibtype{Article}%
\bibitem{Matias:2012xw}
\bibinfo{author}{Joaquim Matias}, \bibinfo{author}{Federico Mescia},
  \bibinfo{author}{Marc Ramon}, \bibinfo{author}{Javier Virto},
  \bibinfo{title}{{Complete Anatomy of $\bar{B}_d -> \bar{K}^{* 0} (-> K
  \pi)l^+l^-$ and its angular distribution}}, \bibinfo{journal}{JHEP}
  \bibinfo{volume}{04} (\bibinfo{year}{2012}) \bibinfo{pages}{104},
  \bibinfo{doi}{\doi{10.1007/JHEP04(2012)104}}.
  \href{http://arxiv.org/abs/1202.4266}{\tt arXiv:1202.4266}.

\bibtype{Article}%
\bibitem{Lee:2006gs}
\bibinfo{author}{Keith S.~M. Lee}, \bibinfo{author}{Zoltan Ligeti},
  \bibinfo{author}{Iain~W. Stewart}, \bibinfo{author}{Frank~J. Tackmann},
  \bibinfo{title}{{Extracting short distance information from $b \to s l^+ l^-$
  effectively}}, \bibinfo{journal}{Phys. Rev. D} \bibinfo{volume}{75}
  (\bibinfo{year}{2007}) \bibinfo{pages}{034016},
  \bibinfo{doi}{\doi{10.1103/PhysRevD.75.034016}}.
  \href{http://arxiv.org/abs/hep-ph/0612156}{\tt arXiv:hep-ph/0612156}.

\bibtype{Article}%
\bibitem{Huber:2020vup}
\bibinfo{author}{Tobias Huber}, \bibinfo{author}{Tobias Hurth},
  \bibinfo{author}{Jack Jenkins}, \bibinfo{author}{Enrico Lunghi},
  \bibinfo{author}{Qin Qin}, \bibinfo{author}{K.~Keri Vos},
  \bibinfo{title}{{Phenomenology of inclusive $ \overline{B}\to
  {X}_s{\mathrm{\ell}}^{+}{\mathrm{\ell}}^{-} $ for the Belle II era}},
  \bibinfo{journal}{JHEP} \bibinfo{volume}{10} (\bibinfo{year}{2020})
  \bibinfo{pages}{088}, \bibinfo{doi}{\doi{10.1007/JHEP10(2020)088}}.
  \href{http://arxiv.org/abs/2007.04191}{\tt arXiv:2007.04191}.

\bibtype{Article}%
\bibitem{Huber:2024rbw}
\bibinfo{author}{Tobias Huber}, \bibinfo{author}{Tobias Hurth},
  \bibinfo{author}{Jack Jenkins}, \bibinfo{author}{Enrico Lunghi},
  \bibinfo{author}{Qin Qin}, \bibinfo{author}{K.~Keri Vos},
  \bibinfo{title}{{Inclusive $ \overline{B}\to {X}_s{\ell}^{+}{\ell}^{-} $ at
  the LHC: theory predictions and new-physics reach}}, \bibinfo{journal}{JHEP}
  \bibinfo{volume}{11} (\bibinfo{year}{2024}) \bibinfo{pages}{130},
  \bibinfo{doi}{\doi{10.1007/JHEP11(2024)130}}.
  \href{http://arxiv.org/abs/2404.03517}{\tt arXiv:2404.03517}.

\bibtype{Article}%
\bibitem{Jager:2012uw}
\bibinfo{author}{S. J\"ager}, \bibinfo{author}{J. Martin~Camalich},
  \bibinfo{title}{{On $B \to V \ell \ell$ at small dilepton invariant mass,
  power corrections, and new physics}}, \bibinfo{journal}{JHEP}
  \bibinfo{volume}{05} (\bibinfo{year}{2013}) \bibinfo{pages}{043},
  \bibinfo{doi}{\doi{10.1007/JHEP05(2013)043}}.
  \href{http://arxiv.org/abs/1212.2263}{\tt arXiv:1212.2263}.

\bibtype{Article}%
\bibitem{Lyon:2014hpa}
\bibinfo{author}{James Lyon}, \bibinfo{author}{Roman Zwicky},
  \bibinfo{title}{{Resonances gone topsy turvy - the charm of QCD or new
  physics in $b \to s \ell^+ \ell^-$?}}  (\bibinfo{year}{2014}).
  \href{http://arxiv.org/abs/1406.0566}{\tt arXiv:1406.0566}.

\bibtype{Article}%
\bibitem{Gubernari:2020eft}
\bibinfo{author}{Nico Gubernari}, \bibinfo{author}{Danny van Dyk},
  \bibinfo{author}{Javier Virto}, \bibinfo{title}{{Non-local matrix elements in
  $B_{(s)}\to \{K^{(*)},\phi\}\ell^+\ell^-$}}, \bibinfo{journal}{JHEP}
  \bibinfo{volume}{02} (\bibinfo{year}{2021}) \bibinfo{pages}{088},
  \bibinfo{doi}{\doi{10.1007/JHEP02(2021)088}}.
  \href{http://arxiv.org/abs/2011.09813}{\tt arXiv:2011.09813}.

\bibtype{Article}%
\bibitem{Gubernari:2022hxn}
\bibinfo{author}{Nico Gubernari}, \bibinfo{author}{M\'eril Reboud},
  \bibinfo{author}{Danny van Dyk}, \bibinfo{author}{Javier Virto},
  \bibinfo{title}{{Improved theory predictions and global analysis of exclusive
  $b \to s\mu^+\mu^-$ processes}}, \bibinfo{journal}{JHEP} \bibinfo{volume}{09}
  (\bibinfo{year}{2022}) \bibinfo{pages}{133},
  \bibinfo{doi}{\doi{10.1007/JHEP09(2022)133}}.
  \href{http://arxiv.org/abs/2206.03797}{\tt arXiv:2206.03797}.

\bibtype{Article}%
\bibitem{Gubernari:2023puw}
\bibinfo{author}{Nico Gubernari}, \bibinfo{author}{M\'eril Reboud},
  \bibinfo{author}{Danny van Dyk}, \bibinfo{author}{Javier Virto},
  \bibinfo{title}{{Dispersive analysis of $B \to K^{(*)}$ and $B_{s} \to \phi$
  form factors}}, \bibinfo{journal}{JHEP} \bibinfo{volume}{12}
  (\bibinfo{year}{2023}) \bibinfo{pages}{153},
  \bibinfo{doi}{\doi{10.1007/JHEP12(2023)153}}, \bibinfo{note}{[Erratum: JHEP
  01, 125 (2025)]}. \href{http://arxiv.org/abs/2305.06301}{\tt
  arXiv:2305.06301}.

\bibtype{Article}%
\bibitem{Altmannshofer:2009ma}
\bibinfo{author}{Wolfgang Altmannshofer}, \bibinfo{author}{Andrzej~J. Buras},
  \bibinfo{author}{David~M. Straub}, \bibinfo{author}{Michael Wick},
  \bibinfo{title}{{New strategies for New Physics search in $B \to K^{*} \nu
  \bar{\nu}$, $B \to K \nu \bar{\nu}$ and $B \to X_{s} \nu \bar{\nu}$ decays}},
  \bibinfo{journal}{JHEP} \bibinfo{volume}{04} (\bibinfo{year}{2009})
  \bibinfo{pages}{022}, \bibinfo{doi}{\doi{10.1088/1126-6708/2009/04/022}}.
  \href{http://arxiv.org/abs/0902.0160}{\tt arXiv:0902.0160}.

\bibtype{Article}%
\bibitem{Buras:2014fpa}
\bibinfo{author}{Andrzej~J. Buras}, \bibinfo{author}{Jennifer Girrbach-Noe},
  \bibinfo{author}{Christoph Niehoff}, \bibinfo{author}{David~M. Straub},
  \bibinfo{title}{{$ B\to {K}^{\left(\ast \right)}\nu \overline{\nu} $ decays
  in the Standard Model and beyond}}, \bibinfo{journal}{JHEP}
  \bibinfo{volume}{02} (\bibinfo{year}{2015}) \bibinfo{pages}{184},
  \bibinfo{doi}{\doi{10.1007/JHEP02(2015)184}}.
  \href{http://arxiv.org/abs/1409.4557}{\tt arXiv:1409.4557}.

\bibtype{Article}%
\bibitem{Browder:2021hbl}
\bibinfo{author}{Thomas~E. Browder}, \bibinfo{author}{Nilendra~G. Deshpande},
  \bibinfo{author}{Rusa Mandal}, \bibinfo{author}{Rahul Sinha},
  \bibinfo{title}{{Impact of $B \to K \nu\bar\nu$ measurements on beyond the
  Standard Model theories}}, \bibinfo{journal}{Phys. Rev. D}
  \bibinfo{volume}{104} (\bibinfo{number}{5}) (\bibinfo{year}{2021})
  \bibinfo{pages}{053007}, \bibinfo{doi}{\doi{10.1103/PhysRevD.104.053007}}.
  \href{http://arxiv.org/abs/2107.01080}{\tt arXiv:2107.01080}.

\bibtype{Article}%
\bibitem{Bause:2021cna}
\bibinfo{author}{Rigo Bause}, \bibinfo{author}{Hector Gisbert},
  \bibinfo{author}{Marcel Golz}, \bibinfo{author}{Gudrun Hiller},
  \bibinfo{title}{{Interplay of dineutrino modes with semileptonic rare
  B-decays}}, \bibinfo{journal}{JHEP} \bibinfo{volume}{12}
  (\bibinfo{year}{2021}) \bibinfo{pages}{061},
  \bibinfo{doi}{\doi{10.1007/JHEP12(2021)061}}.
  \href{http://arxiv.org/abs/2109.01675}{\tt arXiv:2109.01675}.

\bibtype{Article}%
\bibitem{Becirevic:2023aov}
\bibinfo{author}{Damir Be\v{c}irevi\'c}, \bibinfo{author}{Gioacchino Piazza},
  \bibinfo{author}{Olcyr Sumensari}, \bibinfo{title}{{Revisiting $B\rightarrow
  K^{(*)} \nu {\bar{\nu }}$ decays in the Standard Model and beyond}},
  \bibinfo{journal}{Eur. Phys. J. C} \bibinfo{volume}{83} (\bibinfo{number}{3})
  (\bibinfo{year}{2023}) \bibinfo{pages}{252},
  \bibinfo{doi}{\doi{10.1140/epjc/s10052-023-11388-z}}.
  \href{http://arxiv.org/abs/2301.06990}{\tt arXiv:2301.06990}.

\bibtype{Article}%
\bibitem{Buras:2024ewl}
\bibinfo{author}{Andrzej~J. Buras}, \bibinfo{author}{Julia Harz},
  \bibinfo{author}{Martin~A. Mojahed}, \bibinfo{title}{{Disentangling new
  physics in $ K\to \pi \nu \overline{\nu} $ and $ B\to
  K\left({K}^{\ast}\right)\nu \overline{\nu} $ observables}},
  \bibinfo{journal}{JHEP} \bibinfo{volume}{10} (\bibinfo{year}{2024})
  \bibinfo{pages}{087}, \bibinfo{doi}{\doi{10.1007/JHEP10(2024)087}}.
  \href{http://arxiv.org/abs/2405.06742}{\tt arXiv:2405.06742}.

\bibtype{Article}%
\bibitem{Gisbert:2024kob}
\bibinfo{author}{Hector Gisbert}, \bibinfo{author}{Gudrun Hiller},
  \bibinfo{author}{Dominik Suelmann}, \bibinfo{title}{{Effective field theory
  analysis of rare $|\Delta c| = |\Delta u| = 1$ charm decays}},
  \bibinfo{journal}{JHEP} \bibinfo{volume}{12} (\bibinfo{year}{2024})
  \bibinfo{pages}{102}, \bibinfo{doi}{\doi{10.1007/JHEP12(2024)102}}.
  \href{http://arxiv.org/abs/2410.00115}{\tt arXiv:2410.00115}.

\bibtype{Article}%
\bibitem{Bause:2019vpr}
\bibinfo{author}{Rigo Bause}, \bibinfo{author}{Marcel Golz},
  \bibinfo{author}{Gudrun Hiller}, \bibinfo{author}{Andrey Tayduganov},
  \bibinfo{title}{{The new physics reach of null tests with $D \rightarrow \pi
  \ell \ell $ and $D_s \rightarrow K \ell \ell $ decays}},
  \bibinfo{journal}{Eur. Phys. J. C} \bibinfo{volume}{80} (\bibinfo{number}{1})
  (\bibinfo{year}{2020}) \bibinfo{pages}{65},
  \bibinfo{doi}{\doi{10.1140/epjc/s10052-020-7621-7}}, \bibinfo{note}{[Erratum:
  Eur.Phys.J.C 81, 219 (2021)]}. \href{http://arxiv.org/abs/1909.11108}{\tt
  arXiv:1909.11108}.

\bibtype{Article}%
\bibitem{Bause:2020xzj}
\bibinfo{author}{Rigo Bause}, \bibinfo{author}{Hector Gisbert},
  \bibinfo{author}{Marcel Golz}, \bibinfo{author}{Gudrun Hiller},
  \bibinfo{title}{{Rare charm $c\to u\,\nu\bar{\nu}$ dineutrino null tests for
  $e^+e^-$ machines}}, \bibinfo{journal}{Phys. Rev. D} \bibinfo{volume}{103}
  (\bibinfo{number}{1}) (\bibinfo{year}{2021}) \bibinfo{pages}{015033},
  \bibinfo{doi}{\doi{10.1103/PhysRevD.103.015033}}.
  \href{http://arxiv.org/abs/2010.02225}{\tt arXiv:2010.02225}.

\bibtype{Article}%
\bibitem{Fuentes-Martin:2020lea}
\bibinfo{author}{Javier Fuentes-Martin}, \bibinfo{author}{Admir Greljo},
  \bibinfo{author}{Jorge Martin~Camalich}, \bibinfo{author}{Jos\'e~David
  Ruiz-Alvarez}, \bibinfo{title}{{Charm physics confronts high-p$_{T}$ lepton
  tails}}, \bibinfo{journal}{JHEP} \bibinfo{volume}{11} (\bibinfo{year}{2020})
  \bibinfo{pages}{080}, \bibinfo{doi}{\doi{10.1007/JHEP11(2020)080}}.
  \href{http://arxiv.org/abs/2003.12421}{\tt arXiv:2003.12421}.

\bibtype{Article}%
\bibitem{Ciuchini:2022wbq}
\bibinfo{author}{Marco Ciuchini}, \bibinfo{author}{Marco Fedele},
  \bibinfo{author}{Enrico Franco}, \bibinfo{author}{Ayan Paul},
  \bibinfo{author}{Luca Silvestrini}, \bibinfo{author}{Mauro Valli},
  \bibinfo{title}{{Constraints on lepton universality violation from rare B
  decays}}, \bibinfo{journal}{Phys. Rev. D} \bibinfo{volume}{107}
  (\bibinfo{number}{5}) (\bibinfo{year}{2023}) \bibinfo{pages}{055036},
  \bibinfo{doi}{\doi{10.1103/PhysRevD.107.055036}}.
  \href{http://arxiv.org/abs/2212.10516}{\tt arXiv:2212.10516}.

\bibtype{Article}%
\bibitem{Greljo:2022jac}
\bibinfo{author}{Admir Greljo}, \bibinfo{author}{Jakub Salko},
  \bibinfo{author}{Aleks Smolkovi\v{c}}, \bibinfo{author}{Peter Stangl},
  \bibinfo{title}{{Rare b decays meet high-mass Drell-Yan}},
  \bibinfo{journal}{JHEP} \bibinfo{volume}{05} (\bibinfo{year}{2023})
  \bibinfo{pages}{087}, \bibinfo{doi}{\doi{10.1007/JHEP05(2023)087}}.
  \href{http://arxiv.org/abs/2212.10497}{\tt arXiv:2212.10497}.

\bibtype{Article}%
\bibitem{Alguero:2023jeh}
\bibinfo{author}{Marcel Alguer\'o}, \bibinfo{author}{Aritra Biswas},
  \bibinfo{author}{Bernat Capdevila}, \bibinfo{author}{S\'ebastien
  Descotes-Genon}, \bibinfo{author}{Joaquim Matias},
  \bibinfo{author}{Mart\'\i{}n Novoa-Brunet}, \bibinfo{title}{{To (b)e or not
  to (b)e: no electrons at LHCb}}, \bibinfo{journal}{Eur. Phys. J. C}
  \bibinfo{volume}{83} (\bibinfo{number}{7}) (\bibinfo{year}{2023})
  \bibinfo{pages}{648}, \bibinfo{doi}{\doi{10.1140/epjc/s10052-023-11824-0}}.
  \href{http://arxiv.org/abs/2304.07330}{\tt arXiv:2304.07330}.

\bibtype{Article}%
\bibitem{Altmannshofer:2023uci}
\bibinfo{author}{Wolfgang Altmannshofer}, \bibinfo{author}{Sri~Aditya Gadam},
  \bibinfo{author}{Stefano Profumo}, \bibinfo{title}{{Probing new physics with
  $\mu^+ \mu^- \to b s$ at a muon collider}}, \bibinfo{journal}{Phys. Rev. D}
  \bibinfo{volume}{108} (\bibinfo{number}{11}) (\bibinfo{year}{2023})
  \bibinfo{pages}{115033}, \bibinfo{doi}{\doi{10.1103/PhysRevD.108.115033}}.
  \href{http://arxiv.org/abs/2306.15017}{\tt arXiv:2306.15017}.

\bibtype{Article}%
\bibitem{Guadagnoli:2023ddc}
\bibinfo{author}{Diego Guadagnoli}, \bibinfo{author}{Camille Normand},
  \bibinfo{author}{Silvano Simula}, \bibinfo{author}{Ludovico Vittorio},
  \bibinfo{title}{{Insights on the current semi-leptonic B-decay discrepancies
  and how $B_{s} \to \mu^+\mu^- \gamma$ can help}}, \bibinfo{journal}{JHEP}
  \bibinfo{volume}{10} (\bibinfo{year}{2023}) \bibinfo{pages}{102},
  \bibinfo{doi}{\doi{10.1007/JHEP10(2023)102}}.
  \href{http://arxiv.org/abs/2308.00034}{\tt arXiv:2308.00034}.

\bibtype{Article}%
\bibitem{Hurth:2023jwr}
\bibinfo{author}{T. Hurth}, \bibinfo{author}{F. Mahmoudi}, \bibinfo{author}{S.
  Neshatpour}, \bibinfo{title}{{$B$ anomalies in the post $R_{K^{(*)}}$ era}},
  \bibinfo{journal}{Phys. Rev. D} \bibinfo{volume}{108} (\bibinfo{number}{11})
  (\bibinfo{year}{2023}) \bibinfo{pages}{115037},
  \bibinfo{doi}{\doi{10.1103/PhysRevD.108.115037}}.
  \href{http://arxiv.org/abs/2310.05585}{\tt arXiv:2310.05585}.

\bibtype{Article}%
\bibitem{Bordone:2024hui}
\bibinfo{author}{Marzia Bordone}, \bibinfo{author}{Gino isidori},
  \bibinfo{author}{Sandro M\"achler}, \bibinfo{author}{Arianna Tinari},
  \bibinfo{title}{{Short- vs. long-distance physics in $B\to K^{(*)}
  \ell^+\ell^-$: a data-driven analysis}}, \bibinfo{journal}{Eur. Phys. J. C}
  \bibinfo{volume}{84} (\bibinfo{number}{5}) (\bibinfo{year}{2024})
  \bibinfo{pages}{547}, \bibinfo{doi}{\doi{10.1140/epjc/s10052-024-12869-5}}.
  \href{http://arxiv.org/abs/2401.18007}{\tt arXiv:2401.18007}.

\bibtype{Article}%
\bibitem{LHCb:2013ghj}
\bibinfo{author}{R Aaij}, et al. (\bibinfo{collaboration}{LHCb}),
  \bibinfo{title}{{Measurement of Form-Factor-Independent Observables in the
  Decay $B^{0} \to K^{*0} \mu^+ \mu^-$}}, \bibinfo{journal}{Phys. Rev. Lett.}
  \bibinfo{volume}{111} (\bibinfo{year}{2013}) \bibinfo{pages}{191801},
  \bibinfo{doi}{\doi{10.1103/PhysRevLett.111.191801}}.
  \href{http://arxiv.org/abs/1308.1707}{\tt arXiv:1308.1707}.

\bibtype{Article}%
\bibitem{LHCb:2020lmf}
\bibinfo{author}{Roel Aaij}, et al. (\bibinfo{collaboration}{LHCb}),
  \bibinfo{title}{{Measurement of $CP$-Averaged Observables in the
  $B^{0}\rightarrow K^{*0}\mu^{+}\mu^{-}$ Decay}}, \bibinfo{journal}{Phys. Rev.
  Lett.} \bibinfo{volume}{125} (\bibinfo{number}{1}) (\bibinfo{year}{2020})
  \bibinfo{pages}{011802}, \bibinfo{doi}{\doi{10.1103/PhysRevLett.125.011802}}.
  \href{http://arxiv.org/abs/2003.04831}{\tt arXiv:2003.04831}.

\bibtype{Article}%
\bibitem{CMS:2024atz}
\bibinfo{author}{Aram Hayrapetyan}, et al. (\bibinfo{collaboration}{CMS}),
  \bibinfo{title}{{Angular analysis of the $B^0 \to K(892)^{*\,0} \mu^+ \mu^-$
  decay in proton-proton collisions at $\sqrt{s} =13$ TeV}},
  \bibinfo{journal}{Phys. Lett. B} \bibinfo{volume}{864} (\bibinfo{year}{2025})
  \bibinfo{pages}{139406}, \bibinfo{doi}{\doi{10.1016/j.physletb.2025.139406}}.
  \href{http://arxiv.org/abs/2411.11820}{\tt arXiv:2411.11820}.

\bibtype{Article}%
\bibitem{LHCb:2014cxe}
\bibinfo{author}{R. Aaij}, et al. (\bibinfo{collaboration}{LHCb}),
  \bibinfo{title}{{Differential branching fractions and isospin asymmetries of
  $B \to K^{(*)} \mu^+ \mu^-$ decays}}, \bibinfo{journal}{JHEP}
  \bibinfo{volume}{06} (\bibinfo{year}{2014}) \bibinfo{pages}{133},
  \bibinfo{doi}{\doi{10.1007/JHEP06(2014)133}}.
  \href{http://arxiv.org/abs/1403.8044}{\tt arXiv:1403.8044}.

\bibtype{Article}%
\bibitem{LHCb:2021zwz}
\bibinfo{author}{Roel Aaij}, et al. (\bibinfo{collaboration}{LHCb}),
  \bibinfo{title}{{Branching Fraction Measurements of the Rare
  $B^0_s\rightarrow\phi\mu^+\mu^-$ and $B^0_s\rightarrow
  f_2^\prime(1525)\mu^+\mu^-$- Decays}}, \bibinfo{journal}{Phys. Rev. Lett.}
  \bibinfo{volume}{127} (\bibinfo{number}{15}) (\bibinfo{year}{2021})
  \bibinfo{pages}{151801}, \bibinfo{doi}{\doi{10.1103/PhysRevLett.127.151801}}.
  \href{http://arxiv.org/abs/2105.14007}{\tt arXiv:2105.14007}.

\bibtype{Article}%
\bibitem{BaBar:2012obs}
\bibinfo{author}{J.~P. Lees}, et al. (\bibinfo{collaboration}{BaBar}),
  \bibinfo{title}{{Evidence for an excess of $\bar{B} \to D^{(*)}
  \tau^-\bar{\nu}_\tau$ decays}}, \bibinfo{journal}{Phys. Rev. Lett.}
  \bibinfo{volume}{109} (\bibinfo{year}{2012}) \bibinfo{pages}{101802},
  \bibinfo{doi}{\doi{10.1103/PhysRevLett.109.101802}}.
  \href{http://arxiv.org/abs/1205.5442}{\tt arXiv:1205.5442}.

\bibtype{Article}%
\bibitem{BaBar:2013mob}
\bibinfo{author}{J.~P. Lees}, et al. (\bibinfo{collaboration}{BaBar}),
  \bibinfo{title}{{Measurement of an Excess of $\bar{B} \to D^{(*)}\tau^-
  \bar{\nu}_\tau$ Decays and Implications for Charged Higgs Bosons}},
  \bibinfo{journal}{Phys. Rev. D} \bibinfo{volume}{88} (\bibinfo{number}{7})
  (\bibinfo{year}{2013}) \bibinfo{pages}{072012},
  \bibinfo{doi}{\doi{10.1103/PhysRevD.88.072012}}.
  \href{http://arxiv.org/abs/1303.0571}{\tt arXiv:1303.0571}.

\bibtype{Article}%
\bibitem{Belle:2015qfa}
\bibinfo{author}{M. Huschle}, et al. (\bibinfo{collaboration}{Belle}),
  \bibinfo{title}{{Measurement of the branching ratio of $\bar{B} \to
  D^{(\ast)} \tau^- \bar{\nu}_\tau$ relative to $\bar{B} \to D^{(\ast)} \ell^-
  \bar{\nu}_\ell$ decays with hadronic tagging at Belle}},
  \bibinfo{journal}{Phys. Rev. D} \bibinfo{volume}{92} (\bibinfo{number}{7})
  (\bibinfo{year}{2015}) \bibinfo{pages}{072014},
  \bibinfo{doi}{\doi{10.1103/PhysRevD.92.072014}}.
  \href{http://arxiv.org/abs/1507.03233}{\tt arXiv:1507.03233}.

\bibtype{Article}%
\bibitem{Belle:2016dyj}
\bibinfo{author}{S. Hirose}, et al. (\bibinfo{collaboration}{Belle}),
  \bibinfo{title}{{Measurement of the $\tau$ lepton polarization and $R(D^*)$
  in the decay $\bar{B} \to D^* \tau^- \bar{\nu}_\tau$}},
  \bibinfo{journal}{Phys. Rev. Lett.} \bibinfo{volume}{118}
  (\bibinfo{number}{21}) (\bibinfo{year}{2017}) \bibinfo{pages}{211801},
  \bibinfo{doi}{\doi{10.1103/PhysRevLett.118.211801}}.
  \href{http://arxiv.org/abs/1612.00529}{\tt arXiv:1612.00529}.

\bibtype{Article}%
\bibitem{Belle:2017ilt}
\bibinfo{author}{S. Hirose}, et al. (\bibinfo{collaboration}{Belle}),
  \bibinfo{title}{{Measurement of the $\tau$ lepton polarization and $R(D^*)$
  in the decay $\bar{B} \rightarrow D^* \tau^- \bar{\nu}_\tau$ with one-prong
  hadronic $\tau$ decays at Belle}}, \bibinfo{journal}{Phys. Rev. D}
  \bibinfo{volume}{97} (\bibinfo{number}{1}) (\bibinfo{year}{2018})
  \bibinfo{pages}{012004}, \bibinfo{doi}{\doi{10.1103/PhysRevD.97.012004}}.
  \href{http://arxiv.org/abs/1709.00129}{\tt arXiv:1709.00129}.

\bibtype{Article}%
\bibitem{Belle:2019rba}
\bibinfo{author}{G. Caria}, et al. (\bibinfo{collaboration}{Belle}),
  \bibinfo{title}{{Measurement of $\mathcal{R}(D)$ and $\mathcal{R}(D^*)$ with
  a semileptonic tagging method}}, \bibinfo{journal}{Phys. Rev. Lett.}
  \bibinfo{volume}{124} (\bibinfo{number}{16}) (\bibinfo{year}{2020})
  \bibinfo{pages}{161803}, \bibinfo{doi}{\doi{10.1103/PhysRevLett.124.161803}}.
  \href{http://arxiv.org/abs/1910.05864}{\tt arXiv:1910.05864}.

\bibtype{Article}%
\bibitem{LHCb:2023zxo}
\bibinfo{author}{Roel Aaij}, et al. (\bibinfo{collaboration}{LHCb}),
  \bibinfo{title}{{Measurement of the ratios of branching fractions
  $\mathcal{R}(D^{*})$ and $\mathcal{R}(D^{0})$}}, \bibinfo{journal}{Phys. Rev.
  Lett.} \bibinfo{volume}{131} (\bibinfo{year}{2023}) \bibinfo{pages}{111802},
  \bibinfo{doi}{\doi{10.1103/PhysRevLett.131.111802}}.
  \href{http://arxiv.org/abs/2302.02886}{\tt arXiv:2302.02886}.

\bibtype{Article}%
\bibitem{LHCb:2023uiv}
\bibinfo{author}{Roel Aaij}, et al. (\bibinfo{collaboration}{LHCb}),
  \bibinfo{title}{{Test of lepton flavor universality using $B^0 \to D^{*-}
  \tau^+ \nu_\tau$ decays with hadronic $\tau$ channels}},
  \bibinfo{journal}{Phys. Rev. D} \bibinfo{volume}{108} (\bibinfo{number}{1})
  (\bibinfo{year}{2023}) \bibinfo{pages}{012018},
  \bibinfo{doi}{\doi{10.1103/PhysRevD.108.012018}}, \bibinfo{note}{[Erratum:
  Phys.Rev.D 109, 119902 (2024)]}. \href{http://arxiv.org/abs/2305.01463}{\tt
  arXiv:2305.01463}.

\bibtype{Article}%
\bibitem{Belle-II:2024ami}
\bibinfo{author}{I. Adachi}, et al. (\bibinfo{collaboration}{Belle-II}),
  \bibinfo{title}{{Test of lepton flavor universality with a measurement of
  R(D*) using hadronic B tagging at the Belle II experiment}},
  \bibinfo{journal}{Phys. Rev. D} \bibinfo{volume}{110} (\bibinfo{number}{7})
  (\bibinfo{year}{2024}) \bibinfo{pages}{072020},
  \bibinfo{doi}{\doi{10.1103/PhysRevD.110.072020}}.
  \href{http://arxiv.org/abs/2401.02840}{\tt arXiv:2401.02840}.

\bibtype{Article}%
\bibitem{Belle-II:2025yjp}
\bibinfo{author}{I. Adachi}, et al. (\bibinfo{collaboration}{Belle-II}),
  \bibinfo{title}{{Test of lepton flavor universality with measurements of
  $R(D^{+})$ and $R(D^{*+})$ using semileptonic $B$ tagging at the Belle II
  experiment}}  (\bibinfo{year}{2025}).
  \href{http://arxiv.org/abs/2504.11220}{\tt arXiv:2504.11220}.

\bibtype{Article}%
\bibitem{Bryman:2021teu}
\bibinfo{author}{Douglas Bryman}, \bibinfo{author}{Vincenzo Cirigliano},
  \bibinfo{author}{Andreas Crivellin}, \bibinfo{author}{Gianluca Inguglia},
  \bibinfo{title}{{Testing Lepton Flavor Universality with Pion, Kaon, Tau, and
  Beta Decays}}, \bibinfo{journal}{Ann. Rev. Nucl. Part. Sci.}
  \bibinfo{volume}{72} (\bibinfo{year}{2022}) \bibinfo{pages}{69--91},
  \bibinfo{doi}{\doi{10.1146/annurev-nucl-110121-051223}}.
  \href{http://arxiv.org/abs/2111.05338}{\tt arXiv:2111.05338}.

\bibtype{Article}%
\bibitem{NA62:2012lny}
\bibinfo{author}{C. Lazzeroni}, et al. (\bibinfo{collaboration}{NA62}),
  \bibinfo{title}{{Precision Measurement of the Ratio of the Charged Kaon
  Leptonic Decay Rates}}, \bibinfo{journal}{Phys. Lett. B}
  \bibinfo{volume}{719} (\bibinfo{year}{2013}) \bibinfo{pages}{326--336},
  \bibinfo{doi}{\doi{10.1016/j.physletb.2013.01.037}}.
  \href{http://arxiv.org/abs/1212.4012}{\tt arXiv:1212.4012}.

\bibtype{Article}%
\bibitem{PiENu:2015seu}
\bibinfo{author}{A. Aguilar-Arevalo}, et al. (\bibinfo{collaboration}{PiENu}),
  \bibinfo{title}{{Improved Measurement of the $\pi \to \textrm{e} \nu$
  Branching Ratio}}, \bibinfo{journal}{Phys. Rev. Lett.} \bibinfo{volume}{115}
  (\bibinfo{number}{7}) (\bibinfo{year}{2015}) \bibinfo{pages}{071801},
  \bibinfo{doi}{\doi{10.1103/PhysRevLett.115.071801}}.
  \href{http://arxiv.org/abs/1506.05845}{\tt arXiv:1506.05845}.

\bibtype{Article}%
\bibitem{Aebischer:2025qhh}
\bibinfo{author}{Jason Aebischer}, \bibinfo{author}{Andrzej~J. Buras},
  \bibinfo{author}{Jacky Kumar}, \bibinfo{title}{{SMEFT ATLAS: The Landscape
  Beyond the Standard Model}}  (\bibinfo{year}{2025}).
  \href{http://arxiv.org/abs/2507.05926}{\tt arXiv:2507.05926}.

\bibtype{Article}%
\bibitem{deBlas:2025xhe}
\bibinfo{author}{J. de Blas}, \bibinfo{author}{A. Goncalves},
  \bibinfo{author}{V. Miralles}, \bibinfo{author}{L. Reina},
  \bibinfo{author}{L. Silvestrini}, \bibinfo{author}{M. Valli},
  \bibinfo{title}{{Constraining new physics effective interactions via a global
  fit of electroweak, Drell-Yan, Higgs, top, and flavour observables}}
  (\bibinfo{year}{2025}). \href{http://arxiv.org/abs/2507.06191}{\tt
  arXiv:2507.06191}.

\bibtype{Article}%
\bibitem{MartinCamalich:2020dfe}
\bibinfo{author}{Jorge Martin~Camalich}, \bibinfo{author}{Maxim Pospelov},
  \bibinfo{author}{Pham Ngoc~Hoa Vuong}, \bibinfo{author}{Robert Ziegler},
  \bibinfo{author}{Jure Zupan}, \bibinfo{title}{{Quark Flavor Phenomenology of
  the QCD Axion}}, \bibinfo{journal}{Phys. Rev. D} \bibinfo{volume}{102}
  (\bibinfo{number}{1}) (\bibinfo{year}{2020}) \bibinfo{pages}{015023},
  \bibinfo{doi}{\doi{10.1103/PhysRevD.102.015023}}.
  \href{http://arxiv.org/abs/2002.04623}{\tt arXiv:2002.04623}.

\bibtype{Article}%
\bibitem{Belle-II:2023esi}
\bibinfo{author}{I. Adachi}, et al. (\bibinfo{collaboration}{Belle-II}),
  \bibinfo{title}{{Evidence for $B^+ \to K^+ \nu\bar\nu$ decays}},
  \bibinfo{journal}{Phys. Rev. D} \bibinfo{volume}{109} (\bibinfo{number}{11})
  (\bibinfo{year}{2024}) \bibinfo{pages}{112006},
  \bibinfo{doi}{\doi{10.1103/PhysRevD.109.112006}}.
  \href{http://arxiv.org/abs/2311.14647}{\tt arXiv:2311.14647}.

\bibtype{Article}%
\bibitem{Perrevoort:2018ttp}
\bibinfo{author}{Ann-Kathrin Perrevoort} (\bibinfo{collaboration}{Mu3e}),
  \bibinfo{title}{{The Rare and Forbidden: Testing Physics Beyond the Standard
  Model with Mu3e}}, \bibinfo{journal}{SciPost Phys. Proc.} \bibinfo{volume}{1}
  (\bibinfo{year}{2019}) \bibinfo{pages}{052},
  \bibinfo{doi}{\doi{10.21468/SciPostPhysProc.1.052}}.
  \href{http://arxiv.org/abs/1812.00741}{\tt arXiv:1812.00741}.

\bibtype{Article}%
\bibitem{Calibbi:2020jvd}
\bibinfo{author}{Lorenzo Calibbi}, \bibinfo{author}{Diego Redigolo},
  \bibinfo{author}{Robert Ziegler}, \bibinfo{author}{Jure Zupan},
  \bibinfo{title}{{Looking forward to lepton-flavor-violating ALPs}},
  \bibinfo{journal}{JHEP} \bibinfo{volume}{09} (\bibinfo{year}{2021})
  \bibinfo{pages}{173}, \bibinfo{doi}{\doi{10.1007/JHEP09(2021)173}}.
  \href{http://arxiv.org/abs/2006.04795}{\tt arXiv:2006.04795}.

\bibtype{Article}%
\bibitem{Sala:2017ihs}
\bibinfo{author}{Filippo Sala}, \bibinfo{author}{David~M. Straub},
  \bibinfo{title}{{A New Light Particle in B Decays?}}, \bibinfo{journal}{Phys.
  Lett. B} \bibinfo{volume}{774} (\bibinfo{year}{2017})
  \bibinfo{pages}{205--209},
  \bibinfo{doi}{\doi{10.1016/j.physletb.2017.09.072}}.
  \href{http://arxiv.org/abs/1704.06188}{\tt arXiv:1704.06188}.

\bibtype{Article}%
\bibitem{Datta:2017ezo}
\bibinfo{author}{Alakabha Datta}, \bibinfo{author}{Jacky Kumar},
  \bibinfo{author}{Jiajun Liao}, \bibinfo{author}{Danny Marfatia},
  \bibinfo{title}{{New light mediators for the $R_K$ and $R_{K^*}$ puzzles}},
  \bibinfo{journal}{Phys. Rev. D} \bibinfo{volume}{97} (\bibinfo{number}{11})
  (\bibinfo{year}{2018}) \bibinfo{pages}{115038},
  \bibinfo{doi}{\doi{10.1103/PhysRevD.97.115038}}.
  \href{http://arxiv.org/abs/1705.08423}{\tt arXiv:1705.08423}.

\bibtype{Article}%
\bibitem{Altmannshofer:2017bsz}
\bibinfo{author}{Wolfgang Altmannshofer}, \bibinfo{author}{Michael~J. Baker},
  \bibinfo{author}{Stefania Gori}, \bibinfo{author}{Roni Harnik},
  \bibinfo{author}{Maxim Pospelov}, \bibinfo{author}{Emmanuel Stamou},
  \bibinfo{author}{Andrea Thamm}, \bibinfo{title}{{Light resonances and the
  low-q$^{2}$ bin of $ {R}_{K^{*}} $}}, \bibinfo{journal}{JHEP}
  \bibinfo{volume}{03} (\bibinfo{year}{2018}) \bibinfo{pages}{188},
  \bibinfo{doi}{\doi{10.1007/JHEP03(2018)188}}.
  \href{http://arxiv.org/abs/1711.07494}{\tt arXiv:1711.07494}.

\bibtype{Article}%
\bibitem{Crivellin:2022obd}
\bibinfo{author}{Andreas Crivellin}, \bibinfo{author}{Claudio~Andrea Manzari},
  \bibinfo{author}{Wolfgang Altmannshofer}, \bibinfo{author}{Gianluca
  Inguglia}, \bibinfo{author}{Paul Feichtinger}, \bibinfo{author}{Jorge
  Martin~Camalich}, \bibinfo{title}{{Toward excluding a light Z' explanation of
  $b \to s \ell^+ \ell^-$}}, \bibinfo{journal}{Phys. Rev. D}
  \bibinfo{volume}{106} (\bibinfo{number}{3}) (\bibinfo{year}{2022})
  \bibinfo{pages}{L031703}, \bibinfo{doi}{\doi{10.1103/PhysRevD.106.L031703}}.
  \href{http://arxiv.org/abs/2202.12900}{\tt arXiv:2202.12900}.

\bibtype{Article}%
\bibitem{Davidson:1981zd}
\bibinfo{author}{Aharon Davidson}, \bibinfo{author}{Kameshwar~C. Wali},
  \bibinfo{title}{{MINIMAL FLAVOR UNIFICATION VIA MULTIGENERATIONAL
  PECCEI-QUINN SYMMETRY}}, \bibinfo{journal}{Phys. Rev. Lett.}
  \bibinfo{volume}{48} (\bibinfo{year}{1982}) \bibinfo{pages}{11},
  \bibinfo{doi}{\doi{10.1103/PhysRevLett.48.11}}.

\bibtype{Article}%
\bibitem{Wilczek:1982rv}
\bibinfo{author}{Frank Wilczek}, \bibinfo{title}{{Axions and Family Symmetry
  Breaking}}, \bibinfo{journal}{Phys. Rev. Lett.} \bibinfo{volume}{49}
  (\bibinfo{year}{1982}) \bibinfo{pages}{1549--1552},
  \bibinfo{doi}{\doi{10.1103/PhysRevLett.49.1549}}.

\bibtype{Article}%
\bibitem{Calibbi:2016hwq}
\bibinfo{author}{Lorenzo Calibbi}, \bibinfo{author}{Florian Goertz},
  \bibinfo{author}{Diego Redigolo}, \bibinfo{author}{Robert Ziegler},
  \bibinfo{author}{Jure Zupan}, \bibinfo{title}{{Minimal axion model from
  flavor}}, \bibinfo{journal}{Phys. Rev. D} \bibinfo{volume}{95}
  (\bibinfo{number}{9}) (\bibinfo{year}{2017}) \bibinfo{pages}{095009},
  \bibinfo{doi}{\doi{10.1103/PhysRevD.95.095009}}.
  \href{http://arxiv.org/abs/1612.08040}{\tt arXiv:1612.08040}.

\bibtype{Article}%
\bibitem{Ema:2016ops}
\bibinfo{author}{Yohei Ema}, \bibinfo{author}{Koichi Hamaguchi},
  \bibinfo{author}{Takeo Moroi}, \bibinfo{author}{Kazunori Nakayama},
  \bibinfo{title}{{Flaxion: a minimal extension to solve puzzles in the
  standard model}}, \bibinfo{journal}{JHEP} \bibinfo{volume}{01}
  (\bibinfo{year}{2017}) \bibinfo{pages}{096},
  \bibinfo{doi}{\doi{10.1007/JHEP01(2017)096}}.
  \href{http://arxiv.org/abs/1612.05492}{\tt arXiv:1612.05492}.

\bibtype{Article}%
\bibitem{Arias-Aragon:2017eww}
\bibinfo{author}{F. Arias-Aragon}, \bibinfo{author}{L. Merlo},
  \bibinfo{title}{{The Minimal Flavour Violating Axion}},
  \bibinfo{journal}{JHEP} \bibinfo{volume}{10} (\bibinfo{year}{2017})
  \bibinfo{pages}{168}, \bibinfo{doi}{\doi{10.1007/JHEP10(2017)168}},
  \bibinfo{note}{[Erratum: JHEP 11, 152 (2019)]}.
  \href{http://arxiv.org/abs/1709.07039}{\tt arXiv:1709.07039}.

\bibtype{Article}%
\bibitem{Bjorkeroth:2018ipq}
\bibinfo{author}{Fredrik Bj\"orkeroth}, \bibinfo{author}{Luca Di~Luzio},
  \bibinfo{author}{Federico Mescia}, \bibinfo{author}{Enrico Nardi},
  \bibinfo{title}{{$U(1)$ flavour symmetries as Peccei-Quinn symmetries}},
  \bibinfo{journal}{JHEP} \bibinfo{volume}{02} (\bibinfo{year}{2019})
  \bibinfo{pages}{133}, \bibinfo{doi}{\doi{10.1007/JHEP02(2019)133}}.
  \href{http://arxiv.org/abs/1811.09637}{\tt arXiv:1811.09637}.

\bibtype{Article}%
\bibitem{Bonnefoy:2019lsn}
\bibinfo{author}{Q. Bonnefoy}, \bibinfo{author}{E. Dudas}, \bibinfo{author}{S.
  Pokorski}, \bibinfo{title}{{Chiral Froggatt-Nielsen models, gauge anomalies
  and flavourful axions}}, \bibinfo{journal}{JHEP} \bibinfo{volume}{01}
  (\bibinfo{year}{2020}) \bibinfo{pages}{191},
  \bibinfo{doi}{\doi{10.1007/JHEP01(2020)191}}.
  \href{http://arxiv.org/abs/1909.05336}{\tt arXiv:1909.05336}.

\bibtype{Article}%
\bibitem{DiLuzio:2023ndz}
\bibinfo{author}{Luca Di~Luzio}, \bibinfo{author}{Alfredo Walter~Mario
  Guerrera}, \bibinfo{author}{Xavier~Ponce D\'\i{}az}, \bibinfo{author}{Stefano
  Rigolin}, \bibinfo{title}{{On the IR/UV flavour connection in non-universal
  axion models}}, \bibinfo{journal}{JHEP} \bibinfo{volume}{06}
  (\bibinfo{year}{2023}) \bibinfo{pages}{046},
  \bibinfo{doi}{\doi{10.1007/JHEP06(2023)046}}.
  \href{http://arxiv.org/abs/2304.04643}{\tt arXiv:2304.04643}.

\bibtype{Article}%
\bibitem{Chivukula:1987py}
\bibinfo{author}{R.~Sekhar Chivukula}, \bibinfo{author}{Howard Georgi},
  \bibinfo{title}{{Composite Technicolor Standard Model}},
  \bibinfo{journal}{Phys. Lett. B} \bibinfo{volume}{188} (\bibinfo{year}{1987})
  \bibinfo{pages}{99--104}, \bibinfo{doi}{\doi{10.1016/0370-2693(87)90713-1}}.

\bibtype{Article}%
\bibitem{Hall:1990ac}
\bibinfo{author}{L.~J. Hall}, \bibinfo{author}{Lisa Randall},
  \bibinfo{title}{{Weak scale effective supersymmetry}},
  \bibinfo{journal}{Phys. Rev. Lett.} \bibinfo{volume}{65}
  (\bibinfo{year}{1990}) \bibinfo{pages}{2939--2942},
  \bibinfo{doi}{\doi{10.1103/PhysRevLett.65.2939}}.

\bibtype{Article}%
\bibitem{Buras:2000dm}
\bibinfo{author}{A.~J. Buras}, \bibinfo{author}{P. Gambino},
  \bibinfo{author}{M. Gorbahn}, \bibinfo{author}{S. Jager}, \bibinfo{author}{L.
  Silvestrini}, \bibinfo{title}{{Universal unitarity triangle and physics
  beyond the standard model}}, \bibinfo{journal}{Phys. Lett. B}
  \bibinfo{volume}{500} (\bibinfo{year}{2001}) \bibinfo{pages}{161--167},
  \bibinfo{doi}{\doi{10.1016/S0370-2693(01)00061-2}}.
  \href{http://arxiv.org/abs/hep-ph/0007085}{\tt arXiv:hep-ph/0007085}.

\bibtype{Article}%
\bibitem{DAmbrosio:2002vsn}
\bibinfo{author}{G. D'Ambrosio}, \bibinfo{author}{G.~F. Giudice},
  \bibinfo{author}{G. Isidori}, \bibinfo{author}{A. Strumia},
  \bibinfo{title}{{Minimal flavor violation: An Effective field theory
  approach}}, \bibinfo{journal}{Nucl. Phys. B} \bibinfo{volume}{645}
  (\bibinfo{year}{2002}) \bibinfo{pages}{155--187},
  \bibinfo{doi}{\doi{10.1016/S0550-3213(02)00836-2}}.
  \href{http://arxiv.org/abs/hep-ph/0207036}{\tt arXiv:hep-ph/0207036}.

\bibtype{Article}%
\bibitem{Cirigliano:2005ck}
\bibinfo{author}{Vincenzo Cirigliano}, \bibinfo{author}{Benjamin Grinstein},
  \bibinfo{author}{Gino Isidori}, \bibinfo{author}{Mark~B. Wise},
  \bibinfo{title}{{Minimal flavor violation in the lepton sector}},
  \bibinfo{journal}{Nucl. Phys. B} \bibinfo{volume}{728} (\bibinfo{year}{2005})
  \bibinfo{pages}{121--134},
  \bibinfo{doi}{\doi{10.1016/j.nuclphysb.2005.08.037}}.
  \href{http://arxiv.org/abs/hep-ph/0507001}{\tt arXiv:hep-ph/0507001}.

\bibtype{Article}%
\bibitem{Pomarol:1995xc}
\bibinfo{author}{Alex Pomarol}, \bibinfo{author}{Daniele Tommasini},
  \bibinfo{title}{{Horizontal symmetries for the supersymmetric flavor
  problem}}, \bibinfo{journal}{Nucl. Phys. B} \bibinfo{volume}{466}
  (\bibinfo{year}{1996}) \bibinfo{pages}{3--24},
  \bibinfo{doi}{\doi{10.1016/0550-3213(96)00074-0}}.
  \href{http://arxiv.org/abs/hep-ph/9507462}{\tt arXiv:hep-ph/9507462}.

\bibtype{Article}%
\bibitem{Barbieri:1995uv}
\bibinfo{author}{Riccardo Barbieri}, \bibinfo{author}{G.~R. Dvali},
  \bibinfo{author}{Lawrence~J. Hall}, \bibinfo{title}{{Predictions from a U(2)
  flavor symmetry in supersymmetric theories}}, \bibinfo{journal}{Phys. Lett.
  B} \bibinfo{volume}{377} (\bibinfo{year}{1996}) \bibinfo{pages}{76--82},
  \bibinfo{doi}{\doi{10.1016/0370-2693(96)00318-8}}.
  \href{http://arxiv.org/abs/hep-ph/9512388}{\tt arXiv:hep-ph/9512388}.

\bibtype{Article}%
\bibitem{Barbieri:1997tu}
\bibinfo{author}{Riccardo Barbieri}, \bibinfo{author}{Lawrence~J. Hall},
  \bibinfo{author}{Andrea Romanino}, \bibinfo{title}{{Consequences of a U(2)
  flavor symmetry}}, \bibinfo{journal}{Phys. Lett. B} \bibinfo{volume}{401}
  (\bibinfo{year}{1997}) \bibinfo{pages}{47--53},
  \bibinfo{doi}{\doi{10.1016/S0370-2693(97)00372-9}}.
  \href{http://arxiv.org/abs/hep-ph/9702315}{\tt arXiv:hep-ph/9702315}.

\bibtype{Article}%
\bibitem{Barbieri:2011ci}
\bibinfo{author}{Riccardo Barbieri}, \bibinfo{author}{Gino Isidori},
  \bibinfo{author}{Joel Jones-Perez}, \bibinfo{author}{Paolo Lodone},
  \bibinfo{author}{David~M. Straub}, \bibinfo{title}{{$U(2)$ and Minimal
  Flavour Violation in Supersymmetry}}, \bibinfo{journal}{Eur. Phys. J. C}
  \bibinfo{volume}{71} (\bibinfo{year}{2011}) \bibinfo{pages}{1725},
  \bibinfo{doi}{\doi{10.1140/epjc/s10052-011-1725-z}}.
  \href{http://arxiv.org/abs/1105.2296}{\tt arXiv:1105.2296}.

\bibtype{Article}%
\bibitem{Barbieri:2012uh}
\bibinfo{author}{Riccardo Barbieri}, \bibinfo{author}{Dario Buttazzo},
  \bibinfo{author}{Filippo Sala}, \bibinfo{author}{David~M. Straub},
  \bibinfo{title}{{Flavour physics from an approximate $U(2)^3$ symmetry}},
  \bibinfo{journal}{JHEP} \bibinfo{volume}{07} (\bibinfo{year}{2012})
  \bibinfo{pages}{181}, \bibinfo{doi}{\doi{10.1007/JHEP07(2012)181}}.
  \href{http://arxiv.org/abs/1203.4218}{\tt arXiv:1203.4218}.

\bibtype{Article}%
\bibitem{Fuentes-Martin:2019mun}
\bibinfo{author}{Javier Fuentes-Mart{\'\i}n}, \bibinfo{author}{Gino Isidori},
  \bibinfo{author}{Julie Pag{\`e}s}, \bibinfo{author}{Kei Yamamoto},
  \bibinfo{title}{{With or without U(2)? Probing non-standard flavor and
  helicity structures in semileptonic B decays}}, \bibinfo{journal}{Phys. Lett.
  B} \bibinfo{volume}{800} (\bibinfo{year}{2020}) \bibinfo{pages}{135080},
  \bibinfo{doi}{\doi{10.1016/j.physletb.2019.135080}}.
  \href{http://arxiv.org/abs/1909.02519}{\tt arXiv:1909.02519}.

\bibtype{Article}%
\bibitem{Nir:1993mx}
\bibinfo{author}{Yosef Nir}, \bibinfo{author}{Nathan Seiberg},
  \bibinfo{title}{{Should squarks be degenerate?}}, \bibinfo{journal}{Phys.
  Lett. B} \bibinfo{volume}{309} (\bibinfo{year}{1993})
  \bibinfo{pages}{337--343}, \bibinfo{doi}{\doi{10.1016/0370-2693(93)90942-B}}.
  \href{http://arxiv.org/abs/hep-ph/9304307}{\tt arXiv:hep-ph/9304307}.

\bibtype{Article}%
\bibitem{Blum:2009sk}
\bibinfo{author}{Kfir Blum}, \bibinfo{author}{Yuval Grossman},
  \bibinfo{author}{Yosef Nir}, \bibinfo{author}{Gilad Perez},
  \bibinfo{title}{{Combining K0 - anti-K0 mixing and D0 - anti-D0 mixing to
  constrain the flavor structure of new physics}}, \bibinfo{journal}{Phys. Rev.
  Lett.} \bibinfo{volume}{102} (\bibinfo{year}{2009}) \bibinfo{pages}{211802},
  \bibinfo{doi}{\doi{10.1103/PhysRevLett.102.211802}}.
  \href{http://arxiv.org/abs/0903.2118}{\tt arXiv:0903.2118}.

\bibtype{Inproceedings}%
\bibitem{Altmannshofer:2022aml}
\bibinfo{author}{Wolfgang Altmannshofer}, \bibinfo{author}{Jure Zupan},
  \bibinfo{title}{{Snowmass White Paper: Flavor Model Building}}, in:
  \bibinfo{booktitle}{{Snowmass 2021}} \bibinfo{year}{2022}.
  \href{http://arxiv.org/abs/2203.07726}{\tt arXiv:2203.07726}.

\bibtype{Article}%
\bibitem{Altmannshofer:2024jyv}
\bibinfo{author}{Wolfgang Altmannshofer}, \bibinfo{author}{Admir Greljo},
  \bibinfo{title}{{Recent Progress in Flavor Model Building}}
  (\bibinfo{year}{2024}),
  \bibinfo{doi}{\doi{10.1146/annurev-nucl-121423-100950}}.
  \href{http://arxiv.org/abs/2412.04549}{\tt arXiv:2412.04549}.

\bibtype{Article}%
\bibitem{Das:1995df}
\bibinfo{author}{Ashok~K. Das}, \bibinfo{author}{Chung Kao}, \bibinfo{title}{{A
  Two Higgs doublet model for the top quark}}, \bibinfo{journal}{Phys. Lett. B}
  \bibinfo{volume}{372} (\bibinfo{year}{1996}) \bibinfo{pages}{106--112},
  \bibinfo{doi}{\doi{10.1016/0370-2693(96)00031-7}}.
  \href{http://arxiv.org/abs/hep-ph/9511329}{\tt arXiv:hep-ph/9511329}.

\bibtype{Article}%
\bibitem{Blechman:2010cs}
\bibinfo{author}{Andrew~E. Blechman}, \bibinfo{author}{Alexey~A. Petrov},
  \bibinfo{author}{Gagik Yeghiyan}, \bibinfo{title}{{The Flavor puzzle in
  multi-Higgs models}}, \bibinfo{journal}{JHEP} \bibinfo{volume}{11}
  (\bibinfo{year}{2010}) \bibinfo{pages}{075},
  \bibinfo{doi}{\doi{10.1007/JHEP11(2010)075}}.
  \href{http://arxiv.org/abs/1009.1612}{\tt arXiv:1009.1612}.

\bibtype{Article}%
\bibitem{Altmannshofer:2015esa}
\bibinfo{author}{Wolfgang Altmannshofer}, \bibinfo{author}{Stefania Gori},
  \bibinfo{author}{Alexander~L. Kagan}, \bibinfo{author}{Luca Silvestrini},
  \bibinfo{author}{Jure Zupan}, \bibinfo{title}{{Uncovering Mass Generation
  Through Higgs Flavor Violation}}, \bibinfo{journal}{Phys. Rev. D}
  \bibinfo{volume}{93} (\bibinfo{number}{3}) (\bibinfo{year}{2016})
  \bibinfo{pages}{031301}, \bibinfo{doi}{\doi{10.1103/PhysRevD.93.031301}}.
  \href{http://arxiv.org/abs/1507.07927}{\tt arXiv:1507.07927}.

\bibtype{Article}%
\bibitem{Ghosh:2015gpa}
\bibinfo{author}{Diptimoy Ghosh}, \bibinfo{author}{Rick~Sandeepan Gupta},
  \bibinfo{author}{Gilad Perez}, \bibinfo{title}{{Is the Higgs Mechanism of
  Fermion Mass Generation a Fact? A Yukawa-less First-Two-Generation Model}},
  \bibinfo{journal}{Phys. Lett. B} \bibinfo{volume}{755} (\bibinfo{year}{2016})
  \bibinfo{pages}{504--508},
  \bibinfo{doi}{\doi{10.1016/j.physletb.2016.02.059}}.
  \href{http://arxiv.org/abs/1508.01501}{\tt arXiv:1508.01501}.

\bibtype{Article}%
\bibitem{Botella:2016krk}
\bibinfo{author}{F.~J. Botella}, \bibinfo{author}{G.~C. Branco},
  \bibinfo{author}{M.~N. Rebelo}, \bibinfo{author}{J.~I. Silva-Marcos},
  \bibinfo{title}{{What if the masses of the first two quark families are not
  generated by the standard model Higgs boson?}}, \bibinfo{journal}{Phys. Rev.
  D} \bibinfo{volume}{94} (\bibinfo{number}{11}) (\bibinfo{year}{2016})
  \bibinfo{pages}{115031}, \bibinfo{doi}{\doi{10.1103/PhysRevD.94.115031}}.
  \href{http://arxiv.org/abs/1602.08011}{\tt arXiv:1602.08011}.

\bibtype{Article}%
\bibitem{Maayergi:2025ybi}
\bibinfo{author}{Kamal Maayergi}, \bibinfo{author}{Devin G.~E. Walker},
  \bibinfo{author}{Michael~E. Peskin}, \bibinfo{title}{{Sensitivity of Heavy
  Higgs Boson to the Precision Yukawa Coupling Measurements at Higgs
  Factories}}  (\bibinfo{year}{2025}).
  \href{http://arxiv.org/abs/2506.16587}{\tt arXiv:2506.16587}.

\bibtype{Article}%
\bibitem{Altmannshofer:2016zrn}
\bibinfo{author}{Wolfgang Altmannshofer}, \bibinfo{author}{Joshua Eby},
  \bibinfo{author}{Stefania Gori}, \bibinfo{author}{Matteo Lotito},
  \bibinfo{author}{Mario Martone}, \bibinfo{author}{Douglas Tuckler},
  \bibinfo{title}{{Collider Signatures of Flavorful Higgs Bosons}},
  \bibinfo{journal}{Phys. Rev. D} \bibinfo{volume}{94} (\bibinfo{number}{11})
  (\bibinfo{year}{2016}) \bibinfo{pages}{115032},
  \bibinfo{doi}{\doi{10.1103/PhysRevD.94.115032}}.
  \href{http://arxiv.org/abs/1610.02398}{\tt arXiv:1610.02398}.

\bibtype{Article}%
\bibitem{Altmannshofer:2018bch}
\bibinfo{author}{Wolfgang Altmannshofer}, \bibinfo{author}{Brian Maddock},
  \bibinfo{title}{{Flavorful Two Higgs Doublet Models with a Twist}},
  \bibinfo{journal}{Phys. Rev. D} \bibinfo{volume}{98} (\bibinfo{number}{7})
  (\bibinfo{year}{2018}) \bibinfo{pages}{075005},
  \bibinfo{doi}{\doi{10.1103/PhysRevD.98.075005}}.
  \href{http://arxiv.org/abs/1805.08659}{\tt arXiv:1805.08659}.

\bibtype{Article}%
\bibitem{Altmannshofer:2019ogm}
\bibinfo{author}{Wolfgang Altmannshofer}, \bibinfo{author}{Brian Maddock},
  \bibinfo{author}{Douglas Tuckler}, \bibinfo{title}{{Rare Top Decays as Probes
  of Flavorful Higgs Bosons}}, \bibinfo{journal}{Phys. Rev. D}
  \bibinfo{volume}{100} (\bibinfo{number}{1}) (\bibinfo{year}{2019})
  \bibinfo{pages}{015003}, \bibinfo{doi}{\doi{10.1103/PhysRevD.100.015003}}.
  \href{http://arxiv.org/abs/1904.10956}{\tt arXiv:1904.10956}.

\bibtype{Article}%
\bibitem{Altmannshofer:2025pjj}
\bibinfo{author}{Wolfgang Altmannshofer}, \bibinfo{author}{Kevin Toner},
  \bibinfo{title}{{Flavor constraints in a generational three-Higgs-doublet
  model}}, \bibinfo{journal}{Phys. Rev. D} \bibinfo{volume}{111}
  (\bibinfo{number}{7}) (\bibinfo{year}{2025}) \bibinfo{pages}{075009},
  \bibinfo{doi}{\doi{10.1103/PhysRevD.111.075009}}.
  \href{http://arxiv.org/abs/2502.04579}{\tt arXiv:2502.04579}.

\bibtype{Article}%
\bibitem{Das:2025mqs}
\bibinfo{author}{Dipankar Das}, \bibinfo{author}{Miguel Levy},
  \bibinfo{author}{Anugrah~M. Prasad}, \bibinfo{title}{{Flavor puzzle in three
  Higgs-doublet models: Insights from BGL and lessons from flavor data}}
  (\bibinfo{year}{2025}). \href{http://arxiv.org/abs/2502.20296}{\tt
  arXiv:2502.20296}.

\bibtype{Article}%
\bibitem{Porto:2007ed}
\bibinfo{author}{Rafael~A. Porto}, \bibinfo{author}{A. Zee},
  \bibinfo{title}{{The Private Higgs}}, \bibinfo{journal}{Phys. Lett. B}
  \bibinfo{volume}{666} (\bibinfo{year}{2008}) \bibinfo{pages}{491--495},
  \bibinfo{doi}{\doi{10.1016/j.physletb.2008.08.001}}.
  \href{http://arxiv.org/abs/0712.0448}{\tt arXiv:0712.0448}.

\bibtype{Article}%
\bibitem{Hill:2019ldq}
\bibinfo{author}{Christopher~T. Hill}, \bibinfo{author}{Pedro A.~N. Machado},
  \bibinfo{author}{Anders~E. Thomsen}, \bibinfo{author}{Jessica Turner},
  \bibinfo{title}{{Scalar Democracy}}, \bibinfo{journal}{Phys. Rev. D}
  \bibinfo{volume}{100} (\bibinfo{number}{1}) (\bibinfo{year}{2019})
  \bibinfo{pages}{015015}, \bibinfo{doi}{\doi{10.1103/PhysRevD.100.015015}}.
  \href{http://arxiv.org/abs/1902.07214}{\tt arXiv:1902.07214}.

\bibtype{Article}%
\bibitem{Baek:2023cfy}
\bibinfo{author}{S. Baek}, \bibinfo{author}{J. Kersten}, \bibinfo{author}{P.
  Ko}, \bibinfo{author}{L. Velasco-Sevilla}, \bibinfo{title}{{An unfamiliar way
  to generate the hierarchy of standard model fermion masses}},
  \bibinfo{journal}{JHEP} \bibinfo{volume}{02} (\bibinfo{year}{2024})
  \bibinfo{pages}{143}, \bibinfo{doi}{\doi{10.1007/JHEP02(2024)143}}.
  \href{http://arxiv.org/abs/2309.07788}{\tt arXiv:2309.07788}.

\bibtype{Article}%
\bibitem{Weinberg:1972ws}
\bibinfo{author}{Steven Weinberg}, \bibinfo{title}{{Electromagnetic and weak
  masses}}, \bibinfo{journal}{Phys. Rev. Lett.} \bibinfo{volume}{29}
  (\bibinfo{year}{1972}) \bibinfo{pages}{388--392},
  \bibinfo{doi}{\doi{10.1103/PhysRevLett.29.388}}.

\bibtype{Article}%
\bibitem{Altmannshofer:2014qha}
\bibinfo{author}{Wolfgang Altmannshofer}, \bibinfo{author}{Claudia Frugiuele},
  \bibinfo{author}{Roni Harnik}, \bibinfo{title}{{Fermion Hierarchy from
  Sfermion Anarchy}}, \bibinfo{journal}{JHEP} \bibinfo{volume}{12}
  (\bibinfo{year}{2014}) \bibinfo{pages}{180},
  \bibinfo{doi}{\doi{10.1007/JHEP12(2014)180}}.
  \href{http://arxiv.org/abs/1409.2522}{\tt arXiv:1409.2522}.

\bibtype{Article}%
\bibitem{Banks:1987iu}
\bibinfo{author}{Tom Banks}, \bibinfo{title}{{Supersymmetry and the Quark Mass
  Matrix}}, \bibinfo{journal}{Nucl. Phys. B} \bibinfo{volume}{303}
  (\bibinfo{year}{1988}) \bibinfo{pages}{172--188},
  \bibinfo{doi}{\doi{10.1016/0550-3213(88)90222-2}}.

\bibtype{Article}%
\bibitem{Kagan:1989fp}
\bibinfo{author}{Alexander~L. Kagan}, \bibinfo{title}{{Radiative Quark Mass and
  Mixing Hierarchies From Supersymmetric Models With a Fourth Mirror Family}},
  \bibinfo{journal}{Phys. Rev. D} \bibinfo{volume}{40} (\bibinfo{year}{1989})
  \bibinfo{pages}{173}, \bibinfo{doi}{\doi{10.1103/PhysRevD.40.173}}.

\bibtype{Article}%
\bibitem{Arkani-Hamed:1996kxn}
\bibinfo{author}{Nima Arkani-Hamed}, \bibinfo{author}{Hsin-Chia Cheng},
  \bibinfo{author}{L.~J. Hall}, \bibinfo{title}{{A Supersymmetric theory of
  flavor with radiative fermion masses}}, \bibinfo{journal}{Phys. Rev. D}
  \bibinfo{volume}{54} (\bibinfo{year}{1996}) \bibinfo{pages}{2242--2260},
  \bibinfo{doi}{\doi{10.1103/PhysRevD.54.2242}}.
  \href{http://arxiv.org/abs/hep-ph/9601262}{\tt arXiv:hep-ph/9601262}.

\bibtype{Article}%
\bibitem{Borzumati:1999sp}
\bibinfo{author}{Francesca Borzumati}, \bibinfo{author}{Glennys~R. Farrar},
  \bibinfo{author}{Nir Polonsky}, \bibinfo{author}{Scott~D. Thomas},
  \bibinfo{title}{{Soft Yukawa couplings in supersymmetric theories}},
  \bibinfo{journal}{Nucl. Phys. B} \bibinfo{volume}{555} (\bibinfo{year}{1999})
  \bibinfo{pages}{53--115}, \bibinfo{doi}{\doi{10.1016/S0550-3213(99)00328-4}}.
  \href{http://arxiv.org/abs/hep-ph/9902443}{\tt arXiv:hep-ph/9902443}.

\bibtype{Article}%
\bibitem{Baumgart:2014jya}
\bibinfo{author}{Matthew Baumgart}, \bibinfo{author}{Daniel Stolarski},
  \bibinfo{author}{Thomas Zorawski}, \bibinfo{title}{{Split supersymmetry
  radiates flavor}}, \bibinfo{journal}{Phys. Rev. D} \bibinfo{volume}{90}
  (\bibinfo{number}{5}) (\bibinfo{year}{2014}) \bibinfo{pages}{055001},
  \bibinfo{doi}{\doi{10.1103/PhysRevD.90.055001}}.
  \href{http://arxiv.org/abs/1403.6118}{\tt arXiv:1403.6118}.

\bibtype{Article}%
\bibitem{Barr:1979xt}
\bibinfo{author}{Stephen~M. Barr}, \bibinfo{title}{{Light Fermion Mass
  Hierarchy and Grand Unification}}, \bibinfo{journal}{Phys. Rev. D}
  \bibinfo{volume}{21} (\bibinfo{year}{1980}) \bibinfo{pages}{1424},
  \bibinfo{doi}{\doi{10.1103/PhysRevD.21.1424}}.

\bibtype{Article}%
\bibitem{Balakrishna:1987qd}
\bibinfo{author}{B.~S. Balakrishna}, \bibinfo{title}{{Fermion Mass Hierarchy
  From Radiative Corrections}}, \bibinfo{journal}{Phys. Rev. Lett.}
  \bibinfo{volume}{60} (\bibinfo{year}{1988}) \bibinfo{pages}{1602},
  \bibinfo{doi}{\doi{10.1103/PhysRevLett.60.1602}}.

\bibtype{Article}%
\bibitem{Balakrishna:1988ks}
\bibinfo{author}{B.~S. Balakrishna}, \bibinfo{author}{A.~L. Kagan},
  \bibinfo{author}{R.~N. Mohapatra}, \bibinfo{title}{{Quark Mixings and Mass
  Hierarchy From Radiative Corrections}}, \bibinfo{journal}{Phys. Lett. B}
  \bibinfo{volume}{205} (\bibinfo{year}{1988}) \bibinfo{pages}{345--352},
  \bibinfo{doi}{\doi{10.1016/0370-2693(88)91676-0}}.

\bibtype{Article}%
\bibitem{Dobrescu:2008sz}
\bibinfo{author}{Bogdan~A. Dobrescu}, \bibinfo{author}{Patrick~J. Fox},
  \bibinfo{title}{{Quark and lepton masses from top loops}},
  \bibinfo{journal}{JHEP} \bibinfo{volume}{08} (\bibinfo{year}{2008})
  \bibinfo{pages}{100}, \bibinfo{doi}{\doi{10.1088/1126-6708/2008/08/100}}.
  \href{http://arxiv.org/abs/0805.0822}{\tt arXiv:0805.0822}.

\bibtype{Article}%
\bibitem{Graham:2009gr}
\bibinfo{author}{Peter~W. Graham}, \bibinfo{author}{Surjeet Rajendran},
  \bibinfo{title}{{A Domino Theory of Flavor}}, \bibinfo{journal}{Phys. Rev. D}
  \bibinfo{volume}{81} (\bibinfo{year}{2010}) \bibinfo{pages}{033002},
  \bibinfo{doi}{\doi{10.1103/PhysRevD.81.033002}}.
  \href{http://arxiv.org/abs/0906.4657}{\tt arXiv:0906.4657}.

\bibtype{Article}%
\bibitem{Baker:2020vkh}
\bibinfo{author}{Michael~J. Baker}, \bibinfo{author}{Peter Cox},
  \bibinfo{author}{Raymond~R. Volkas}, \bibinfo{title}{{Has the Origin of the
  Third-Family Fermion Masses been Determined?}}, \bibinfo{journal}{JHEP}
  \bibinfo{volume}{04} (\bibinfo{year}{2021}) \bibinfo{pages}{151},
  \bibinfo{doi}{\doi{10.1007/JHEP04(2021)151}}.
  \href{http://arxiv.org/abs/2012.10458}{\tt arXiv:2012.10458}.

\bibtype{Article}%
\bibitem{Froggatt:1978nt}
\bibinfo{author}{C.~D. Froggatt}, \bibinfo{author}{Holger~Bech Nielsen},
  \bibinfo{title}{{Hierarchy of Quark Masses, Cabibbo Angles and CP
  Violation}}, \bibinfo{journal}{Nucl. Phys. B} \bibinfo{volume}{147}
  (\bibinfo{year}{1979}) \bibinfo{pages}{277--298},
  \bibinfo{doi}{\doi{10.1016/0550-3213(79)90316-X}}.

\bibtype{Article}%
\bibitem{Davidson:1979wr}
\bibinfo{author}{Aharon Davidson}, \bibinfo{author}{Mehmet Koca},
  \bibinfo{author}{Kameshwar~C. Wali}, \bibinfo{title}{{U(1) as the Minimal
  Horizontal Gauge Symmetry}}, \bibinfo{journal}{Phys. Rev. Lett.}
  \bibinfo{volume}{43} (\bibinfo{year}{1979}) \bibinfo{pages}{92},
  \bibinfo{doi}{\doi{10.1103/PhysRevLett.43.92}}.

\bibtype{Article}%
\bibitem{Leurer:1992wg}
\bibinfo{author}{Miriam Leurer}, \bibinfo{author}{Yosef Nir},
  \bibinfo{author}{Nathan Seiberg}, \bibinfo{title}{{Mass matrix models}},
  \bibinfo{journal}{Nucl. Phys. B} \bibinfo{volume}{398} (\bibinfo{year}{1993})
  \bibinfo{pages}{319--342}, \bibinfo{doi}{\doi{10.1016/0550-3213(93)90112-3}}.
  \href{http://arxiv.org/abs/hep-ph/9212278}{\tt arXiv:hep-ph/9212278}.

\bibtype{Article}%
\bibitem{Leurer:1993gy}
\bibinfo{author}{Miriam Leurer}, \bibinfo{author}{Yosef Nir},
  \bibinfo{author}{Nathan Seiberg}, \bibinfo{title}{{Mass matrix models: The
  Sequel}}, \bibinfo{journal}{Nucl. Phys. B} \bibinfo{volume}{420}
  (\bibinfo{year}{1994}) \bibinfo{pages}{468--504},
  \bibinfo{doi}{\doi{10.1016/0550-3213(94)90074-4}}.
  \href{http://arxiv.org/abs/hep-ph/9310320}{\tt arXiv:hep-ph/9310320}.

\bibtype{Article}%
\bibitem{Ben-Hamo:1994dha}
\bibinfo{author}{Valerie Ben-Hamo}, \bibinfo{author}{Yosef Nir},
  \bibinfo{title}{{Implications of horizontal symmetries on baryon number
  violation in supersymmetric models}}, \bibinfo{journal}{Phys. Lett. B}
  \bibinfo{volume}{339} (\bibinfo{year}{1994}) \bibinfo{pages}{77--82},
  \bibinfo{doi}{\doi{10.1016/0370-2693(94)91135-5}}.
  \href{http://arxiv.org/abs/hep-ph/9408315}{\tt arXiv:hep-ph/9408315}.

\bibtype{Article}%
\bibitem{Fedele:2020fvh}
\bibinfo{author}{Marco Fedele}, \bibinfo{author}{Alessio Mastroddi},
  \bibinfo{author}{Mauro Valli}, \bibinfo{title}{{Minimal Froggatt-Nielsen
  textures}}, \bibinfo{journal}{JHEP} \bibinfo{volume}{03}
  (\bibinfo{year}{2021}) \bibinfo{pages}{135},
  \bibinfo{doi}{\doi{10.1007/JHEP03(2021)135}}.
  \href{http://arxiv.org/abs/2009.05587}{\tt arXiv:2009.05587}.

\bibtype{Article}%
\bibitem{Cornella:2023zme}
\bibinfo{author}{Claudia Cornella}, \bibinfo{author}{David Curtin},
  \bibinfo{author}{Ethan~T. Neil}, \bibinfo{author}{Jedidiah~O. Thompson},
  \bibinfo{title}{{Mapping and probing Froggatt-Nielsen solutions to the quark
  flavor puzzle}}, \bibinfo{journal}{Phys. Rev. D} \bibinfo{volume}{111}
  (\bibinfo{number}{1}) (\bibinfo{year}{2025}) \bibinfo{pages}{015042},
  \bibinfo{doi}{\doi{10.1103/PhysRevD.111.015042}}.
  \href{http://arxiv.org/abs/2306.08026}{\tt arXiv:2306.08026}.

\bibtype{Article}%
\bibitem{Barbieri:2015yvd}
\bibinfo{author}{Riccardo Barbieri}, \bibinfo{author}{Gino Isidori},
  \bibinfo{author}{Andrea Pattori}, \bibinfo{author}{Fabrizio Senia},
  \bibinfo{title}{{Anomalies in $B$-decays and $U(2)$ flavour symmetry}},
  \bibinfo{journal}{Eur. Phys. J. C} \bibinfo{volume}{76} (\bibinfo{number}{2})
  (\bibinfo{year}{2016}) \bibinfo{pages}{67},
  \bibinfo{doi}{\doi{10.1140/epjc/s10052-016-3905-3}}.
  \href{http://arxiv.org/abs/1512.01560}{\tt arXiv:1512.01560}.

\bibtype{Article}%
\bibitem{Antusch:2023shi}
\bibinfo{author}{Stefan Antusch}, \bibinfo{author}{Admir Greljo},
  \bibinfo{author}{Ben~A. Stefanek}, \bibinfo{author}{Anders~Eller Thomsen},
  \bibinfo{title}{{U(2) Is Right for Leptons and Left for Quarks}},
  \bibinfo{journal}{Phys. Rev. Lett.} \bibinfo{volume}{132}
  (\bibinfo{number}{15}) (\bibinfo{year}{2024}) \bibinfo{pages}{151802},
  \bibinfo{doi}{\doi{10.1103/PhysRevLett.132.151802}}.
  \href{http://arxiv.org/abs/2311.09288}{\tt arXiv:2311.09288}.

\bibtype{Article}%
\bibitem{Greljo:2024zrj}
\bibinfo{author}{Admir Greljo}, \bibinfo{author}{Anders~Eller Thomsen},
  \bibinfo{author}{Hector Tiblom}, \bibinfo{title}{{Flavor hierarchies from
  SU(2) flavor and quark-lepton unification}}, \bibinfo{journal}{JHEP}
  \bibinfo{volume}{08} (\bibinfo{year}{2024}) \bibinfo{pages}{143},
  \bibinfo{doi}{\doi{10.1007/JHEP08(2024)143}}.
  \href{http://arxiv.org/abs/2406.02687}{\tt arXiv:2406.02687}.

\bibtype{Article}%
\bibitem{Bai:2009ij}
\bibinfo{author}{Yang Bai}, \bibinfo{author}{Gustavo Burdman},
  \bibinfo{author}{Christopher~T. Hill}, \bibinfo{title}{{Topological
  Interactions in Warped Extra Dimensions}}, \bibinfo{journal}{JHEP}
  \bibinfo{volume}{02} (\bibinfo{year}{2010}) \bibinfo{pages}{049},
  \bibinfo{doi}{\doi{10.1007/JHEP02(2010)049}}.
  \href{http://arxiv.org/abs/0911.1358}{\tt arXiv:0911.1358}.

\bibtype{Article}%
\bibitem{Burdman:2012sb}
\bibinfo{author}{Gustavo Burdman}, \bibinfo{author}{Nayara Fonseca},
  \bibinfo{author}{Leonardo de Lima}, \bibinfo{title}{{Full-hierarchy Quiver
  Theories of Electroweak Symmetry Breaking and Fermion Masses}},
  \bibinfo{journal}{JHEP} \bibinfo{volume}{01} (\bibinfo{year}{2013})
  \bibinfo{pages}{094}, \bibinfo{doi}{\doi{10.1007/JHEP01(2013)094}}.
  \href{http://arxiv.org/abs/1210.5568}{\tt arXiv:1210.5568}.

\bibtype{Article}%
\bibitem{Giudice:2016yja}
\bibinfo{author}{Gian~F. Giudice}, \bibinfo{author}{Matthew McCullough},
  \bibinfo{title}{{A Clockwork Theory}}, \bibinfo{journal}{JHEP}
  \bibinfo{volume}{02} (\bibinfo{year}{2017}) \bibinfo{pages}{036},
  \bibinfo{doi}{\doi{10.1007/JHEP02(2017)036}}.
  \href{http://arxiv.org/abs/1610.07962}{\tt arXiv:1610.07962}.

\bibtype{Article}%
\bibitem{vonGersdorff:2017iym}
\bibinfo{author}{Gero von Gersdorff}, \bibinfo{title}{{Natural Fermion
  Hierarchies from Random Yukawa Couplings}}, \bibinfo{journal}{JHEP}
  \bibinfo{volume}{09} (\bibinfo{year}{2017}) \bibinfo{pages}{094},
  \bibinfo{doi}{\doi{10.1007/JHEP09(2017)094}}.
  \href{http://arxiv.org/abs/1705.05430}{\tt arXiv:1705.05430}.

\bibtype{Article}%
\bibitem{Patel:2017pct}
\bibinfo{author}{Ketan~M. Patel}, \bibinfo{title}{{Clockwork mechanism for
  flavor hierarchies}}, \bibinfo{journal}{Phys. Rev. D} \bibinfo{volume}{96}
  (\bibinfo{number}{11}) (\bibinfo{year}{2017}) \bibinfo{pages}{115013},
  \bibinfo{doi}{\doi{10.1103/PhysRevD.96.115013}}.
  \href{http://arxiv.org/abs/1711.05393}{\tt arXiv:1711.05393}.

\bibtype{Article}%
\bibitem{Ibarra:2017tju}
\bibinfo{author}{Alejandro Ibarra}, \bibinfo{author}{Ashwani Kushwaha},
  \bibinfo{author}{Sudhir~K. Vempati}, \bibinfo{title}{{Clockwork for Neutrino
  Masses and Lepton Flavor Violation}}, \bibinfo{journal}{Phys. Lett. B}
  \bibinfo{volume}{780} (\bibinfo{year}{2018}) \bibinfo{pages}{86--92},
  \bibinfo{doi}{\doi{10.1016/j.physletb.2018.02.047}}.
  \href{http://arxiv.org/abs/1711.02070}{\tt arXiv:1711.02070}.

\bibtype{Article}%
\bibitem{Alonso:2018bcg}
\bibinfo{author}{Rodrigo Alonso}, \bibinfo{author}{Adrian Carmona},
  \bibinfo{author}{Barry~M. Dillon}, \bibinfo{author}{Jernej~F. Kamenik},
  \bibinfo{author}{Jorge Martin~Camalich}, \bibinfo{author}{Jure Zupan},
  \bibinfo{title}{{A clockwork solution to the flavor puzzle}},
  \bibinfo{journal}{JHEP} \bibinfo{volume}{10} (\bibinfo{year}{2018})
  \bibinfo{pages}{099}, \bibinfo{doi}{\doi{10.1007/JHEP10(2018)099}}.
  \href{http://arxiv.org/abs/1807.09792}{\tt arXiv:1807.09792}.

\bibtype{Article}%
\bibitem{Hong:2019bki}
\bibinfo{author}{Sungwoo Hong}, \bibinfo{author}{Gowri Kurup},
  \bibinfo{author}{Maxim Perelstein}, \bibinfo{title}{{Clockwork Neutrinos}},
  \bibinfo{journal}{JHEP} \bibinfo{volume}{10} (\bibinfo{year}{2019})
  \bibinfo{pages}{073}, \bibinfo{doi}{\doi{10.1007/JHEP10(2019)073}}.
  \href{http://arxiv.org/abs/1903.06191}{\tt arXiv:1903.06191}.

\bibtype{Article}%
\bibitem{AbreudeSouza:2019ixc}
\bibinfo{author}{Fernando Abreu~de Souza}, \bibinfo{author}{Gero von
  Gersdorff}, \bibinfo{title}{{A Random Clockwork of Flavor}},
  \bibinfo{journal}{JHEP} \bibinfo{volume}{02} (\bibinfo{year}{2020})
  \bibinfo{pages}{186}, \bibinfo{doi}{\doi{10.1007/JHEP02(2020)186}}.
  \href{http://arxiv.org/abs/1911.08476}{\tt arXiv:1911.08476}.

\bibtype{Article}%
\bibitem{vonGersdorff:2020ods}
\bibinfo{author}{Gero von Gersdorff}, \bibinfo{title}{{Realistic GUT Yukawa
  couplings from a random clockwork model}}, \bibinfo{journal}{Eur. Phys. J. C}
  \bibinfo{volume}{80} (\bibinfo{number}{12}) (\bibinfo{year}{2020})
  \bibinfo{pages}{1176}, \bibinfo{doi}{\doi{10.1140/epjc/s10052-020-08764-4}}.
  \href{http://arxiv.org/abs/2005.14207}{\tt arXiv:2005.14207}.

\bibtype{Article}%
\bibitem{Kang:2020cxo}
\bibinfo{author}{Yoo-Jin Kang}, \bibinfo{author}{Soonbin Kim},
  \bibinfo{author}{Hyun~Min Lee}, \bibinfo{title}{{The Clockwork Standard
  Model}}, \bibinfo{journal}{JHEP} \bibinfo{volume}{09} (\bibinfo{year}{2020})
  \bibinfo{pages}{005}, \bibinfo{doi}{\doi{10.1007/JHEP09(2020)005}}.
  \href{http://arxiv.org/abs/2006.03043}{\tt arXiv:2006.03043}.

\bibtype{Article}%
\bibitem{Babu:2020tnf}
\bibinfo{author}{K.~S. Babu}, \bibinfo{author}{Shaikh Saad},
  \bibinfo{title}{{Flavor Hierarchies from Clockwork in SO(10) GUT}},
  \bibinfo{journal}{Phys. Rev. D} \bibinfo{volume}{103} (\bibinfo{number}{1})
  (\bibinfo{year}{2021}) \bibinfo{pages}{015009},
  \bibinfo{doi}{\doi{10.1103/PhysRevD.103.015009}}.
  \href{http://arxiv.org/abs/2007.16085}{\tt arXiv:2007.16085}.

\bibtype{Article}%
\bibitem{Smolkovic:2019jow}
\bibinfo{author}{Aleks Smolkovi\v{c}}, \bibinfo{author}{Michele Tammaro},
  \bibinfo{author}{Jure Zupan}, \bibinfo{title}{{Anomaly free Froggatt-Nielsen
  models of flavor}}, \bibinfo{journal}{JHEP} \bibinfo{volume}{10}
  (\bibinfo{year}{2019}) \bibinfo{pages}{188},
  \bibinfo{doi}{\doi{10.1007/JHEP10(2019)188}}, \bibinfo{note}{[Erratum: JHEP
  02, 033 (2022)]}. \href{http://arxiv.org/abs/1907.10063}{\tt
  arXiv:1907.10063}.

\bibtype{Article}%
\bibitem{Altmannshofer:2021qwx}
\bibinfo{author}{Wolfgang Altmannshofer}, \bibinfo{author}{Sri~Aditya Gadam},
  \bibinfo{title}{{Supersymmetric flavor clockwork model}},
  \bibinfo{journal}{Phys. Rev. D} \bibinfo{volume}{104} (\bibinfo{number}{3})
  (\bibinfo{year}{2021}) \bibinfo{pages}{035030},
  \bibinfo{doi}{\doi{10.1103/PhysRevD.104.035030}}.
  \href{http://arxiv.org/abs/2106.09869}{\tt arXiv:2106.09869}.

\bibtype{Article}%
\bibitem{Arkani-Hamed:1999ylh}
\bibinfo{author}{Nima Arkani-Hamed}, \bibinfo{author}{Martin Schmaltz},
  \bibinfo{title}{{Hierarchies without symmetries from extra dimensions}},
  \bibinfo{journal}{Phys. Rev. D} \bibinfo{volume}{61} (\bibinfo{year}{2000})
  \bibinfo{pages}{033005}, \bibinfo{doi}{\doi{10.1103/PhysRevD.61.033005}}.
  \href{http://arxiv.org/abs/hep-ph/9903417}{\tt arXiv:hep-ph/9903417}.

\bibtype{Article}%
\bibitem{Randall:1999ee}
\bibinfo{author}{Lisa Randall}, \bibinfo{author}{Raman Sundrum},
  \bibinfo{title}{{A Large mass hierarchy from a small extra dimension}},
  \bibinfo{journal}{Phys. Rev. Lett.} \bibinfo{volume}{83}
  (\bibinfo{year}{1999}) \bibinfo{pages}{3370--3373},
  \bibinfo{doi}{\doi{10.1103/PhysRevLett.83.3370}}.
  \href{http://arxiv.org/abs/hep-ph/9905221}{\tt arXiv:hep-ph/9905221}.

\bibtype{Article}%
\bibitem{Arkani-Hamed:2001kyx}
\bibinfo{author}{Nima Arkani-Hamed}, \bibinfo{author}{Andrew~G. Cohen},
  \bibinfo{author}{Howard Georgi}, \bibinfo{title}{{(De)constructing
  dimensions}}, \bibinfo{journal}{Phys. Rev. Lett.} \bibinfo{volume}{86}
  (\bibinfo{year}{2001}) \bibinfo{pages}{4757--4761},
  \bibinfo{doi}{\doi{10.1103/PhysRevLett.86.4757}}.
  \href{http://arxiv.org/abs/hep-th/0104005}{\tt arXiv:hep-th/0104005}.

\bibtype{Article}%
\bibitem{Hill:2000mu}
\bibinfo{author}{Christopher~T. Hill}, \bibinfo{author}{Stefan Pokorski},
  \bibinfo{author}{Jing Wang}, \bibinfo{title}{{Gauge Invariant Effective
  Lagrangian for Kaluza-Klein Modes}}, \bibinfo{journal}{Phys. Rev. D}
  \bibinfo{volume}{64} (\bibinfo{year}{2001}) \bibinfo{pages}{105005},
  \bibinfo{doi}{\doi{10.1103/PhysRevD.64.105005}}.
  \href{http://arxiv.org/abs/hep-th/0104035}{\tt arXiv:hep-th/0104035}.

\bibtype{Article}%
\bibitem{Grossman:1999ra}
\bibinfo{author}{Yuval Grossman}, \bibinfo{author}{Matthias Neubert},
  \bibinfo{title}{{Neutrino masses and mixings in nonfactorizable geometry}},
  \bibinfo{journal}{Phys. Lett. B} \bibinfo{volume}{474} (\bibinfo{year}{2000})
  \bibinfo{pages}{361--371},
  \bibinfo{doi}{\doi{10.1016/S0370-2693(00)00054-X}}.
  \href{http://arxiv.org/abs/hep-ph/9912408}{\tt arXiv:hep-ph/9912408}.

\bibtype{Article}%
\bibitem{Gherghetta:2000qt}
\bibinfo{author}{Tony Gherghetta}, \bibinfo{author}{Alex Pomarol},
  \bibinfo{title}{{Bulk fields and supersymmetry in a slice of AdS}},
  \bibinfo{journal}{Nucl. Phys. B} \bibinfo{volume}{586} (\bibinfo{year}{2000})
  \bibinfo{pages}{141--162},
  \bibinfo{doi}{\doi{10.1016/S0550-3213(00)00392-8}}.
  \href{http://arxiv.org/abs/hep-ph/0003129}{\tt arXiv:hep-ph/0003129}.

\bibtype{Article}%
\bibitem{Huber:2000ie}
\bibinfo{author}{Stephan~J. Huber}, \bibinfo{author}{Qaisar Shafi},
  \bibinfo{title}{{Fermion masses, mixings and proton decay in a
  Randall-Sundrum model}}, \bibinfo{journal}{Phys. Lett. B}
  \bibinfo{volume}{498} (\bibinfo{year}{2001}) \bibinfo{pages}{256--262},
  \bibinfo{doi}{\doi{10.1016/S0370-2693(00)01399-X}}.
  \href{http://arxiv.org/abs/hep-ph/0010195}{\tt arXiv:hep-ph/0010195}.

\bibtype{Article}%
\bibitem{Agashe:2004cp}
\bibinfo{author}{Kaustubh Agashe}, \bibinfo{author}{Gilad Perez},
  \bibinfo{author}{Amarjit Soni}, \bibinfo{title}{{Flavor structure of warped
  extra dimension models}}, \bibinfo{journal}{Phys. Rev. D}
  \bibinfo{volume}{71} (\bibinfo{year}{2005}) \bibinfo{pages}{016002},
  \bibinfo{doi}{\doi{10.1103/PhysRevD.71.016002}}.
  \href{http://arxiv.org/abs/hep-ph/0408134}{\tt arXiv:hep-ph/0408134}.

\bibtype{Article}%
\bibitem{Blanke:2008zb}
\bibinfo{author}{Monika Blanke}, \bibinfo{author}{Andrzej~J. Buras},
  \bibinfo{author}{Bjoern Duling}, \bibinfo{author}{Stefania Gori},
  \bibinfo{author}{Andreas Weiler}, \bibinfo{title}{{$\Delta$ F=2 Observables
  and Fine-Tuning in a Warped Extra Dimension with Custodial Protection}},
  \bibinfo{journal}{JHEP} \bibinfo{volume}{03} (\bibinfo{year}{2009})
  \bibinfo{pages}{001}, \bibinfo{doi}{\doi{10.1088/1126-6708/2009/03/001}}.
  \href{http://arxiv.org/abs/0809.1073}{\tt arXiv:0809.1073}.

\bibtype{Article}%
\bibitem{Cacciapaglia:2007fw}
\bibinfo{author}{Giacomo Cacciapaglia}, \bibinfo{author}{Csaba Csaki},
  \bibinfo{author}{Jamison Galloway}, \bibinfo{author}{Guido Marandella},
  \bibinfo{author}{John Terning}, \bibinfo{author}{Andreas Weiler},
  \bibinfo{title}{{A GIM Mechanism from Extra Dimensions}},
  \bibinfo{journal}{JHEP} \bibinfo{volume}{04} (\bibinfo{year}{2008})
  \bibinfo{pages}{006}, \bibinfo{doi}{\doi{10.1088/1126-6708/2008/04/006}}.
  \href{http://arxiv.org/abs/0709.1714}{\tt arXiv:0709.1714}.

\bibtype{Article}%
\bibitem{Kaplan:1991dc}
\bibinfo{author}{David~B. Kaplan}, \bibinfo{title}{{Flavor at SSC energies: A
  New mechanism for dynamically generated fermion masses}},
  \bibinfo{journal}{Nucl. Phys. B} \bibinfo{volume}{365} (\bibinfo{year}{1991})
  \bibinfo{pages}{259--278},
  \bibinfo{doi}{\doi{10.1016/S0550-3213(05)80021-5}}.

\bibtype{Article}%
\bibitem{Bellazzini:2014yua}
\bibinfo{author}{Brando Bellazzini}, \bibinfo{author}{Csaba Cs\'aki},
  \bibinfo{author}{Javi Serra}, \bibinfo{title}{{Composite Higgses}},
  \bibinfo{journal}{Eur. Phys. J. C} \bibinfo{volume}{74} (\bibinfo{number}{5})
  (\bibinfo{year}{2014}) \bibinfo{pages}{2766},
  \bibinfo{doi}{\doi{10.1140/epjc/s10052-014-2766-x}}.
  \href{http://arxiv.org/abs/1401.2457}{\tt arXiv:1401.2457}.

\bibtype{Book}%
\bibitem{Panico:2015jxa}
\bibinfo{author}{Giuliano Panico}, \bibinfo{author}{Andrea Wulzer},
  \bibinfo{title}{{The Composite Nambu-Goldstone Higgs}},
  \bibinfo{comment}{vol.} \bibinfo{volume}{913}, \bibinfo{publisher}{Springer}
  \bibinfo{year}{2016}, \bibinfo{doi}{\doi{10.1007/978-3-319-22617-0}}.
  \href{http://arxiv.org/abs/1506.01961}{\tt arXiv:1506.01961}.

\bibtype{Article}%
\bibitem{Csaki:2008zd}
\bibinfo{author}{Csaba Csaki}, \bibinfo{author}{Adam Falkowski},
  \bibinfo{author}{Andreas Weiler}, \bibinfo{title}{{The Flavor of the
  Composite Pseudo-Goldstone Higgs}}, \bibinfo{journal}{JHEP}
  \bibinfo{volume}{09} (\bibinfo{year}{2008}) \bibinfo{pages}{008},
  \bibinfo{doi}{\doi{10.1088/1126-6708/2008/09/008}}.
  \href{http://arxiv.org/abs/0804.1954}{\tt arXiv:0804.1954}.

\bibtype{Article}%
\bibitem{Redi:2011zi}
\bibinfo{author}{Michele Redi}, \bibinfo{author}{Andreas Weiler},
  \bibinfo{title}{{Flavor and CP Invariant Composite Higgs Models}},
  \bibinfo{journal}{JHEP} \bibinfo{volume}{11} (\bibinfo{year}{2011})
  \bibinfo{pages}{108}, \bibinfo{doi}{\doi{10.1007/JHEP11(2011)108}}.
  \href{http://arxiv.org/abs/1106.6357}{\tt arXiv:1106.6357}.

\bibtype{Article}%
\bibitem{Niehoff:2015iaa}
\bibinfo{author}{Christoph Niehoff}, \bibinfo{author}{Peter Stangl},
  \bibinfo{author}{David~M. Straub}, \bibinfo{title}{{Direct and indirect
  signals of natural composite Higgs models}}, \bibinfo{journal}{JHEP}
  \bibinfo{volume}{01} (\bibinfo{year}{2016}) \bibinfo{pages}{119},
  \bibinfo{doi}{\doi{10.1007/JHEP01(2016)119}}.
  \href{http://arxiv.org/abs/1508.00569}{\tt arXiv:1508.00569}.

\bibtype{Article}%
\bibitem{Niehoff:2016zso}
\bibinfo{author}{Christoph Niehoff}, \bibinfo{author}{Peter Stangl},
  \bibinfo{author}{David~M. Straub}, \bibinfo{title}{{Electroweak symmetry
  breaking and collider signatures in the next-to-minimal composite Higgs
  model}}, \bibinfo{journal}{JHEP} \bibinfo{volume}{04} (\bibinfo{year}{2017})
  \bibinfo{pages}{117}, \bibinfo{doi}{\doi{10.1007/JHEP04(2017)117}}.
  \href{http://arxiv.org/abs/1611.09356}{\tt arXiv:1611.09356}.

\bibtype{Article}%
\bibitem{Glioti:2024hye}
\bibinfo{author}{Alfredo Glioti}, \bibinfo{author}{Riccardo Rattazzi},
  \bibinfo{author}{Lorenzo Ricci}, \bibinfo{author}{Luca Vecchi},
  \bibinfo{title}{{Exploring the Flavor Symmetry Landscape}}
  (\bibinfo{year}{2024}). \href{http://arxiv.org/abs/2402.09503}{\tt
  arXiv:2402.09503}.

\bibtype{Article}%
\bibitem{Agashe:2025tge}
\bibinfo{author}{Kaustubh Agashe}, \bibinfo{author}{Lorenzo Ricci},
  \bibinfo{author}{Raman Sundrum}, \bibinfo{title}{{Partial Compositeness: from
  Anarchy to Symmetry}}  (\bibinfo{year}{2025}).
  \href{http://arxiv.org/abs/2507.05332}{\tt arXiv:2507.05332}.

\bibtype{Article}%
\bibitem{Li:1981nk}
\bibinfo{author}{Xiaoyuan Li}, \bibinfo{author}{Ernest Ma},
  \bibinfo{title}{{Gauge Model of Generation Nonuniversality}},
  \bibinfo{journal}{Phys. Rev. Lett.} \bibinfo{volume}{47}
  (\bibinfo{year}{1981}) \bibinfo{pages}{1788},
  \bibinfo{doi}{\doi{10.1103/PhysRevLett.47.1788}}.

\bibtype{Article}%
\bibitem{LHCb:2022qnv}
\bibinfo{author}{R. Aaij}, et al. (\bibinfo{collaboration}{LHCb}),
  \bibinfo{title}{{Test of lepton universality in $b \rightarrow s \ell^+
  \ell^-$ decays}}, \bibinfo{journal}{Phys. Rev. Lett.} \bibinfo{volume}{131}
  (\bibinfo{number}{5}) (\bibinfo{year}{2023}) \bibinfo{pages}{051803},
  \bibinfo{doi}{\doi{10.1103/PhysRevLett.131.051803}}.
  \href{http://arxiv.org/abs/2212.09152}{\tt arXiv:2212.09152}.

\bibtype{Article}%
\bibitem{Bordone:2017bld}
\bibinfo{author}{Marzia Bordone}, \bibinfo{author}{Claudia Cornella},
  \bibinfo{author}{Javier Fuentes-Martin}, \bibinfo{author}{Gino Isidori},
  \bibinfo{title}{{A three-site gauge model for flavor hierarchies and flavor
  anomalies}}, \bibinfo{journal}{Phys. Lett. B} \bibinfo{volume}{779}
  (\bibinfo{year}{2018}) \bibinfo{pages}{317--323},
  \bibinfo{doi}{\doi{10.1016/j.physletb.2018.02.011}}.
  \href{http://arxiv.org/abs/1712.01368}{\tt arXiv:1712.01368}.

\bibtype{Article}%
\bibitem{Davighi:2023iks}
\bibinfo{author}{Joe Davighi}, \bibinfo{author}{Gino Isidori},
  \bibinfo{title}{{Non-universal gauge interactions addressing the inescapable
  link between Higgs and flavour}}, \bibinfo{journal}{JHEP}
  \bibinfo{volume}{07} (\bibinfo{year}{2023}) \bibinfo{pages}{147},
  \bibinfo{doi}{\doi{10.1007/JHEP07(2023)147}}.
  \href{http://arxiv.org/abs/2303.01520}{\tt arXiv:2303.01520}.

\bibtype{Article}%
\bibitem{Davighi:2023evx}
\bibinfo{author}{Joe Davighi}, \bibinfo{author}{Ben~A. Stefanek},
  \bibinfo{title}{{Deconstructed hypercharge: a natural model of flavour}},
  \bibinfo{journal}{JHEP} \bibinfo{volume}{11} (\bibinfo{year}{2023})
  \bibinfo{pages}{100}, \bibinfo{doi}{\doi{10.1007/JHEP11(2023)100}}.
  \href{http://arxiv.org/abs/2305.16280}{\tt arXiv:2305.16280}.

\bibtype{Article}%
\bibitem{Davighi:2023xqn}
\bibinfo{author}{Joe Davighi}, \bibinfo{author}{Alastair Gosnay},
  \bibinfo{author}{David~J. Miller}, \bibinfo{author}{Sophie Renner},
  \bibinfo{title}{{Phenomenology of a Deconstructed Electroweak Force}},
  \bibinfo{journal}{JHEP} \bibinfo{volume}{05} (\bibinfo{year}{2024})
  \bibinfo{pages}{085}, \bibinfo{doi}{\doi{10.1007/JHEP05(2024)085}}.
  \href{http://arxiv.org/abs/2312.13346}{\tt arXiv:2312.13346}.

\bibtype{Article}%
\bibitem{Barbieri:2023qpf}
\bibinfo{author}{Riccardo Barbieri}, \bibinfo{author}{Gino Isidori},
  \bibinfo{title}{{Minimal flavour deconstruction}}, \bibinfo{journal}{JHEP}
  \bibinfo{volume}{05} (\bibinfo{year}{2024}) \bibinfo{pages}{033},
  \bibinfo{doi}{\doi{10.1007/JHEP05(2024)033}}.
  \href{http://arxiv.org/abs/2312.14004}{\tt arXiv:2312.14004}.

\bibtype{Article}%
\bibitem{Capdevila:2024gki}
\bibinfo{author}{Bernat Capdevila}, \bibinfo{author}{Andreas Crivellin},
  \bibinfo{author}{Javier~M. Lizana}, \bibinfo{author}{Stefan Pokorski},
  \bibinfo{title}{{SU(2)$_{L}$ deconstruction and flavour (non)-universality}},
  \bibinfo{journal}{JHEP} \bibinfo{volume}{08} (\bibinfo{year}{2024})
  \bibinfo{pages}{031}, \bibinfo{doi}{\doi{10.1007/JHEP08(2024)031}}.
  \href{http://arxiv.org/abs/2401.00848}{\tt arXiv:2401.00848}.

\bibtype{Article}%
\bibitem{Fuentes-Martin:2024fpx}
\bibinfo{author}{Javier Fuentes-Mart\'\i{}n}, \bibinfo{author}{Javier~M.
  Lizana}, \bibinfo{title}{{Deconstructing flavor anomalously}},
  \bibinfo{journal}{JHEP} \bibinfo{volume}{07} (\bibinfo{year}{2024})
  \bibinfo{pages}{117}, \bibinfo{doi}{\doi{10.1007/JHEP07(2024)117}}.
  \href{http://arxiv.org/abs/2402.09507}{\tt arXiv:2402.09507}.

\bibtype{Article}%
\bibitem{FernandezNavarro:2024hnv}
\bibinfo{author}{Mario Fern\'andez~Navarro}, \bibinfo{author}{Stephen~F. King},
  \bibinfo{author}{Avelino Vicente}, \bibinfo{title}{{Minimal complete
  tri-hypercharge theories of flavour}}, \bibinfo{journal}{JHEP}
  \bibinfo{volume}{07} (\bibinfo{year}{2024}) \bibinfo{pages}{147},
  \bibinfo{doi}{\doi{10.1007/JHEP07(2024)147}}.
  \href{http://arxiv.org/abs/2404.12442}{\tt arXiv:2404.12442}.

\bibtype{Article}%
\bibitem{Davighi:2022fer}
\bibinfo{author}{Joe Davighi}, \bibinfo{author}{Joseph Tooby-Smith},
  \bibinfo{title}{{Electroweak flavour unification}}, \bibinfo{journal}{JHEP}
  \bibinfo{volume}{09} (\bibinfo{year}{2022}) \bibinfo{pages}{193},
  \bibinfo{doi}{\doi{10.1007/JHEP09(2022)193}}.
  \href{http://arxiv.org/abs/2201.07245}{\tt arXiv:2201.07245}.

\bibtype{Article}%
\bibitem{Farina:2013mla}
\bibinfo{author}{Marco Farina}, \bibinfo{author}{Duccio Pappadopulo},
  \bibinfo{author}{Alessandro Strumia}, \bibinfo{title}{{A modified naturalness
  principle and its experimental tests}}, \bibinfo{journal}{JHEP}
  \bibinfo{volume}{08} (\bibinfo{year}{2013}) \bibinfo{pages}{022},
  \bibinfo{doi}{\doi{10.1007/JHEP08(2013)022}}.
  \href{http://arxiv.org/abs/1303.7244}{\tt arXiv:1303.7244}.

\bibtype{Article}%
\bibitem{Fuentes-Martin:2020bnh}
\bibinfo{author}{Javier Fuentes-Mart{\'\i}n}, \bibinfo{author}{Peter Stangl},
  \bibinfo{title}{{Third-family quark-lepton unification with a fundamental
  composite Higgs}}, \bibinfo{journal}{Phys. Lett. B} \bibinfo{volume}{811}
  (\bibinfo{year}{2020}) \bibinfo{pages}{135953},
  \bibinfo{doi}{\doi{10.1016/j.physletb.2020.135953}}.
  \href{http://arxiv.org/abs/2004.11376}{\tt arXiv:2004.11376}.

\bibtype{Article}%
\bibitem{Fuentes-Martin:2022xnb}
\bibinfo{author}{Javier Fuentes-Martin}, \bibinfo{author}{Gino Isidori},
  \bibinfo{author}{Javier~M. Lizana}, \bibinfo{author}{Nudzeim Selimovic},
  \bibinfo{author}{Ben~A. Stefanek}, \bibinfo{title}{{Flavor hierarchies,
  flavor anomalies, and Higgs mass from a warped extra dimension}},
  \bibinfo{journal}{Phys. Lett. B} \bibinfo{volume}{834} (\bibinfo{year}{2022})
  \bibinfo{pages}{137382}, \bibinfo{doi}{\doi{10.1016/j.physletb.2022.137382}}.
  \href{http://arxiv.org/abs/2203.01952}{\tt arXiv:2203.01952}.

\bibtype{Article}%
\bibitem{Covone:2024elw}
\bibinfo{author}{Sebastiano Covone}, \bibinfo{author}{Joe Davighi},
  \bibinfo{author}{Gino Isidori}, \bibinfo{author}{Marko Pesut},
  \bibinfo{title}{{Flavour deconstructing the composite Higgs}},
  \bibinfo{journal}{JHEP} \bibinfo{volume}{01} (\bibinfo{year}{2025})
  \bibinfo{pages}{041}, \bibinfo{doi}{\doi{10.1007/JHEP01(2025)041}}.
  \href{http://arxiv.org/abs/2407.10950}{\tt arXiv:2407.10950}.

\bibtype{Inbook}%
\bibitem{Feruglio:2017spp}
\bibinfo{author}{Ferruccio Feruglio}, \bibinfo{title}{{Are neutrino masses
  modular forms?}} \bibinfo{year}{2019} pp. \bibinfo{pages}{227--266},
  \bibinfo{doi}{\doi{10.1142/9789813238053_0012}}.
  \href{http://arxiv.org/abs/1706.08749}{\tt arXiv:1706.08749}.

\bibtype{Article}%
\bibitem{Ding:2023htn}
\bibinfo{author}{Gui-Jun Ding}, \bibinfo{author}{Stephen~F. King},
  \bibinfo{title}{{Neutrino mass and mixing with modular symmetry}},
  \bibinfo{journal}{Rept. Prog. Phys.} \bibinfo{volume}{87}
  (\bibinfo{number}{8}) (\bibinfo{year}{2024}) \bibinfo{pages}{084201},
  \bibinfo{doi}{\doi{10.1088/1361-6633/ad52a3}}.
  \href{http://arxiv.org/abs/2311.09282}{\tt arXiv:2311.09282}.

\bibtype{Article}%
\bibitem{Ding:2020zxw}
\bibinfo{author}{Gui-Jun Ding}, \bibinfo{author}{Ferruccio Feruglio},
  \bibinfo{author}{Xiang-Gan Liu}, \bibinfo{title}{{Automorphic Forms and
  Fermion Masses}}, \bibinfo{journal}{JHEP} \bibinfo{volume}{01}
  (\bibinfo{year}{2021}) \bibinfo{pages}{037},
  \bibinfo{doi}{\doi{10.1007/JHEP01(2021)037}}.
  \href{http://arxiv.org/abs/2010.07952}{\tt arXiv:2010.07952}.

\bibtype{Article}%
\bibitem{Qu:2024rns}
\bibinfo{author}{Bu-Yao Qu}, \bibinfo{author}{Gui-Jun Ding},
  \bibinfo{title}{{Non-holomorphic modular flavor symmetry}},
  \bibinfo{journal}{JHEP} \bibinfo{volume}{08} (\bibinfo{year}{2024})
  \bibinfo{pages}{136}, \bibinfo{doi}{\doi{10.1007/JHEP08(2024)136}}.
  \href{http://arxiv.org/abs/2406.02527}{\tt arXiv:2406.02527}.

\bibtype{Article}%
\bibitem{Ding:2024inn}
\bibinfo{author}{Gui-Jun Ding}, \bibinfo{author}{Jun-Nan Lu},
  \bibinfo{author}{S.~T. Petcov}, \bibinfo{author}{Bu-Yao Qu},
  \bibinfo{title}{{Non-holomorphic modular S$_{4}$ lepton flavour models}},
  \bibinfo{journal}{JHEP} \bibinfo{volume}{01} (\bibinfo{year}{2025})
  \bibinfo{pages}{191}, \bibinfo{doi}{\doi{10.1007/JHEP01(2025)191}}.
  \href{http://arxiv.org/abs/2408.15988}{\tt arXiv:2408.15988}.

\bibtype{Article}%
\bibitem{Okada:2020ukr}
\bibinfo{author}{Hiroshi Okada}, \bibinfo{author}{Morimitsu Tanimoto},
  \bibinfo{title}{{Modular invariant flavor model of $A_4$ and hierarchical
  structures at nearby fixed points}}, \bibinfo{journal}{Phys. Rev. D}
  \bibinfo{volume}{103} (\bibinfo{number}{1}) (\bibinfo{year}{2021})
  \bibinfo{pages}{015005}, \bibinfo{doi}{\doi{10.1103/PhysRevD.103.015005}}.
  \href{http://arxiv.org/abs/2009.14242}{\tt arXiv:2009.14242}.

\bibtype{Article}%
\bibitem{Feruglio:2021dte}
\bibinfo{author}{Ferruccio Feruglio}, \bibinfo{author}{Valerio Gherardi},
  \bibinfo{author}{Andrea Romanino}, \bibinfo{author}{Arsenii Titov},
  \bibinfo{title}{{Modular invariant dynamics and fermion mass hierarchies
  around $\tau = i$}}, \bibinfo{journal}{JHEP} \bibinfo{volume}{05}
  (\bibinfo{year}{2021}) \bibinfo{pages}{242},
  \bibinfo{doi}{\doi{10.1007/JHEP05(2021)242}}.
  \href{http://arxiv.org/abs/2101.08718}{\tt arXiv:2101.08718}.

\bibtype{Article}%
\bibitem{Novichkov:2021evw}
\bibinfo{author}{P.~P. Novichkov}, \bibinfo{author}{J.~T. Penedo},
  \bibinfo{author}{S.~T. Petcov}, \bibinfo{title}{{Fermion mass hierarchies,
  large lepton mixing and residual modular symmetries}},
  \bibinfo{journal}{JHEP} \bibinfo{volume}{04} (\bibinfo{year}{2021})
  \bibinfo{pages}{206}, \bibinfo{doi}{\doi{10.1007/JHEP04(2021)206}}.
  \href{http://arxiv.org/abs/2102.07488}{\tt arXiv:2102.07488}.

\bibtype{Article}%
\bibitem{Chen:2025tby}
\bibinfo{author}{Mu-Chun Chen}, \bibinfo{author}{Xueqi Li},
  \bibinfo{author}{Xiang-Gan Liu}, \bibinfo{author}{Michael Ratz},
  \bibinfo{title}{{Modular Flavor Symmetries and Fermion Mass Hierarchies}}
  (\bibinfo{year}{2025}). \href{http://arxiv.org/abs/2506.23343}{\tt
  arXiv:2506.23343}.

\bibtype{Article}%
\bibitem{King:2020qaj}
\bibinfo{author}{Simon J.~D. King}, \bibinfo{author}{Stephen~F. King},
  \bibinfo{title}{{Fermion mass hierarchies from modular symmetry}},
  \bibinfo{journal}{JHEP} \bibinfo{volume}{09} (\bibinfo{year}{2020})
  \bibinfo{pages}{043}, \bibinfo{doi}{\doi{10.1007/JHEP09(2020)043}}.
  \href{http://arxiv.org/abs/2002.00969}{\tt arXiv:2002.00969}.

\bibtype{Article}%
\bibitem{Chen:2019ewa}
\bibinfo{author}{Mu-Chun Chen}, \bibinfo{author}{Sa\'ul Ramos-S\'anchez},
  \bibinfo{author}{Michael Ratz}, \bibinfo{title}{{A note on the predictions of
  models with modular flavor symmetries}}, \bibinfo{journal}{Phys. Lett. B}
  \bibinfo{volume}{801} (\bibinfo{year}{2020}) \bibinfo{pages}{135153},
  \bibinfo{doi}{\doi{10.1016/j.physletb.2019.135153}}.
  \href{http://arxiv.org/abs/1909.06910}{\tt arXiv:1909.06910}.

\bibtype{Article}%
\bibitem{Lu:2019vgm}
\bibinfo{author}{Jun-Nan Lu}, \bibinfo{author}{Xiang-Gan Liu},
  \bibinfo{author}{Gui-Jun Ding}, \bibinfo{title}{{Modular symmetry origin of
  texture zeros and quark lepton unification}}, \bibinfo{journal}{Phys. Rev. D}
  \bibinfo{volume}{101} (\bibinfo{number}{11}) (\bibinfo{year}{2020})
  \bibinfo{pages}{115020}, \bibinfo{doi}{\doi{10.1103/PhysRevD.101.115020}}.
  \href{http://arxiv.org/abs/1912.07573}{\tt arXiv:1912.07573}.

\bibtype{Article}%
\bibitem{Okada:2019uoy}
\bibinfo{author}{Hiroshi Okada}, \bibinfo{author}{Morimitsu Tanimoto},
  \bibinfo{title}{{Towards unification of quark and lepton flavors in $A_4$
  modular invariance}}, \bibinfo{journal}{Eur. Phys. J. C} \bibinfo{volume}{81}
  (\bibinfo{number}{1}) (\bibinfo{year}{2021}) \bibinfo{pages}{52},
  \bibinfo{doi}{\doi{10.1140/epjc/s10052-021-08845-y}}.
  \href{http://arxiv.org/abs/1905.13421}{\tt arXiv:1905.13421}.

\bibtype{Article}%
\bibitem{Liu:2020akv}
\bibinfo{author}{Xiang-Gan Liu}, \bibinfo{author}{Chang-Yuan Yao},
  \bibinfo{author}{Gui-Jun Ding}, \bibinfo{title}{{Modular invariant quark and
  lepton models in double covering of $S_4$ modular group}},
  \bibinfo{journal}{Phys. Rev. D} \bibinfo{volume}{103} (\bibinfo{number}{5})
  (\bibinfo{year}{2021}) \bibinfo{pages}{056013},
  \bibinfo{doi}{\doi{10.1103/PhysRevD.103.056013}}.
  \href{http://arxiv.org/abs/2006.10722}{\tt arXiv:2006.10722}.

\bibtype{Article}%
\bibitem{Baur:2019kwi}
\bibinfo{author}{Alexander Baur}, \bibinfo{author}{Hans~Peter Nilles},
  \bibinfo{author}{Andreas Trautner}, \bibinfo{author}{Patrick K.~S.
  Vaudrevange}, \bibinfo{title}{{Unification of Flavor, CP, and Modular
  Symmetries}}, \bibinfo{journal}{Phys. Lett. B} \bibinfo{volume}{795}
  (\bibinfo{year}{2019}) \bibinfo{pages}{7--14},
  \bibinfo{doi}{\doi{10.1016/j.physletb.2019.03.066}}.
  \href{http://arxiv.org/abs/1901.03251}{\tt arXiv:1901.03251}.

\bibtype{Article}%
\bibitem{Baur:2019iai}
\bibinfo{author}{Alexander Baur}, \bibinfo{author}{Hans~Peter Nilles},
  \bibinfo{author}{Andreas Trautner}, \bibinfo{author}{Patrick K.~S.
  Vaudrevange}, \bibinfo{title}{{A String Theory of Flavor and $CP$}},
  \bibinfo{journal}{Nucl. Phys. B} \bibinfo{volume}{947} (\bibinfo{year}{2019})
  \bibinfo{pages}{114737},
  \bibinfo{doi}{\doi{10.1016/j.nuclphysb.2019.114737}}.
  \href{http://arxiv.org/abs/1908.00805}{\tt arXiv:1908.00805}.

\bibtype{Article}%
\bibitem{Nilles:2020nnc}
\bibinfo{author}{Hans~Peter Nilles}, \bibinfo{author}{Sa\'ul Ramos-S\'anchez},
  \bibinfo{author}{Patrick K.~S. Vaudrevange}, \bibinfo{title}{{Eclectic Flavor
  Groups}}, \bibinfo{journal}{JHEP} \bibinfo{volume}{02} (\bibinfo{year}{2020})
  \bibinfo{pages}{045}, \bibinfo{doi}{\doi{10.1007/JHEP02(2020)045}}.
  \href{http://arxiv.org/abs/2001.01736}{\tt arXiv:2001.01736}.

\bibtype{Article}%
\bibitem{Nilles:2020kgo}
\bibinfo{author}{Hans~Peter Nilles}, \bibinfo{author}{Saul Ramos-Sanchez},
  \bibinfo{author}{Patrick K.~S. Vaudrevange}, \bibinfo{title}{{Lessons from
  eclectic flavor symmetries}}, \bibinfo{journal}{Nucl. Phys. B}
  \bibinfo{volume}{957} (\bibinfo{year}{2020}) \bibinfo{pages}{115098},
  \bibinfo{doi}{\doi{10.1016/j.nuclphysb.2020.115098}}.
  \href{http://arxiv.org/abs/2004.05200}{\tt arXiv:2004.05200}.

\bibtype{Article}%
\bibitem{Baur:2022hma}
\bibinfo{author}{Alexander Baur}, \bibinfo{author}{Hans~Peter Nilles},
  \bibinfo{author}{Saul Ramos-Sanchez}, \bibinfo{author}{Andreas Trautner},
  \bibinfo{author}{Patrick K.~S. Vaudrevange}, \bibinfo{title}{{The first
  string-derived eclectic flavor model with realistic phenomenology}},
  \bibinfo{journal}{JHEP} \bibinfo{volume}{09} (\bibinfo{year}{2022})
  \bibinfo{pages}{224}, \bibinfo{doi}{\doi{10.1007/JHEP09(2022)224}}.
  \href{http://arxiv.org/abs/2207.10677}{\tt arXiv:2207.10677}.

\bibtype{Article}%
\bibitem{Feruglio:2024ytl}
\bibinfo{author}{Ferruccio Feruglio}, \bibinfo{author}{Matteo Parriciatu},
  \bibinfo{author}{Alessandro Strumia}, \bibinfo{author}{Arsenii Titov},
  \bibinfo{title}{{Solving the strong CP problem without axions}},
  \bibinfo{journal}{JHEP} \bibinfo{volume}{08} (\bibinfo{year}{2024})
  \bibinfo{pages}{214}, \bibinfo{doi}{\doi{10.1007/JHEP08(2024)214}}.
  \href{http://arxiv.org/abs/2406.01689}{\tt arXiv:2406.01689}.

\bibtype{Article}%
\bibitem{Gaiotto:2014kfa}
\bibinfo{author}{Davide Gaiotto}, \bibinfo{author}{Anton Kapustin},
  \bibinfo{author}{Nathan Seiberg}, \bibinfo{author}{Brian Willett},
  \bibinfo{title}{{Generalized Global Symmetries}}, \bibinfo{journal}{JHEP}
  \bibinfo{volume}{02} (\bibinfo{year}{2015}) \bibinfo{pages}{172},
  \bibinfo{doi}{\doi{10.1007/JHEP02(2015)172}}.
  \href{http://arxiv.org/abs/1412.5148}{\tt arXiv:1412.5148}.

\bibtype{Article}%
\bibitem{Brennan:2023mmt}
\bibinfo{author}{T.~Daniel Brennan}, \bibinfo{author}{Sungwoo Hong},
  \bibinfo{title}{{Introduction to Generalized Global Symmetries in QFT and
  Particle Physics}}  (\bibinfo{year}{2023}).
  \href{http://arxiv.org/abs/2306.00912}{\tt arXiv:2306.00912}.

\bibtype{Article}%
\bibitem{Cordova:2022ieu}
\bibinfo{author}{Clay Cordova}, \bibinfo{author}{Kantaro Ohmori},
  \bibinfo{title}{{Noninvertible Chiral Symmetry and Exponential Hierarchies}},
  \bibinfo{journal}{Phys. Rev. X} \bibinfo{volume}{13} (\bibinfo{number}{1})
  (\bibinfo{year}{2023}) \bibinfo{pages}{011034},
  \bibinfo{doi}{\doi{10.1103/PhysRevX.13.011034}}.
  \href{http://arxiv.org/abs/2205.06243}{\tt arXiv:2205.06243}.

\bibtype{Article}%
\bibitem{Cordova:2022fhg}
\bibinfo{author}{Clay Cordova}, \bibinfo{author}{Sungwoo Hong},
  \bibinfo{author}{Seth Koren}, \bibinfo{author}{Kantaro Ohmori},
  \bibinfo{title}{{Neutrino Masses from Generalized Symmetry Breaking}},
  \bibinfo{journal}{Phys. Rev. X} \bibinfo{volume}{14} (\bibinfo{number}{3})
  (\bibinfo{year}{2024}) \bibinfo{pages}{031033},
  \bibinfo{doi}{\doi{10.1103/PhysRevX.14.031033}}.
  \href{http://arxiv.org/abs/2211.07639}{\tt arXiv:2211.07639}.

\bibtype{Article}%
\bibitem{Kobayashi:2024yqq}
\bibinfo{author}{Tatsuo Kobayashi}, \bibinfo{author}{Hajime Otsuka},
  \bibinfo{title}{{Non-invertible flavor symmetries in magnetized extra
  dimensions}}, \bibinfo{journal}{JHEP} \bibinfo{volume}{11}
  (\bibinfo{year}{2024}) \bibinfo{pages}{120},
  \bibinfo{doi}{\doi{10.1007/JHEP11(2024)120}}.
  \href{http://arxiv.org/abs/2408.13984}{\tt arXiv:2408.13984}.

\bibtype{Article}%
\bibitem{Kobayashi:2024cvp}
\bibinfo{author}{Tatsuo Kobayashi}, \bibinfo{author}{Hajime Otsuka},
  \bibinfo{author}{Morimitsu Tanimoto}, \bibinfo{title}{{Yukawa textures from
  non-invertible symmetries}}, \bibinfo{journal}{JHEP} \bibinfo{volume}{12}
  (\bibinfo{year}{2024}) \bibinfo{pages}{117},
  \bibinfo{doi}{\doi{10.1007/JHEP12(2024)117}}.
  \href{http://arxiv.org/abs/2409.05270}{\tt arXiv:2409.05270}.

\bibtype{Article}%
\bibitem{Cordova:2022qtz}
\bibinfo{author}{Clay Cordova}, \bibinfo{author}{Seth Koren},
  \bibinfo{title}{{Higher Flavor Symmetries in the Standard Model}},
  \bibinfo{journal}{Annalen Phys.} \bibinfo{volume}{535} (\bibinfo{number}{8})
  (\bibinfo{year}{2023}) \bibinfo{pages}{2300031},
  \bibinfo{doi}{\doi{10.1002/andp.202300031}}.
  \href{http://arxiv.org/abs/2212.13193}{\tt arXiv:2212.13193}.

\bibtype{Inproceedings}%
\bibitem{Grossman:2017thq}
\bibinfo{author}{Yuval Grossman}, \bibinfo{author}{Philip Tanedo},
  \bibinfo{title}{{Just a taste: lectures on flavor physics.}}, in:
  \bibinfo{booktitle}{{Theoretical Advanced Study Institute in Elementary
  Particle Physics}: {Anticipating the Next Discoveries in Particle Physics}}
  \bibinfo{year}{2018}, pp. \bibinfo{pages}{109--295},
  \bibinfo{doi}{\doi{10.1142/9789813233348_0004}}.
  \href{http://arxiv.org/abs/1711.03624}{\tt arXiv:1711.03624}.

\bibtype{Article}%
\bibitem{Zupan:2019uoi}
\bibinfo{author}{Jure Zupan}, \bibinfo{title}{{Introduction to flavour
  physics}}, \bibinfo{journal}{CERN Yellow Rep. School Proc.}
  \bibinfo{volume}{6} (\bibinfo{year}{2019}) \bibinfo{pages}{181--212},
  \bibinfo{doi}{\doi{10.23730/CYRSP-2019-006.181}}.
  \href{http://arxiv.org/abs/1903.05062}{\tt arXiv:1903.05062}.

\bibtype{Article}%
\bibitem{Silvestrini:2019sey}
\bibinfo{author}{Luca Silvestrini}, \bibinfo{title}{{Effective Theories for
  Quark Flavour Physics}}  (\bibinfo{year}{2019}),
  \bibinfo{doi}{\doi{10.1093/oso/9780198855743.003.0008}}.
  \href{http://arxiv.org/abs/1905.00798}{\tt arXiv:1905.00798}.

\bibtype{Article}%
\bibitem{Gori:2019ybw}
\bibinfo{author}{Stefania Gori}, \bibinfo{title}{{TASI lectures on flavor
  physics}}, \bibinfo{journal}{PoS} \bibinfo{volume}{TASI2018}
  (\bibinfo{year}{2019}) \bibinfo{pages}{013}.

\bibtype{Article}%
\bibitem{Altmannshofer:2024ykf}
\bibinfo{author}{Wolfgang Altmannshofer}, \bibinfo{title}{{TASI 2022 lectures
  on flavor physics}}, \bibinfo{journal}{PoS} \bibinfo{volume}{TASI2022}
  (\bibinfo{year}{2024}) \bibinfo{pages}{001},
  \bibinfo{doi}{\doi{10.22323/1.439.0001}}.

\bibtype{Inproceedings}%
\bibitem{Isidori:2025iyu}
\bibinfo{author}{Gino Isidori}, \bibinfo{title}{{Flavour Physics and CP
  Violation}} \bibinfo{year}{2025}. \href{http://arxiv.org/abs/2503.14042}{\tt
  arXiv:2503.14042}.

\end{thebibliography*}

\end{document}